\documentclass[12pt]{article}

\usepackage{subfigure}
\usepackage{amssymb}
\usepackage{amsfonts}
\usepackage{bm}
\usepackage{color}
\usepackage{calc}
\usepackage[latin1]{inputenc}
\usepackage{amsfonts}
\usepackage{amsmath}
\usepackage[english]{babel}
\usepackage[T1]{fontenc}
\usepackage{verbatim}
\usepackage{subfig}
\usepackage{enumerate}
\usepackage{graphicx}
\usepackage{natbib}
\usepackage{url}
\usepackage{algorithmic,algorithm}

\def\Bmp#1{ \begin{minipage}{#1} }
\def\Bmpc#1{ \begin{minipage}[c]{#1} }
\def\Bmpt#1{ \begin{minipage}[t]{#1} }
\def\Bmpb#1{ \begin{minipage}[b]{#1} }
\def\Emp{ \end{minipage} }

\newcommand{\B}{\mathcal{B}}

\newcommand{\W}{\mathcal{W}}

\def\E{{\mathcal{E}}}

\def\H{{\mathcal{H}}}

\def\O{\mbox{\textit{O}}}
\def\P{{\mathcal{P}}}
\def\R{{\mathcal{R}}}

\def\K{{\mathcal{K}}}

\def\tf0{\tilde{\varphi}_{0}}

\def\RR{{\mathbb{R}}}

\def\u{{\bf u}}
\def\0{{\bf 0}}

\def\bnabla{\boldsymbol{\nabla}}

\newcommand{\Avec}{\mathbf{A}}
\newcommand{\nvec}{\mathbf{n}}
\newcommand{\xvec}{\mathbf{x}}
\newcommand{\uvec}{\mathbf{u}}
\newcommand{\kvec}{\mathbf{k}}
\newcommand{\vvec}{\mathbf{v}}

\newcommand{\wvec}{\bm{\omega}}
\newcommand{\laplacian}{\Delta}
\newcommand{\rot}{\bnabla\times}
\newcommand{\tuvec}{\widetilde{\mathbf{u}}}

\newcommand{\tuvecE}{\widetilde{\mathbf{u}}_{\E_0}}

\newcommand{\tuvecKE}{\widetilde{\mathbf{u}}_{\K_0,\E_0}}
\newcommand{\M}[1]{\mathcal{S}_{#1}}
\newcommand{\e}[1]{\textrm{e}^{#1}}

\newcommand{\tr}{\operatorname{tr}}

\newcommand{\argmax}{\operatorname{argmax}}

\newcommand{\Id}{\operatorname{Id}}
\newcommand{\Ker}{\operatorname{Ker}}

\renewcommand{\Re}[1]{#1_{\textrm{Re}}}
\renewcommand{\Im}[1]{#1_{\textrm{Im}}}

\newtheorem{problem}{Problem}[section]
\newtheorem{definition}{Definition}[section]

\begin{document}
\title{Extreme Vortex States and the Growth of Enstrophy in 3D Incompressible Flows}

\author{Diego Ayala$^{1,2}$ and Bartosz Protas$^{2,}$\thanks{Email address for correspondence: bprotas@mcmaster.ca}
\\ \\ 
$^1$ Department of Mathematics, University of Michigan, \\ 
Ann Arbor, MI  48109, USA
\\ \\
$^2$ Department of Mathematics and Statistics, McMaster University \\
Hamilton, Ontario, L8S 4K1, Canada
}

\date{\today}
\maketitle

\begin{abstract}
  In this investigation we study extreme vortex states defined as
  incompressible velocity fields with prescribed enstrophy $\E_0$
  which maximize the instantaneous rate of growth of enstrophy
  $d\E/dt$.  We provide {an analytic} characterization of these
  extreme vortex states in the limit of vanishing enstrophy $\E_0$
  and, in particular, show that the Taylor-Green vortex is in fact a
  local maximizer of $d\E / dt$ {in this limit}. For finite
  values of enstrophy, the extreme vortex states are computed
  numerically by solving a constrained variational optimization
  problem using a suitable gradient method.  In combination with a
  continuation approach, this allows us to construct an entire family
  of maximizing vortex states parameterized by their enstrophy. We
  also confirm the findings of the seminal study by \cite{ld08} that
  these extreme vortex states saturate (up to a numerical prefactor)
  the fundamental bound $d\E / dt < C \, \E^3$, for some constant $C >
  0$.  The time evolution corresponding to these extreme vortex states
  leads to a larger growth of enstrophy than the growth achieved by
  any of the commonly used initial conditions with the same enstrophy
  $\E_0$.  However, based on several different diagnostics, there is
  no evidence of any tendency towards singularity formation in finite
  time.  Finally, we discuss possible physical reasons why the
  initially large growth of enstrophy is not sustained for longer
  times.
\end{abstract}

\begin{flushleft}
\end{flushleft}


\section{Introduction}
\label{sec:intro}

The objective of this investigation is to study three-dimensional (3D)
flows of viscous incompressible fluids which are constructed to
exhibit extreme growth of enstrophy. It is motivated by the question
whether the solutions to the 3D incompressible Navier-Stokes system on
unbounded or periodic domains corresponding to smooth initial data may
develop a singularity in finite time \citep{d09}. By formation of a
``singularity'' we mean the situation when some norms of the solution
corresponding to smooth initial data have become unbounded after a
finite time. This so-called ``blow-up problem'' is one of the key open
questions in mathematical fluid mechanics and, in fact, its importance
for mathematics in general has been recognized by the Clay Mathematics
Institute as one of its ``millennium problems'' \citep{f00}. 
Questions concerning global-in-time existence of smooth solutions remain open
also for a number of other flow models including the 3D Euler
equations \citep{gbk08} and some of the ``active scalar'' equations
\citep{k10}.

While the blow-up problem is fundamentally a question in
mathematical analysis, a lot of computational studies have been
carried out since the mid-'90s in order to shed light on the
hydrodynamic mechanisms which might lead to singularity formation in
finite time. Given that such flows evolving near the edge of
regularity involve formation of very small flow structures, these
computations typically require the use of state-of-the-art
computational resources available at a given time. The computational
studies focused on the possibility of finite-time blow-up in the 3D
Navier-Stokes and/or Euler system include \cite{bmonmu83,ps90,b91,k93,p01,bk08,oc08,o08,ghdg08, gbk08,h09,opc12,bb12,opmc14},
all of which considered problems defined on domains periodic in all
three dimensions. Recent investigations by
\cite{dggkpv13,k13,gdgkpv14,k13b} focused on the time evolution of
vorticity moments and compared it with the predictions derived from
analysis based on rigorous bounds.  We also mention the studies by
\cite{mbf08} and \cite{sc09}, along with the references found therein,
in which various complexified forms of the Euler equation were
investigated. The idea of this approach is that, since the solutions
to complexified equations have singularities in the complex plane,
singularity formation in the real-valued problem is manifested by the
collapse of the complex-plane singularities onto the real axis.

Overall, the outcome of these investigations is rather inconclusive:
while for the Navier-Stokes flows most of recent computations do not
offer support for finite-time blow-up, the evidence appears split in
the case of the Euler system.  In particular, the recent studies by
\cite{bb12} and \cite{opc12} hinted at the possibility of singularity
formation in finite time. In this connection we also mention the
recent investigations by \cite{lh14a,lh14b} in which blow-up was
observed in axisymmetric Euler flows in a bounded (tubular) domain.

A common feature of all of the aforementioned investigations was that
the initial data for the Navier-Stokes or Euler system was chosen in
an ad-hoc manner, based on some heuristic arguments. On the other
hand, in the present study we pursue a fundamentally different
approach, proposed originally by \cite{ld08} and employed also by
\cite{ap11a,ap13a,ap13b} for a range of related problems, in which the
initial data leading to the most singular behaviour is sought
systematically via solution of a suitable variational optimization
problem. We carefully analyze the time evolution induced by the
extreme vortex states first identified by \cite{ld08} and compare it
to the time evolution corresponding to a number of other candidate
initial conditions considered in the literature
\citep{bmonmu83,k93,p01,cb05,opc12}. We demonstrate that the
Taylor-Green vortex, studied in the context of the blow-up problem by
\cite{tg37,bmonmu83,b91,cb05}, is in fact a particular member of the
family of extreme vortex states maximizing the instantaneous rate of
enstrophy production in the limit of vanishing enstrophy.  In
addition, based on these findings, we identify the set of initial
data, parameterized by its energy and enstrophy, for which one can a
priori guarantee global-in-time existence of smooth solutions. This
result therefore offers a physically appealing interpretation
  of an ``abstract'' mathematical theorem concerning global existence
of classical solutions corresponding to ``small'' initial data
\citep{lady69}. We also emphasize that, in order to establish a
direct link with the results of the mathematical analysis
discussed below, in our investigation we follow a rather different
strategy than in most of the studies referenced above.  While these
earlier studies relied on data from a relatively small number of
simulations performed at a high (at the given time) resolution, in the
present investigation we explore a broad range of cases, each of which
is however computed at a more moderate resolution (or, equivalently,
Reynolds number).  With such an approach to the use of available
computational resources, we are able to reveal trends resulting from
the variation of parameters which otherwise would be hard to detect.
Systematic computations conducted in this way thus allow us to probe
the sharpness of the mathematical analysis relevant to the problem.

The question of regularity of solution to the Navier-Stokes system is
{usually addressed using ``energy'' methods which rely on finding
  upper bounds (with respect to time) on certain quantities of
  interest, typically taken as suitable Sobolev norms of the solution.
  A key intermediate step is obtaining bounds on the rate of growth of
  the quantity of interest, a problem which can be studied with ODE
  methods.}  While {for the Navier-Stokes system} different norms
of the velocity gradient or vorticity can be used to study the
regularity of solutions, the use of enstrophy $\E$ (see equation
\eqref{eq:EnsDef_3D} {below}) is privileged by the well-known
result of \citet{ft89}, where it was established that if the uniform
bound
\begin{equation}\label{eq:RegCrit_FoiasTemam}
\mathop{\sup}_{0 \leq t \leq T} \E(\u(t))  < \infty
\end{equation}
holds, then the regularity of the solution $\u(t)$ is guaranteed up to
time $T$ (to be precise, the solution remains in a suitable
Gevrey class). From the computational point of view, the enstrophy
$\E(t) := \E(\u(t))$ is thus a convenient indicator of the regularity
of solutions, because in the light of \eqref{eq:RegCrit_FoiasTemam},
singularity formation must manifest itself by the enstrophy becoming
infinite. 

While characterization of the maximum possible finite-time growth of
enstrophy in the 3D Navier-Stokes flows is the ultimate objective of
this research program, analogous questions can also be posed in the
context of more tractable problems involving the one-dimensional (1D)
Burgers equation and the two-dimensional (2D) Navier-Stokes equation.
Although global-in-time existence of the classical (smooth) solutions
is well known for both these problems \citep{kl04}, questions
concerning the sharpness of the corresponding estimates for the
instantaneous and finite-time growth of various quantities are
relevant, because these estimates are obtained using essentially the
same methods as {employed to derive} their 3D counterparts. Since
in 2D flows on unbounded or periodic domains the enstrophy may not
increase ($d\E/dt \leq 0$), the relevant quantity in this case is the
palinstrophy $\P(\uvec) := \frac{1}{2}\int_\Omega |
\bnabla\wvec(\xvec,t) |^2 \,d\xvec$, where $\wvec :=\rot\uvec$ is the
vorticity (which reduces to a pseudo-scalar in 2D). Different
questions concerning sharpness of estimates addressed in our research
program are summarized together with the results obtained to date in
Table \ref{tab:estimates}. We remark that the best finite-time
estimate for the 1D Burgers equation was found {\em not} to be sharp
using the initial data obtained from both the instantaneous and the
finite-time variational optimization problems \citep{ap11a}. On the
other hand, in 2D the bounds on both the \emph{instantaneous} and
\emph{finite-time} growth of palinstrophy were found to be sharp and,
somewhat surprisingly, both estimates were realized by the same family
of incompressible vector fields parameterized by energy $\K$ and
palinstrophy $\P$, obtained as the solution of an {\em instantaneous}
optimization problem \citep{ap13a}.  It is worth mentioning that while
the estimate for the instantaneous rate of growth of palinstrophy
$d\P/dt \leq C\K^{1/2}\P^{3/2}/\nu$ (see Table \ref{tab:estimates})
was found to be sharp with respect to variations in palinstrophy, the
estimate is in fact not sharp with respect to the prefactor
$C_{\uvec,\nu} = \K^{1/2}/\nu$ \citep{ados16}, with the correct
prefactor being of the form $\widetilde{C}_{\uvec,\nu} =
\sqrt{\log\left(\K^{1/2}/\nu\right)}$. We add that what distinguishes
the 2D problem, in regard to both the instantaneous and finite-time
bounds, is that the RHS of these bounds are expressed in terms of two
quantities, namely, energy $\K$ and enstrophy $\E$, in contrast to the
enstrophy alone appearing in the 1D and 3D estimates. As a result, the
2D instantaneous optimization problem had to be solved subject to {\em
  two} constraints.

In the present investigation we advance the research program 
summarized in Table \ref{tab:estimates} by assessing to what extent 
the finite-time growth of enstrophy predicted by the analytic estimates 
\eqref{eq:Et_estimate_E0} and \eqref{eq:Evs_t_fixE} can be 
actually realized by flow evolution starting from different initial 
conditions, including the extreme vortex states found by \cite{ld08} 
to saturate the instantaneous estimate \eqref{eq:dEdt_estimate_E}. 
The key finding is that, at least for the range of modest enstrophy 
values we considered, the growth of enstrophy corresponding to 
this initial data, which has the form of two colliding axisymmetric 
vortex rings, is rapidly depleted and there is no indication of 
singularity formation in finite time. Thus, should finite-time 
singularity be possible in the Navier-Stokes system, it is 
unlikely to result from initial conditions instantaneously 
maximizing the rate of growth of enstrophy. We also provide 
a comprehensive characterization of the extreme vortex states 
which realize estimate \eqref{eq:dEdt_estimate_E} together 
with the resulting flow evolutions. 

The structure of the paper is as follows: in the next section we
present analytic estimates {on} the instantaneous and finite-time
growth of enstrophy in 3D flows. In \S \ref{sec:3D_InstOpt} we
formulate the variational optimization problems which will be solved
to find the vortex states with the largest rate of enstrophy
production and in \S \ref{sec:3D_InstOpt_E0to0} we provide an
asymptotic representation for these optimal states in the limit of
vanishing enstrophy. In \S \ref{sec:3D_InstOpt_E} we present
numerically computed extreme vortex states corresponding to
intermediate and large enstrophy values, while in \S
\ref{sec:timeEvolution} we analyze the temporal evolution
corresponding to different initial data in order to compare it with
the predictions of estimates \eqref{eq:Et_estimate_E0} and
\eqref{eq:Evs_t_fixE}. Our findings are discussed in \S
\ref{sec:discuss}, whereas conclusions and outlook are deferred to \S
\ref{sec:final}.

\begin{table}
  \begin{center}
    \hspace*{-1.1cm}
    \begin{tabular}{l|c|c}      
      &  \Bmp{3.0cm} \small \begin{center} {\sc Estimate} \\ \smallskip \end{center} \Emp   
      & \Bmp{3.5cm} \small \begin{center} {\sc Realizability }  \end{center} \Emp \\  
      \hline
      \Bmp{2.5cm}  \small {\begin{center} \smallskip 1D Burgers  \\ instantaneous \smallskip \end{center}} \Emp &  
      \small {$\frac{d\E}{dt} \leq \frac{3}{2}\left(\frac{1}{\pi^2\nu}\right)^{1/3}\E^{5/3}$}  & 
      \Bmp{3.5cm} \footnotesize {\begin{center} \smallskip {\sc Yes} \\ \citep{ld08}  \smallskip  \end{center}} \Emp \\ 
      \hline 
      \Bmp{3.0cm} \small {\begin{center} \smallskip 1D Burgers  \\ finite-time \smallskip \end{center}} \Emp &  
      \small {$\max_{t \in [0,T]} \E(t) \leq \left[\E_0^{1/3} + \frac{1}{16}\left(\frac{1}{\pi^2 \nu}\right)^{4/3}\E_0\right]^{3}$} &  
      \Bmp{3.5cm} \small {\begin{center} \smallskip {\sc No} \\ \citep{ap11a} \smallskip  \end{center}} \Emp \\ 
      \hline 
      \Bmp{3.0cm} \small {\begin{center} \smallskip 2D Navier-Stokes  \\ instantaneous \smallskip\end{center}} \Emp &  
      \Bmp{7.0cm} \smallskip \centering \small $\frac{d\P(t)}{dt}  \le -\nu\frac{\P^2}{\E} + \frac{C_1}{\nu} \E\,\P$ \\ \smallskip $\frac{d\P(t)}{dt} \le \frac{C_2}{\nu} \K^{1/2}\P^{3/2}$ \smallskip \Emp& 
      \Bmp{3.5cm} \small {\begin{center} \smallskip {\sc Yes} \\ \citep{ap13a}  \smallskip  \end{center}} \Emp  \\ 
      \hline 
      \Bmp{3.0cm} \small \begin{center} \smallskip 2D Navier-Stokes \\ finite-time \smallskip\end{center} \Emp &  
      \Bmp{7.0cm} \smallskip \centering \small $\max_{t>0} \P(t) \le \P_0 + \frac{C_1}{2\nu^2}\E_0^2$ \\ \smallskip $\max_{t>0} \P(t) \le \left(\P_0^{1/2} + \frac{C_2}{4\nu^2}\K_0^{1/2}\E_0\right)^2$ \smallskip \Emp  & 
      \Bmp{3.5cm} \small {\begin{center} \smallskip {\sc Yes} \\ \citep{ap13a}  \smallskip  \end{center}} \Emp \\ 
      \hline 
      \Bmp{3.0cm} \small {\begin{center} \smallskip 3D Navier-Stokes  \\ instantaneous  \smallskip \end{center}} \Emp &  
      \small {$\frac{d\E(t)}{dt} \le \frac{27}{8\pi^4 \nu^3} \E(t)^3$} & \Bmp{3.5cm} \small {\begin{center} \smallskip {\sc Yes} \\ \citep{ld08} \smallskip  \end{center}} \Emp  \\ 
      \hline 
      \Bmp{3.0cm} \small \begin{center} \smallskip\smallskip 3D Navier-Stokes  \\ finite-time \smallskip \end{center} \Emp &  
      \Bmp{7.0cm} \smallskip \centering \small $\E(t) \le \frac{\E(0)}{\sqrt{1 - 4 \frac{C \E(0)^2}{\nu^3} t}}$ \\ \smallskip $\frac{1}{\E(0)} - \frac{1}{\E(t)} \leq  \frac{27}{(2\pi\nu)^4}\left[\K(0) - \K(t) \right]$ \smallskip \Emp & \Bmp{3.5cm} \centering {???} \smallskip\smallskip \Emp \\
\hline
    \end{tabular}
  \end{center}
  \caption{Summary of selected estimates for the instantaneous 
    rate of growth and the growth over finite time of enstrophy and palinstrophy 
    in 1D Burgers, 2D and 3D Navier-Stokes systems. {The quantities $\K$ and 
      $\E$ are defined in \eqref{eq:EnerDef_3D} and \eqref{eq:EnsDef_3D}.}}
  \label{tab:estimates}
\end{table}

\section{Bounds on the Growth of Enstrophy in 3D Navier-Stokes Flows}
\label{sec:Bounds3DNS}

We consider the incompressible Navier-Stokes system defined on the 3D
unit cube $\Omega = [0,1]^3$ with periodic boundary conditions
\begin{subequations}\label{eq:NSE3D}
\begin{alignat}{2}
\partial_t\uvec + \uvec\cdot\bnabla\uvec + \bnabla p - \nu\laplacian\uvec & = 0 & &\qquad\mbox{in} \,\,\Omega\times(0,T), \\
\bnabla\cdot\uvec & = 0 & & \qquad\mbox{in} \,\,\Omega\times[0,T), \\
\uvec(\xvec,0) & = \uvec_0(\xvec), &   &
\end{alignat}
\end{subequations}
where the vector $\uvec = [u_1, u_2, u_3]$ is the velocity field, $p$
is the pressure and $\nu>0$ is the coefficient of kinematic viscosity
(hereafter we will set $\nu=0.01$ which is the same value as used in
the seminal study by \cite{ld08}). The velocity gradient
  $\bnabla\uvec$ is the tensor with components $[\bnabla\uvec]_{ij} =
  \partial_j u_i$, $i,j=1,2,3$.  The fluid density $\rho$ is assumed
to be constant and equal to unity ($\rho=1$). The relevant properties
of solutions to system \eqref{eq:NSE3D} can be studied using energy
methods, with the energy $\K(\uvec)$ and its rate of growth given by
\begin{eqnarray}
  \K(\uvec) & := & \frac{1}{2}\int_\Omega |\uvec(\xvec,t)|^2 \,d\xvec, \label{eq:EnerDef_3D}\\
  \frac{d\K(\uvec)}{dt} & = & -\nu\int_\Omega |\nabla\uvec|^2 \, d\xvec, \label{eq:dK/dt_3D}
\end{eqnarray}
where ``$:=$'' means ``equal to by definition''. The enstrophy
$\E(\uvec)$ and its rate of growth are given by
\begin{eqnarray}
\E(\uvec) & := & \frac{1}{2}\int_\Omega | \rot\uvec(\xvec,t) |^2 \,d\xvec, \label{eq:EnsDef_3D}\\
\frac{d\E(\uvec)}{dt} & = & -\nu\int_\Omega |\laplacian\uvec|^2\,d\xvec  + 
\int_{\Omega} \uvec\cdot\nabla\uvec\cdot\laplacian\uvec\, d\xvec =: \R(\uvec). \label{eq:dEdt}
\end{eqnarray}
For incompressible flows
with periodic boundary conditions we also have the following identity
\citep{dg95}
\begin{equation}
\int_{\Omega} |\rot\uvec|^2\,d\xvec = \int_{\Omega} |\nabla\uvec|^2\,d\xvec.
\label{eq:duL2}
\end{equation}
Hence, combining \eqref{eq:EnerDef_3D}--\eqref{eq:duL2}, the energy
and enstrophy satisfy the system of ordinary differential equations
\begin{subequations}
\begin{align}
\frac{d\K(\uvec)}{dt} & =  -2\nu\E(\uvec), \label{eq:dKdt_system}\\
\frac{d\E(\uvec)}{dt} & =  \R(\uvec). \label{eq:dEdt_system}
\end{align}
\end{subequations}

A standard approach at this point is to try to upper-bound
$d\E / dt$ and using standard techniques of functional analysis it is
possible to obtain the following well-known estimate in terms of $\K$
and $\E$ \citep{d09}
\begin{equation}\label{eq:dEdt_estimate_KE}
\frac{d\E}{dt} \leq -\nu \frac{\E^2}{\K} + \frac{c}{\nu^3} \E^3
\end{equation} 
for $c$ an absolute constant. A related estimate 
expressed entirely in terms of the enstrophy $\E$ is given by
\begin{equation}
\frac{d\E}{dt} \leq \frac{27}{8\,\pi^4\,\nu^3} \E^3. 
\label{eq:dEdt_estimate_E}
\end{equation} 
By simply integrating the differential inequality in \eqref{eq:dEdt_estimate_E} with respect to time we obtain the finite-time bound
\begin{equation}
\E(t) \leq \frac{\E(0)}{\sqrt{1 - \frac{27}{4\,\pi^4\,\nu^3}\,\E(0)^2\, t}}
\label{eq:Et_estimate_E0}
\end{equation}
which clearly becomes infinite at time $t_0 = 4\,\pi^4\,\nu^3 /
[27\,\E(0)^2]$. Thus, based on estimate \eqref{eq:Et_estimate_E0}, it
is not possible to establish the boundedness of the enstrophy $\E(t)$
globally in time and hence the regularity of solutions.  Therefore, the
question about the finite-time singularity formation can be recast in
terms of whether or not estimate \eqref{eq:Et_estimate_E0} can be
saturated. By this we mean the existence of initial data with
enstrophy $\E_0 := \E(0)> 0$ such that the resulting time evolution
realizes the largest growth of enstrophy $\E(t)$ allowed by the
right-hand side (RHS) of estimate \eqref{eq:Et_estimate_E0}. A
systematic search for such most singular initial data using
variational optimization methods is the key theme of this study.
Although different notions of sharpness of an estimate can be defined,
e.g., sharpness with respect to constants or exponents in the case of
estimates in the form of power laws, the precise notion of sharpness
considered in this study is the following
\begin{definition}\label{def:NotionSharpness}
  Given a parameter $p\in\mathbb{R}$ and maps
  $f,g:\mathbb{R}\to\mathbb{R}$, the estimate
\begin{displaymath}
f(p) \leq g(p) 
\end{displaymath}
is declared sharp in the limit $p \to p_0\in\mathbb{R}$ if and only if
\begin{displaymath}
\lim_{p \to p_0} \frac{f(p)}{g(p)} \sim \beta, \quad \beta \in \RR.
\end{displaymath}
\end{definition}
\noindent
From this definition, the sharpness of estimates in the form $g(p) =
C\, p^{\alpha}$ for some $C \in \RR_+$ and $\alpha \in \RR$ can be
addressed in the limit $p \rightarrow \infty$ by studying the
adequacy of the exponent $\alpha$.

The question of sharpness of estimate \eqref{eq:dEdt_estimate_E} was
addressed in the seminal study by \cite{ld08}, see also \cite{l06},
who constructed a family of divergence-free velocity fields saturating
this estimate.  More precisely, these vector fields were parameterized
by their enstrophy and for sufficiently large values of $\E$ the
corresponding rate of growth $d\E/dt$ was found to be proportional to
$\E^3$. Therefore, in agreement with definition
\ref{def:NotionSharpness}, estimate \eqref{eq:dEdt_estimate_E} was
declared sharp up to a numerical prefactor.  However, the sharpness of
the instantaneous estimate alone does not allow us to conclude about
the possibility of singularity formation, because for this situation
to occur, a sufficiently large enstrophy growth rate would need to be
sustained over a {\em finite} time window $[0,t_0)$. In fact, assuming
the instantaneous rate of growth of enstrophy in the form $d\E / dt =
C \, \E^{\alpha}$ for some $C>0$, any exponent $\alpha > 2$ will
produce blow-up of $\E(t)$ in finite time if the rate of growth
is sustained. The fact that there is no blow-up for $\alpha \le
  2$ follows from Gr\"onwall's lemma and the fact that one factor of
$\E$ in \eqref{eq:dEdt_estimate_E} can be bounded in terms of the
initial energy using \eqref{eq:dK/dt_3D} as follows
\begin{equation}
\int_0^t \E(s)\, ds = \frac{1}{2\nu} \left[ \K(0) - \K(t)\right] \leq \frac{1}{2\nu} \K(0).
\label{eq:Kt}
\end{equation}
This relation also leads to an alternative form of the estimate for
the finite-time growth of enstrophy, namely
\begin{align}
\frac{d\E}{dt} & \leq  \frac{27}{8\pi^4\nu^3}\E^3\quad\Longrightarrow \nonumber \\
\int_{\E(0)}^{\E(t)}\E^{-2}\,d\E & \leq  \frac{27}{8\pi^4\nu^3}\int_0^t\E(s)\,ds \quad\Longrightarrow \nonumber \\
\frac{1}{\E(0)} - \frac{1}{\E(t)} & \leq  \frac{27}{(2\pi\nu)^4}\left[\K(0) - \K(t) \right] \label{eq:Evs_t_fixE}
\end{align}
which is more convenient than \eqref{eq:Et_estimate_E0} from
the computational point of view and will be used in the present study.
We note, however, that since the RHS of this inequality cannot
be expressed entirely in terms of properties of the initial data,
this is {\em not} in fact an a priori estimate. Estimate
\eqref{eq:Evs_t_fixE} also allows us to obtain a condition on the size
of the initial data, given in terms of its energy $\K(0)$ and
enstrophy $\E(0)$, which guarantees that smooth solutions will exist
globally in time, namely,
\begin{equation}\label{eq:Cond_for_globalReg}
\mathop{\max}_{t \geq 0} \E(t) \leq \frac{\E(0)}{1 - \frac{27}{(2\pi\nu)^4}\K(0)\E(0)} 
\end{equation}
from which it follows that
\begin{equation}\label{eq:K0E0}
\K(0)\E(0) < \frac{(2\pi\nu)^4}{27}.
\end{equation}
Thus, flows with energy and enstrophy satisfying inequality
\eqref{eq:K0E0} are guaranteed to be smooth for all time, in agreement
with {the} regularity results {available under the
  assumption of} small initial data {\citep{lady69}}.

\section{Instantaneously Optimal Growth of Enstrophy} 
\label{sec:3D_InstOpt}

Sharpness of instantaneous estimate \eqref{eq:dEdt_estimate_E},
  in the sense of definition \ref{def:NotionSharpness}, can be probed
  by constructing a family of ``extreme vortex states'' $\tuvecE$
  which, for each $\E_0 > 0$, have prescribed enstrophy $\E(\tuvecE) =
  \E_0$ and produce the largest possible rate of growth of enstrophy
  $\R(\tuvecE)$. Given the form of \eqref{eq:dEdt}, the fields
  $\tuvecE$ can be expected to exhibit (at least piecewise) smooth
  dependence on $\E_0$ and we will refer to the mapping $\E_0
  \longmapsto \tuvecE$ as a ``{maximizing} branch''.  Thus,
  information about the sharpness of estimate
  \eqref{eq:dEdt_estimate_E} can be deduced by analyzing the relation
  $\E_0$ versus $\R(\tuvecE)$ obtained for a possibly broad range of
  enstrophy values. A {maximizing} branch is constructed by finding,
  for different values of $\E_0$, the extreme vortex states $\tuvecE$
  as solutions of a variational optimization problem defined below.

Hereafter, $H^2(\Omega)$ will denote the Sobolev space of functions
with square-integrable second derivatives endowed with the inner
product \citep{af05}
\begin{equation}
\forall\,\mathbf{z}_1, \mathbf{z}_2 \in H^2(\Omega) \qquad 
\Big\langle \mathbf{z}_1, \mathbf{z}_2 \Big\rangle_{H^2(\Omega)}
= \int_{\Omega} \mathbf{z}_1 \cdot \mathbf{z}_2 
+ \ell_1^2 \,\bnabla \mathbf{z}_1 \colon \bnabla \mathbf{z}_2
+ \ell_2^4 \,\Delta \mathbf{z}_1 \cdot \Delta \mathbf{z}_2  \, d\xvec,  \label{eq:ipH2} 
\end{equation}
where $\ell_1,\ell_2\in \RR_+$ are parameters with the meaning of
length scales (the reasons for introducing these parameters in the
definition of the inner product will become clear below). The inner
product in the space $L_2(\Omega)$ is obtained from \eqref{eq:ipH2} by
setting $\ell_1 = \ell_2 = 0$.  The notation $H^2_0(\Omega)$ will
refer to the Sobolev space $H^2(\Omega)$ of functions with zero mean.
For every fixed value $\E_0$ of enstrophy we will look for a
divergence-free vector field $\tuvecE$ maximizing the objective
function $\R \; : \; H^2_0(\Omega) \rightarrow \RR$ defined in
\eqref{eq:dEdt}. We thus have the following
\begin{problem}\label{pb:maxdEdt_E}
  Given $\E_0\in\mathbb{R}_+$ and the objective functional $\R$ from
  equation \eqref{eq:dEdt}, find
\begin{eqnarray*}
\tuvecE & = & \mathop{\arg\max}_{\uvec\in\M{\E_0}} \, \R(\uvec) \\
\M{\E_0} & = & \left\{\uvec\in H_0^2(\Omega)\,\colon\,\nabla\cdot\uvec = 0, \; \E(\uvec) = \E_0 \right\}
\end{eqnarray*} 
\end{problem}
\noindent
which will be solved for enstrophy $\E_0$ spanning a broad
range of values. This approach was originally proposed and
investigated by \cite{ld08}. In the present study we extend and
generalize these results by first showing how other fields considered
in the context of the blow-up problem for both the Euler and
Navier-Stokes system, namely the Taylor-Green vortex, also arise from
variational problem \ref{pb:maxdEdt_E}. We then thoroughly analyze the
time evolution corresponding to our extreme vortex states and compare
it with the predictions of the finite-time estimates
\eqref{eq:Et_estimate_E0} and \eqref{eq:Evs_t_fixE}. As discussed
at the end of this section, some important aspects of our
approach to solving problem \ref{pb:maxdEdt_E} are also quite
different from the method adopted by \cite{ld08}.

The smoothness requirement in the statement of problem
\ref{pb:maxdEdt_E} ($\uvec \in H_0^2(\Omega)$) follows from the
definition of the objective functional $\R$ in equation
\eqref{eq:dEdt}, where both the viscous term $\nu\int_\Omega
|\laplacian\uvec|^2\,d\xvec$ and the cubic term $\int_{\Omega}
\uvec\cdot\bnabla\uvec\cdot\laplacian\uvec\, d\xvec$ contain
derivatives of order up to two. The constraint manifold $\M{\E_0}$ can
be interpreted as an intersection of the manifold (a subspace) $\M{0}
\in H_0^2(\Omega)$ of divergence-free fields and the manifold
$\mathcal{S}'_{\E_0} \in H_0^2(\Omega)$ of fields with prescribed
enstrophy $\E_0$.  The structure of these constraint manifolds is
reflected in the definition of the corresponding projections
$\mathbb{P}_{\mathcal{S}}:H_0^2\to\mathcal{S}$ (without a subscript,
$\mathcal{S}$ refers to a generic manifold) which is given for 
each of the two constraints as follows:
\begin{itemize}
\item ({\em div-free})-constraint: the projection of a field $\uvec$
  onto the subspace of solenoidal fields $\mathcal{S}_0$ is performed
  using the Helmholtz decomposition;  accordingly, every zero-mean
  vector field $\uvec \in H_0^2(\Omega)$ can be decomposed
  uniquely as
\begin{displaymath}
\uvec = \bnabla\phi + \rot\Avec,  
\end{displaymath}
where $\phi$ and $\Avec$ are scalar and vector potentials,
respectively; it follows from the identity
$\bnabla\cdot(\rot\Avec)\equiv0$, valid for any sufficiently smooth
vector field $\Avec$, that the projection
$\mathbb{P}_{\mathcal{S}_0}(\uvec)$ is given simply by
$\rot\Avec$ and is therefore calculated as
\begin{equation}
\mathbb{P}_{\mathcal{S}_0}(\uvec) = \uvec - \bnabla\left[\laplacian^{-1}(\bnabla\cdot\uvec)\right], 
\label{eq:Phodge}
\end{equation}
where $\laplacian^{-1}$ is the inverse Laplacian associated with the
periodic boundary conditions; the operator
$\mathbb{P}_{\mathcal{S}_0}$ is also known as the Leray-Helmholtz
projector.

\item $(\E_0)$-constraint: the projection onto the manifold
  $\mathcal{S}'_{\E_0}$ is calculated by the normalization
\begin{equation}\label{eq:FixE0_3D}
\mathbb{P}_{\mathcal{S}'_{\E_0}}(\uvec) = \sqrt{\frac{\E_0}{\E\left(\uvec\right)}}\,\uvec.
\end{equation}
\end{itemize}
Thus, composing \eqref{eq:Phodge} with \eqref{eq:FixE0_3D}, the
projection onto the manifold $\M{\E_0}$ defined in problem
\ref{pb:maxdEdt_E} is constructed as
\begin{equation}
\mathbb{P}_{\mathcal{S}_{\E_0}}(\uvec) = \mathbb{P}_{\mathcal{S}'_{\E_0}}\Big(  \mathbb{P}_{\mathcal{S}_0} (\uvec)\Big).
\label{eq:P}
\end{equation}
This approach, which was already successfully employed by
\cite{ap11a,ap13a}, allows one to enforce the enstrophy
  constraint essentially with the machine precision.

For a given value of $\E_0$, the maximizer $\tuvecE$ can be
found as $\tuvecE = \lim_{n\rightarrow \infty} \uvec_{\E_0}^{(n)}$
using the following iterative procedure representing a discretization
of a gradient flow projected on $\mathcal{S}_{\E_0}$
\begin{equation}
\begin{aligned}
\uvec_{\E_0}^{(n+1)} & =  \mathbb{P}_{\mathcal{S}_{\E_0}}\left(\;\uvec^{(n)}_{\E_0} + \tau_n \nabla\R\left(\uvec^{(n)}_{\E_0}\right)\;\right), \\ 
\uvec_{\E_0}^{(1)} & =  \uvec^0,
\end{aligned}
\label{eq:desc}
\end{equation}
where $\uvec^{(n)}_{\E_0}$ is an approximation of the maximizer
obtained at the $n$-th iteration, $\uvec^0$ is the initial guess and
$\tau_n$ is the length of the step in the direction of the gradient.
It is ensured that the maximizers $\tuvecE$ obtained for
  different values of $\E_0$ lie on the same {maximizing} branch by
  using the continuation approach, where the maximizer $\tuvecE$ is
  {employed} as the initial guess $\uvec^0$ to compute
  $\widetilde{\mathbf{u}}_{\E_0+\Delta\E}$ at the next enstrophy level
  for some sufficiently small $\Delta\E > 0$. As will be demonstrated
  in \S \ref{sec:3D_InstOpt_E0to0}, in the limit $\E_0 \rightarrow 0$
  optimization problem \ref{pb:maxdEdt_E} admits a discrete family of
  closed-form solutions and each of these vortex states is the
  limiting (initial) member $\widetilde{\mathbf{u}}_{0}$ of the
  corresponding {maximizing} branch. As such, these limiting extreme
  vortex states are used as the initial guesses $\uvec^0$ for the
  calculation of $\widetilde{\mathbf{u}}_{\Delta\E}$, i.e., they serve
  as ``seeds'' for the calculation of an entire {maximizing} branch
  (as discussed in \S \ref{sec:discuss}, while there exist
  alternatives to the continuation approach, this technique {in fact
    results} in the fastest convergence of iterations \eqref{eq:desc}
  and also ensures that all computed extreme vortex states lie on a
  single branch). The procedure outlined above is summarized as
  Algorithm \ref{alg:optimAlg}, {whereas all} details are presented
  below.

\begin{algorithm}[t!]
\begin{algorithmic}
\STATE set $\E_0 = 0$
\STATE set $\tuvecE = \widetilde{\mathbf{u}}_{0}$
\REPEAT 
\STATE \COMMENT{------------------------ loop over increasing enstrophy values $\E_0$ ------------------------}
\STATE $\uvec_{\E_0}^{(0)} = \tuvecE$
\STATE $\E_0 = \E_0 + \Delta \E$
\STATE $n = 0$
\STATE compute $\R_0 = \R\left(\uvec_{\E_0}^{(0)}\right)$
\REPEAT 

\STATE \COMMENT{------------------------------ optimization iterations \eqref{eq:desc}  ------------------------------}

\STATE compute the $L_2$ gradient $\nabla^{L_2}\R\left(\uvec_{\E_0}^{(n)}\right)$, see equation \eqref{eq:gradRL2}

\STATE compute the Sobolev gradient $\nabla\R\left(\uvec_{\E_0}^{(n)}\right)$, see equation \eqref{eq:gradRH2}

\STATE compute the step size $\tau_n$, see equation \eqref{eq:tau_n}

\STATE set $\uvec_{\E_0}^{(n+1)} = \mathbb{P}_{\mathcal{S}_{\E_0}}\left(\;\uvec_{\E_0}^{(n)} + \tau_n \nabla\R\left(\uvec_{\E_0}^{(n)}\right)\;\right)$, see equations \eqref{eq:Phodge}--\eqref{eq:P}

\STATE set $\R_1 = \R\left(\uvec_{\E_0}^{(n+1)}\right)$

\STATE compute the \texttt{relative error} $ = (\R_1 - \R_0)/\R_0$

\STATE set $\R_0 = \R_1$

\STATE set $n=n+1$

\UNTIL{ \ \texttt{relative error} < $\epsilon$}
\UNTIL {\ $\E_0 > \E_{\text{max}}$}

\end{algorithmic}
\caption{
 Computation of a maximizing branch using continuation approach.  \newline
     \textbf{Input:} \newline
 \hspace*{0.22cm} $\widetilde{\mathbf{u}}_{0}$ --- limiting extreme vortex state (corresponding to $\E_0 \rightarrow 0$, see Table \ref{tab:E0}) \newline
 \hspace*{0.22cm} $\E_{\text{max}}$ --- maximum enstrophy \newline
 \hspace*{0.22cm}    $\Delta \E$ --- (adjustable) enstrophy increment \newline
 \hspace*{0.22cm}    $\epsilon$ --- tolerance in the solution of optimization problem  \ref{pb:maxdEdt_E} via iterations \eqref{eq:desc} \newline
 \hspace*{0.22cm}    $\ell_1,\ell_2$ --- adjustable length scales defining inner product \eqref{eq:ipH2}, see also \eqref{eq:gradRH2} \newline
 \textbf{Output:} \newline
 \hspace*{0.22cm}    branch of extreme vortex states $\tuvecE$, \ $0 \le \E_0 \le \E_{\text{max}}$
}
\label{alg:optimAlg}
\end{algorithm}

A key step of Algorithm \ref{alg:optimAlg} is the evaluation
of the gradient $\nabla\R(\uvec)$ of the objective functional
$\R(\uvec)$, cf.  \eqref{eq:dEdt}, representing its
(infinite-dimensional) sensitivity to perturbations of the velocity
field $\uvec$, and it is essential that the gradient be characterized
by the required regularity, namely, $\nabla\R(\uvec) \in H^2(\Omega)$.
This is, in fact, guaranteed by the Riesz representation theorem
\citep{l69} applicable because the G\^ateaux differential
$\R'(\uvec;\cdot) : H_0^2(\Omega) \rightarrow \RR$, defined as
$\R'(\uvec;\uvec') := \lim_{\epsilon \rightarrow 0}
\epsilon^{-1}\left[\R(\uvec+\epsilon \uvec') - \R(\uvec)\right]$ for
some perturbation $\uvec' \in H_0^2(\Omega)$, is a bounded linear
functional on $H_0^2(\Omega)$.  The G\^ateaux differential can be
computed directly to give
\begin{equation}
\R'(\uvec;\uvec') = \int_{\Omega}\left[\uvec'\cdot\bnabla\uvec\cdot\laplacian\uvec + 
\uvec\cdot\bnabla\uvec'\cdot\laplacian\uvec + 
\uvec\cdot\bnabla\uvec\cdot\laplacian\uvec' \right]\,d\xvec 
-2\nu\int_{\Omega}\laplacian^2\uvec\cdot\uvec'\,d\xvec
\label{eq:dR}
\end{equation}
from which, by the Riesz representation theorem, we obtain
\begin{equation}
\R'(\uvec;\uvec') 
= \Big\langle \nabla\R(\uvec), \uvec' \Big\rangle_{H^2(\Omega)}
= \Big\langle \nabla^{L_2}\R(\uvec), \uvec' \Big\rangle_{L_2(\Omega)}
\label{eq:riesz}
\end{equation}
with the Riesz representers $\nabla\R(\uvec)$ and
$\nabla^{L_2}\R(\uvec)$ being the gradients computed with
respect to the $H^2$ and $L_2$ topology, respectively, and the
inner products defined in \eqref{eq:ipH2}. We remark that, while the
$H^2$ gradient is used exclusively in the actual computations, cf.
\eqref{eq:desc}, the $L_2$ gradient is computed first as an
intermediate step.  Identifying the G\^ateaux differential
  \eqref{eq:dR} with the $L_2$ inner product and performing
  integration by parts yields
\begin{equation}
\nabla^{L_2}\R(\uvec) = \laplacian\left( \uvec\cdot\bnabla\uvec \right) + (\bnabla\uvec)^T\laplacian\uvec - 
\uvec\cdot\bnabla(\laplacian\uvec) - 2\nu\laplacian^2\uvec.
\label{eq:gradRL2}
\end{equation}
Similarly, identifying the G\^ateaux differential \eqref{eq:dR} with
the $H^2$ inner product \eqref{eq:ipH2}, integrating by parts and
using \eqref{eq:gradRL2}, we obtain the required $H^2$ gradient
$\nabla\R$ as a solution of the following elliptic boundary-value
problem
\begin{equation}
\begin{aligned}
&\left[ \Id \, - \,\ell_1^2 \,\Delta + \,\ell_2^4 \,\Delta^2 \right] \nabla\R
= \nabla^{L_2} \R  \qquad \text{in} \ \Omega, \\
& \text{Periodic Boundary Conditions}.
\end{aligned}
\label{eq:gradRH2}
\end{equation}
The gradient fields $\nabla^{L_2}\R(\uvec)$ and
  $\nabla\R(\uvec)$ can be interpreted as infinite-dimensional
  sensitivities of the objective function $\R(\uvec)$,
  cf.~\eqref{eq:dEdt}, with respect to perturbations of the field
  $\uvec$. While these two gradients may point towards the same local
  maximizer, they represent distinct ``directions'', since they are
  defined with respect to different topologies ($L_2$ vs.~$H^2$). As
shown by \citet{pbh04}, extraction of gradients in spaces of smoother
functions such as $H^2(\Omega)$ can be interpreted as low-pass
filtering of the $L_2$ gradients with parameters $\ell_1$ and $\ell_2$
acting as cut-off length-scales and the choice of their numerical
values will be discussed in \S \ref{sec:3D_InstOpt_E}.

The step size $\tau_n$ in algorithm \eqref{eq:desc} is computed
as
\begin{equation}\label{eq:tau_n}
\tau_n = \mathop{\argmax}_{\tau>0} \left\{ \R\left[\mathbb{P}_{\mathcal{S}_{\E_0}}
\left( \;\uvec^{(n)} + \tau\,\nabla\R(\uvec^{(n)}) \;\right)\right] \right\}
\end{equation}
which is done using a suitable derivative-free line-search
algorithm \citep{r06}. Equation \eqref{eq:tau_n} can be interpreted as
a modification of a standard line search method where the optimization
is performed following an arc (a geodesic) lying on the constraint
manifold $\mathcal{S}_{\E_0}$, rather than a straight line. This
approach was already successfully employed to solve similar problems
in \citet{ap11a,ap13a}. 

It ought to be emphasized here that the approach presented above in
which the projections \eqref{eq:Phodge}--\eqref{eq:FixE0_3D} and
gradients \eqref{eq:gradRL2}--\eqref{eq:gradRH2} are obtained based on
the infinite-dimensional (continuous) formulation to be discretized
only at the final stage is fundamentally different from the method
employed in the original study by \cite{ld08} in which the
optimization problem was solved in a fully discrete setting (the two
approaches are referred to as ``optimize-then-discretize'' and
``discretize-then-optimize'', respectively, cf.~\cite{g03}). A
  practical advantage of the continuous (``optimize-then-discretize'')
  formulation used in the present work is that the expressions
  representing the sensitivity of the objective functional $\R$, i.e.
  the gradients $\nabla^{L_2}\R$ and $\nabla\R$, are independent of
  the specific discretization approach chosen to evaluate them.  This
  should be contrasted with the discrete
  (``discretize-then-optimize'') formulation, where a change of the
  discretization method would require rederivation of the
    gradient expressions.  In addition, the continuous formulation
  allows us to strictly enforce the regularity of maximizers required
  in problem \ref{pb:maxdEdt_E}. Finally and perhaps most importantly,
  the continuous formulation of the maximization problem makes it
  possible to obtain elegant closed-form solutions of the problem in
  the limit $\E_0 \rightarrow 0$, which is done in \S
  \ref{sec:3D_InstOpt_E0to0} below. These analytical solutions will
  then be used in \S \ref{sec:3D_InstOpt_E} to guide the computation
  of maximizing branches by numerically solving problem
  \ref{pb:maxdEdt_E} for a broad range of $\E_0$, as outlined in
  Algorithm \ref{alg:optimAlg}.

\section{Extreme Vortex States in the Limit $\E_0 \to 0$}
\label{sec:3D_InstOpt_E0to0}

It is possible to find analytic solutions to problem
\ref{pb:maxdEdt_E} in the limit $\E_0 \to 0$ using perturbation
methods. To simplify the notation, in this section we will drop the
subscript $\E_0$ when referring to the optimal field. The
Euler-Lagrange system representing the first-order optimality
conditions in optimization problem \ref{pb:maxdEdt_E} is given by
\citep{l69}
\begin{subequations}\label{eq:KKT_E}
\begin{align} 
\B(\tuvec,\tuvec) - 2\nu\laplacian^2\tuvec - \lambda\laplacian\tuvec - \bnabla q & = 0 \qquad\mbox{in}\,\,\Omega , \label{eq:KKT_E_gradR}\\
\nabla\cdot\tuvec & = 0 \qquad\mbox{in}\,\,\Omega , \label{eq:KKT_E_divConstr}\\
\E(\tuvec) - \E_0 & = 0, \label{eq:KKT_E_E0Constr}
\end{align}
\end{subequations}
where $\lambda\in\mathbb{R}$ and $q:\Omega\to\mathbb{R}$ are the Lagrange 
multipliers associated with the constraints defining the manifold $\M{\E_0}$, 
and $\B(\uvec,\vvec)$, given by
\begin{displaymath}
\B(\uvec,\vvec) :=  \laplacian\left( \uvec\cdot\bnabla\vvec \right) + (\bnabla\uvec)^T\laplacian\vvec - 
\uvec\cdot\bnabla(\laplacian\vvec),
\end{displaymath}
is the bilinear form from equation \eqref{eq:gradRL2}.  Using the
formal series expansions with $\alpha > 0$
\begin{subequations}\label{eq:series3D}
\begin{align}
\tuvec & = \uvec_0 + \E_0^{\alpha}\uvec_1 + \E_0^{2\alpha}\uvec_2 + \ldots, \\
\lambda & = \lambda_0 + \E_0^{\alpha}\lambda_1 + \E_0^{2\alpha}\lambda_2 + \ldots, \\
q & = q_0 + \E_0^{\alpha}q_1 + \E_0^{2\alpha}q_2 + \ldots
\end{align} 
\end{subequations}
in \eqref{eq:KKT_E} and collecting terms proportional to different
powers of $\E_0^{\alpha}$, it follows from \eqref{eq:KKT_E_gradR}
that, at every order $m=1,2,\dots$ in $\E_0^{\alpha}$, we have
\begin{equation} 
\E_0^{m\alpha}: \qquad\sum_{j=0}^m \B(\uvec_j,\uvec_{m-j}) - 2\nu\laplacian^2\uvec_m -
\sum_{j=0}^m\lambda_j\laplacian\uvec_{m-j} - \nabla q_m = 0 \quad\mbox{in}\,\,\Omega.
\end{equation}
Similarly, equation \eqref{eq:KKT_E_divConstr} leads to
\begin{equation}\label{eq:Incompressible_Uk} 
\nabla\cdot\uvec_m = 0 \quad\mbox{in}\,\,\Omega
\end{equation}
at every order $m$ in $\E_0^{\alpha}$. It then follows from
equation \eqref{eq:KKT_E_E0Constr} that
\begin{eqnarray*}
\E(\uvec) & = & \E(\uvec_0) -\big\langle\uvec_0,\laplacian\uvec_1\big\rangle_{L_2}\E_0^{\alpha} + 
                    \left[ \E(\uvec_1) - \big\langle\uvec_0,\laplacian\uvec_2\big\rangle_{L_2}\right]\E_0^{2\alpha} + \ldots \\
              & = & \E_0,
\end{eqnarray*}
which, for $\alpha \neq 0$, forces $\E(\uvec_0) = 0$. Hence, $\uvec_0
\equiv 0$, $\alpha = 1/2$ and $\E(\uvec_1) = 1$. The systems at orders
$\E_0^{1/2}$ and $\E_0^1$ are given by:
\begin{subequations}\label{eq:maxdEdt_Asympt_1}
\begin{align}
\E^{1/2}_0:\quad\qquad\qquad\qquad\qquad\qquad 2\nu\laplacian^2\uvec_1 + \lambda_0\laplacian\uvec_1 + 
\nabla q_1 & = 0 \quad\mbox{in}\,\,\Omega , \label{eq:maxdEdt_Asympt_1_PDE}\\
\nabla\cdot\uvec_1 & = 0 \quad\mbox{in}\,\,\Omega , \label{eq:maxdEdt_Asympt_1_Div0Constr} \\
\E(\uvec_1) & = 1, \label{eq:maxdEdt_Asympt_1_E0Constr} 
\end{align} 
\end{subequations}
\begin{subequations}\label{eq:maxdEdt_Asympt_2}
\begin{align}
\E_0:\qquad 2\nu\laplacian^2\uvec_2 + \lambda_0\laplacian\uvec_2  + \nabla q_2 - \B(\uvec_1,\uvec_1) + 
\lambda_1\laplacian\uvec_1 & = 0 \quad\mbox{in}\,\,\Omega, \\
\nabla\cdot\uvec_2 & = 0 \quad\mbox{in}\,\,\Omega , \\
\langle \laplacian\uvec_1, \uvec_2 \rangle_{L_2} & = 0, 
\end{align} 
\end{subequations}
where the fact that $\B(\uvec_0,\uvec_j) = 0$ for all $j$ has been
used. While continuing this process to larger values of $m$ may
lead to some interesting insights, for the purpose of this
investigation it is sufficient to truncate expansions
\eqref{eq:series3D} at the order $\O(\E_0)$. The corresponding
approximation of the objective functional \eqref{eq:dEdt} then
becomes
\begin{equation}\label{eq:R03D}
\R(\tuvec) = - \nu\E_0\int_{\Omega} \left| \laplacian \uvec_1 \right|^2 \, d\xvec + \O(\E_0^{3/2}).
\end{equation}
It is worth noting that, in the light of relation
  \eqref{eq:R03D}, the maximum rate of growth of enstrophy in the
  limit of small $\E_0$ is in fact negative, meaning that, for
  sufficiently small $\E_0$, the enstrophy itself is a decreasing
  function for all times. This observation is consistent with the
  small-data regularity result discussed in Introduction.

As regards problem \eqref{eq:maxdEdt_Asympt_1} defining the triplet
$\{\uvec_1, q_1,\lambda_0\}$, taking the divergence of equation
\eqref{eq:maxdEdt_Asympt_1_PDE} and using the condition
$\nabla\cdot\uvec_1=0$ leads to the Laplace equation $\laplacian q_1 =
0$ in $\Omega$. Since for zero-mean functions defined on
$\Omega$, $\Ker(\laplacian) = \{ 0 \}$, it follows that $q_1
\equiv 0$ and equation \eqref{eq:maxdEdt_Asympt_1_PDE} is reduced to
the eigenvalue problem
\begin{equation}\label{eq:maxdEdt_smallE0_eig}
2\nu\laplacian\uvec_1 + \lambda_0\uvec_1 = 0, 
\end{equation}
with $\uvec_1$ satisfying the incompressibility condition
\eqref{eq:maxdEdt_Asympt_1_Div0Constr}.  Direct calculation
using equation \eqref{eq:maxdEdt_smallE0_eig} and condition
\eqref{eq:maxdEdt_Asympt_1_E0Constr} leads to an asymptotic expression
for the objective functional in the limit of small enstrophy
\begin{displaymath}\label{eq:R0_lambda0_3D}
\R(\tuvec) \approx - \lambda_0\E_0.
\end{displaymath}
Solutions to the eigenvalue problem in equation
\eqref{eq:maxdEdt_smallE0_eig} can be found using the Fourier
expansion of $\uvec_1$ given as (with hats denoting Fourier
  coefficients)
\begin{displaymath}
\uvec_1(\xvec) = \sum_{\kvec\in\W}\widehat{\uvec}_1(\kvec)\e{2\pi i\kvec\cdot\xvec}, 
\end{displaymath}
where $\W\subseteq\mathbb{Z}^3$ is a set of wavevectors $\kvec$ for
which $\widehat{\uvec}_1(\kvec) \neq 0$.  The eigenvalue problem
\eqref{eq:maxdEdt_smallE0_eig} then becomes
\begin{eqnarray*}
\left[-2\nu(2\pi)^2|\kvec|^2 + \lambda_0\right]\widehat{\uvec}_1(\kvec) & = & 0 \qquad\forall\,\kvec\in\W, \\
\widehat{\uvec}_1(\kvec)\cdot\kvec & = & 0 \qquad\forall\,\kvec\in\W,
\end{eqnarray*}
with solutions obtained by choosing, for any $k\in\mathbb{Z}\setminus\{0\}$, 
a set of wavevectors with the following structure
\begin{equation}\label{eq:defW}
\W_k = \left\{\kvec\in\mathbb{Z}^3\colon|\kvec|^2 = k \right\} 
\end{equation}
and $\widehat{\uvec}_1(\kvec)$ with an appropriate form satisfying the
incompressibility condition $\widehat{\uvec}_1\cdot\kvec = 0$. For
the solutions to equation \eqref{eq:maxdEdt_smallE0_eig} constructed in
such manner it then follows that $\lambda_0 = 2\nu(2\pi)^2|\kvec|^2$
and the optimal asymptotic value of $\R$ is given by
\begin{equation}\label{eq:R0_kvec_3D}
\R(\tuvec) \approx - 8\pi^2\nu|\kvec|^2\E_0.
\end{equation}

Since the fields $\uvec_1$ are real-valued, their Fourier modes
must satisfy $\widehat{\uvec}_1(-\kvec) =
\overline{\widehat{\uvec}_1(\kvec)}$, where $\overline{z}$ denotes the
complex conjugate (C.C.) of $z\in\mathbb{C}$. Depending on the choice
of $\W_k$, a number of different solutions of
\eqref{eq:maxdEdt_Asympt_1} can be constructed and below we focus on
the following three most relevant cases characterized by the largest
values of $\R(\tuvec)$:
\begin{enumerate}
\renewcommand{\theenumi}{\roman{enumi}}
\item $\W_1 = \{ \kvec_1, \kvec_2, \kvec_3, -\kvec_1, -\kvec_2,
  -\kvec_3 \}$, where $\kvec_i = \mathbf{e}_i$, $i=1,2,3$, is the
  $i^{\textrm{th}}$ unit vector of the canonical basis of
  $\mathbb{R}^3$; the most general solution can then be
  constructed as
\begin{equation}\label{eq:uvec_3D_k1}
\uvec_1(\xvec) = \mathbf{A}\e{2\pi i\kvec_1\cdot\xvec} + 
                 \mathbf{B}\e{2\pi i\kvec_2\cdot\xvec} +
                 \mathbf{C}\e{2\pi i\kvec_3\cdot\xvec} + \textrm{C.C.}
\end{equation}
with the complex-valued constant vectors $\mathbf{A} = [0,A_2,A_3]$,
$\mathbf{B} = [B_1,0,B_3]$ and $\mathbf{C} = [C_1,C_2,0]$ suitably
chosen so that $\E(\uvec_1) = 1$; hereafter we will use the values $A_2 =
A_3 = \ldots = C_2 = 1/(48\pi^2)$; it follows that $|\kvec|^2 =
1$ $\forall\,\,\kvec\in\W_1$, and the optimal asymptotic value of
$\R$ obtained from equation \eqref{eq:R0_kvec_3D} is given by
\begin{equation}\label{eq:R0_kvec_3D_k1}
  \R(\tuvec) \approx - 8\pi^2\nu\E_0,
\end{equation}
\label{c1}

\item $\W_2 = \W \cup (-\W)$, where $-\W$ denotes the set whose
  elements are the negatives of the elements of set $\W$, for $\W = \{
  \kvec_1 + \kvec_2, \kvec_1 - \kvec_2, \kvec_1 + \kvec_3, \kvec_1 -
  \kvec_3, \kvec_2 + \kvec_3, \kvec_2 - \kvec_3 \}$; the most general
  solution can be then constructed as
\begin{eqnarray}
\uvec_1(\xvec) & = & \mathbf{A}\e{2\pi i[1,1,0]\cdot\xvec} + 
                     \mathbf{B}\e{2\pi i[1,-1,0]\cdot\xvec} +
                     \mathbf{C}\e{2\pi i[1,0,1]\cdot\xvec} + \nonumber \\
               &   & \mathbf{D}\e{2\pi i[1,0,-1]\cdot\xvec} +
                     \mathbf{E}\e{2\pi i[0,1,1]\cdot\xvec} + 
                     \mathbf{F}\e{2\pi i[0,1,-1]\cdot\xvec} + \textrm{C.C.} 
\label{eq:uvec_3D_k2}
\end{eqnarray}
with the constants
$\mathbf{A},\mathbf{B},\ldots,\mathbf{F}\in\mathbb{C}^3$ suitably
chosen so that $\mathbf{A}\cdot[1,1,0] = 0$, $\mathbf{B}\cdot[1,-1,0]
= 0 ,\ldots,\mathbf{F}\cdot[0,1,-1] = 0$, which ensures that
incompressibility condition \eqref{eq:maxdEdt_Asympt_1_Div0Constr} is
satisfied, and that $\E(\uvec_1) = 1$; in this case, $|\kvec|^2 =
2$, $\forall\,\kvec\in\W_2$, and the optimal asymptotic value of $\R$ is
\begin{equation}\label{eq:R0_kvec_3D_k2}
\R(\tuvec) \approx - 16\pi^2\nu\E_0,
\end{equation}
\label{c2}

\item $\W_3 = \W \cup (-\W)$ for $\W = \{
  \kvec_1+\kvec_2+\kvec_3,-\kvec_1+\kvec_2+\kvec_3,\kvec_1-\kvec_2+\kvec_3,\kvec_1+\kvec_2-\kvec_3
  \}$; the most general solution can then be constructed as
\begin{eqnarray}
\uvec_1(\xvec) & = & \mathbf{A}\e{2\pi i[1,1,1]\cdot\xvec} + 
                     \mathbf{B}\e{2\pi i[-1,1,1]\cdot\xvec} + \nonumber \\
               &   & \mathbf{C}\e{2\pi i[1,-1,1]\cdot\xvec} + 
                     \mathbf{D}\e{2\pi i[1,1,-1]\cdot\xvec} + \textrm{C.C.} 
\label{eq:uvec_3D_k3}
\end{eqnarray}
with the constants
$\mathbf{A},\mathbf{B},\mathbf{C},\mathbf{D}\in\mathbb{C}^3$ suitably
chosen so that $\mathbf{A}\cdot[1,1,1] = 0$, $\mathbf{B}\cdot[-1,1,1]
= 0$, $\mathbf{C}\cdot[1,-1,1] = 0$ and $\mathbf{D}\cdot[1,1,-1] = 0$,
which ensures that incompressibility condition
\eqref{eq:maxdEdt_Asympt_1_Div0Constr} is satisfied, and that
$\E(\uvec_1) = 1$; in this case, $|\kvec|^2 =3$,
$\forall\,\kvec\in\W_3$, and the optimal asymptotic value of $\R$ is
\begin{equation}\label{eq:R0_kvec_3D_k3}
\R(\tuvec) \approx - 24\pi^2\nu\E_0.
\end{equation}
\label{c3}
\end{enumerate}
The three constructions of the extremal field $\uvec_1$ given in
\eqref{eq:uvec_3D_k1}, \eqref{eq:uvec_3D_k2} and \eqref{eq:uvec_3D_k3}
are all defined up to arbitrary shifts in all three directions,
reflections with respect to different planes and rotations by angles
which are multiples of $\pi / 2$ about the different axes. As a
  result of this nonuniqueness, there is some freedom in choosing the
  constants $\mathbf{A},\ldots,\mathbf{F}$. Given that the optimal
  asymptotic value of $\R$ depends exclusively on the wavevector
magnitude $|\kvec|$, cf.~\eqref{eq:R0_kvec_3D}, any combination of
constants $\mathbf{A},\ldots,\mathbf{F}$ will produce the {\em same}
optimal rate of growth of enstrophy. Thus, to fix attention, in our
analysis we will set $\mathbf{A}=\mathbf{B}=\mathbf{C}$ in case (i),
$\mathbf{A}=\mathbf{B}=\ldots=\mathbf{F}$ in case (\ref{c2}) and
$\mathbf{A}=\ldots=\mathbf{D}$ in case (\ref{c3}). With these choices,
the contribution from each component of the field $\uvec_1$ to the
total enstrophy is the same. The maximum (i.e., least negative) value
of $\R$ can be thus obtained by choosing the smallest possible
$|\kvec|^2$. This maximum is achieved in case (\ref{c1}) with the
wavevectors $\kvec_1 = [1,0,0]$, $\kvec_2 = [0,1,0]$, $\kvec_3 =
[0,0,1]$, and $-\kvec_1$, $-\kvec_2$ and $-\kvec_3$, for which
$|\kvec|^2 = 1$.  Because of this maximization property, this is the
field we will focus on in our analysis in \S \ref{sec:3D_InstOpt_E}
and \S\ref{sec:timeEvolution}.

\begin{figure}
\begin{center}
\subfigure[$|\kvec|^2 = 1$]{\includegraphics[width=0.32\textwidth]{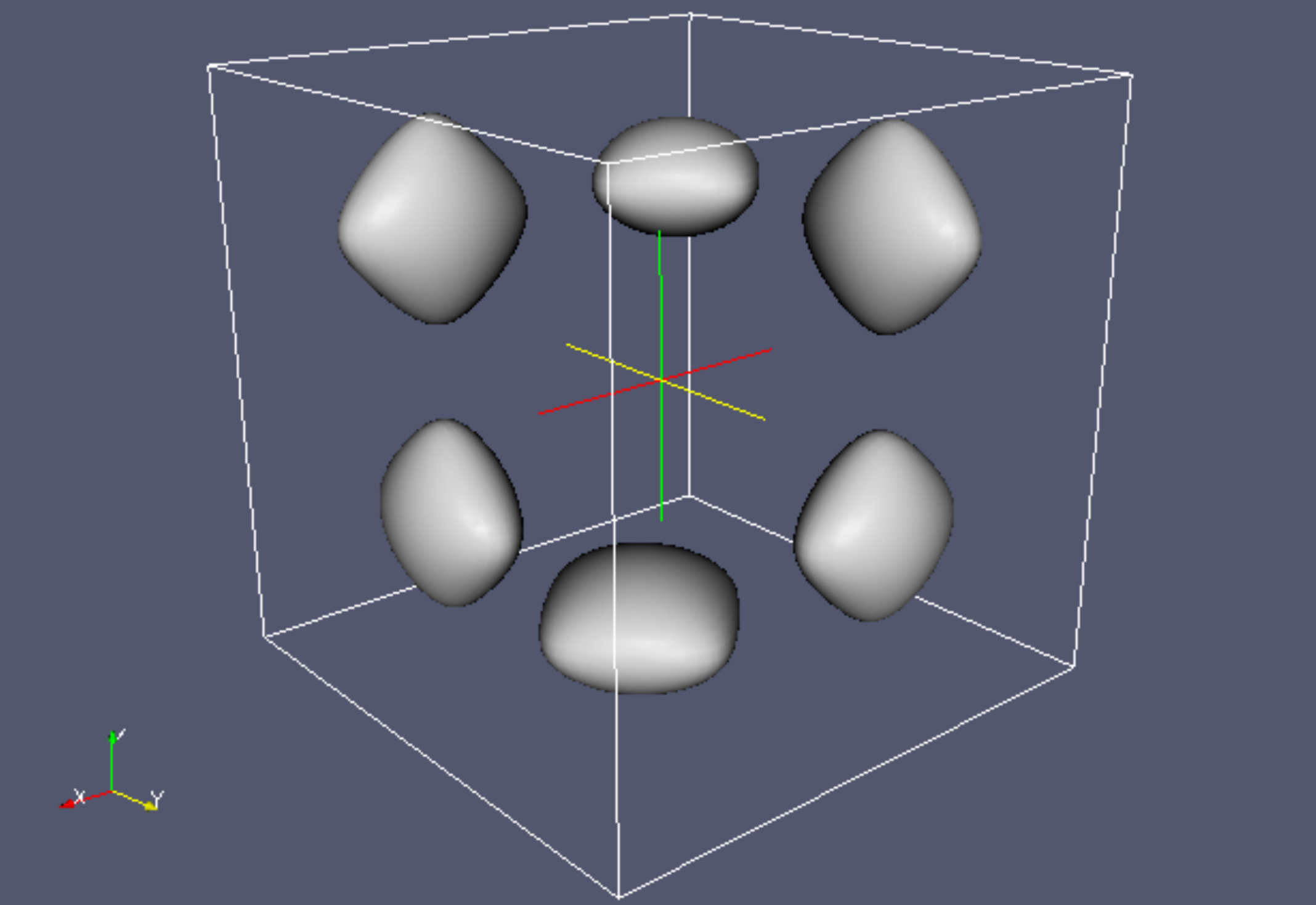}}
\subfigure[$|\kvec|^2 = 2$]{\includegraphics[width=0.32\textwidth]{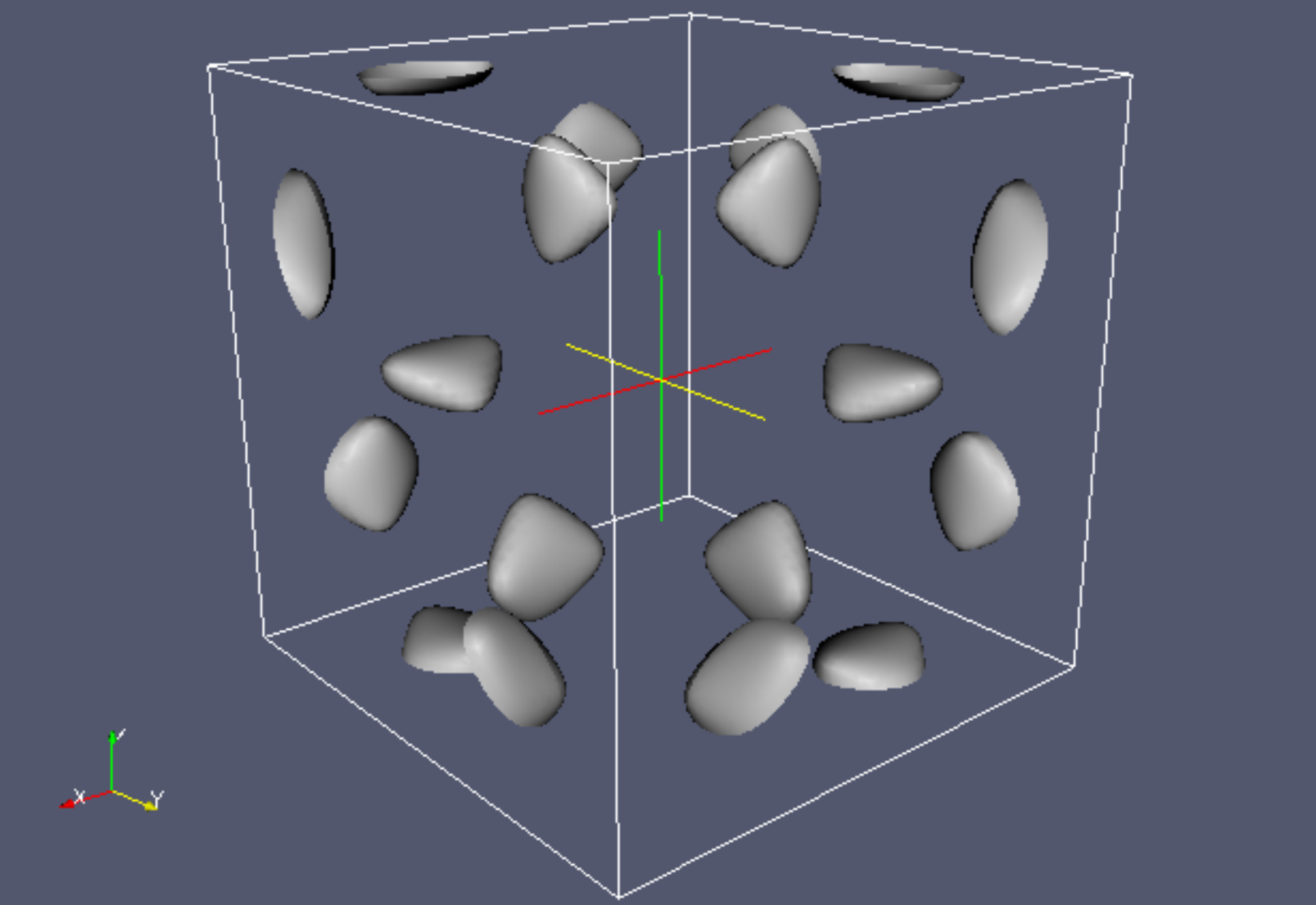}}
\subfigure[$|\kvec|^2 = 3$]{\includegraphics[width=0.32\textwidth]{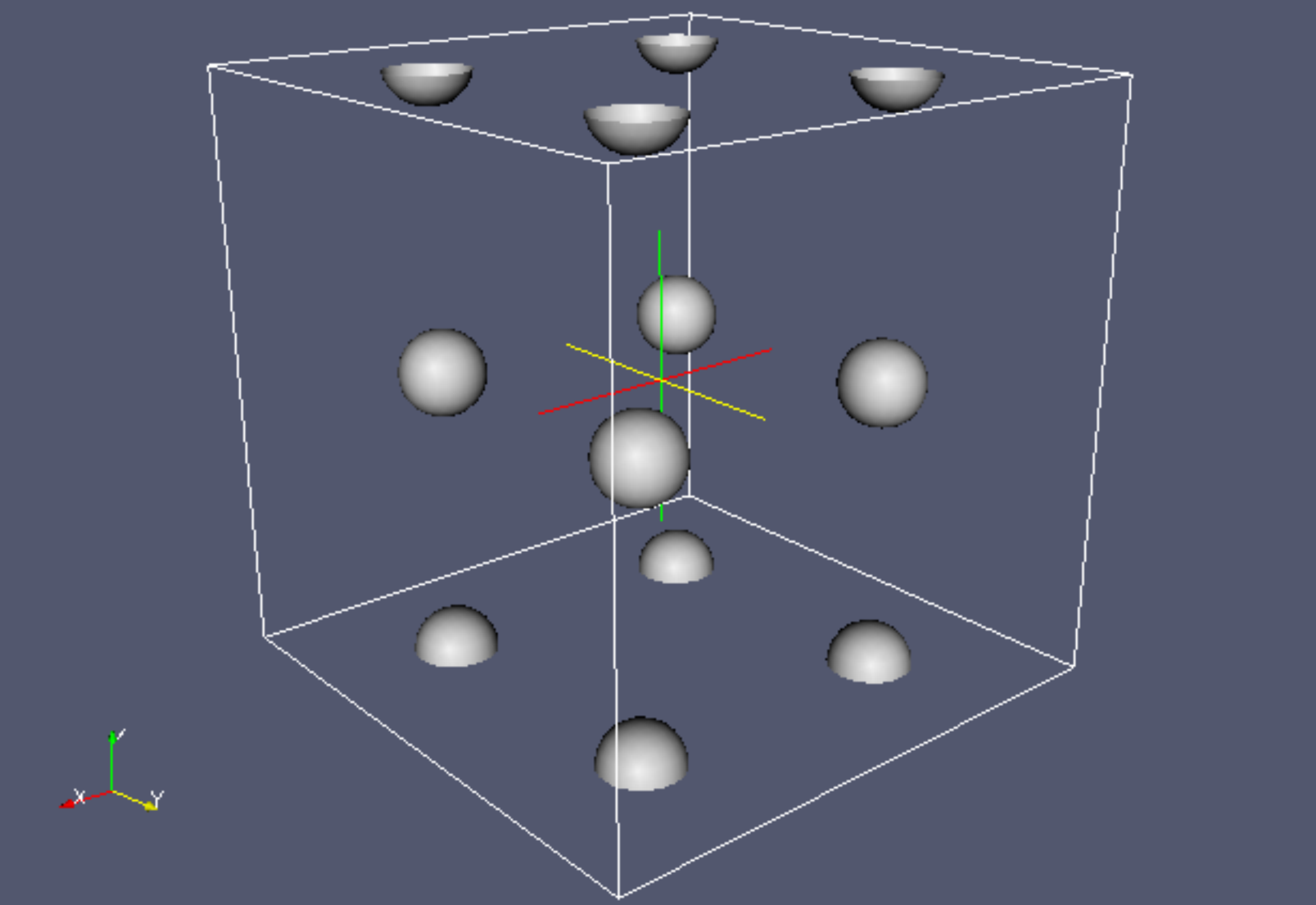}}
\subfigure[staggered ABC flow]{\includegraphics[width=0.32\textwidth]{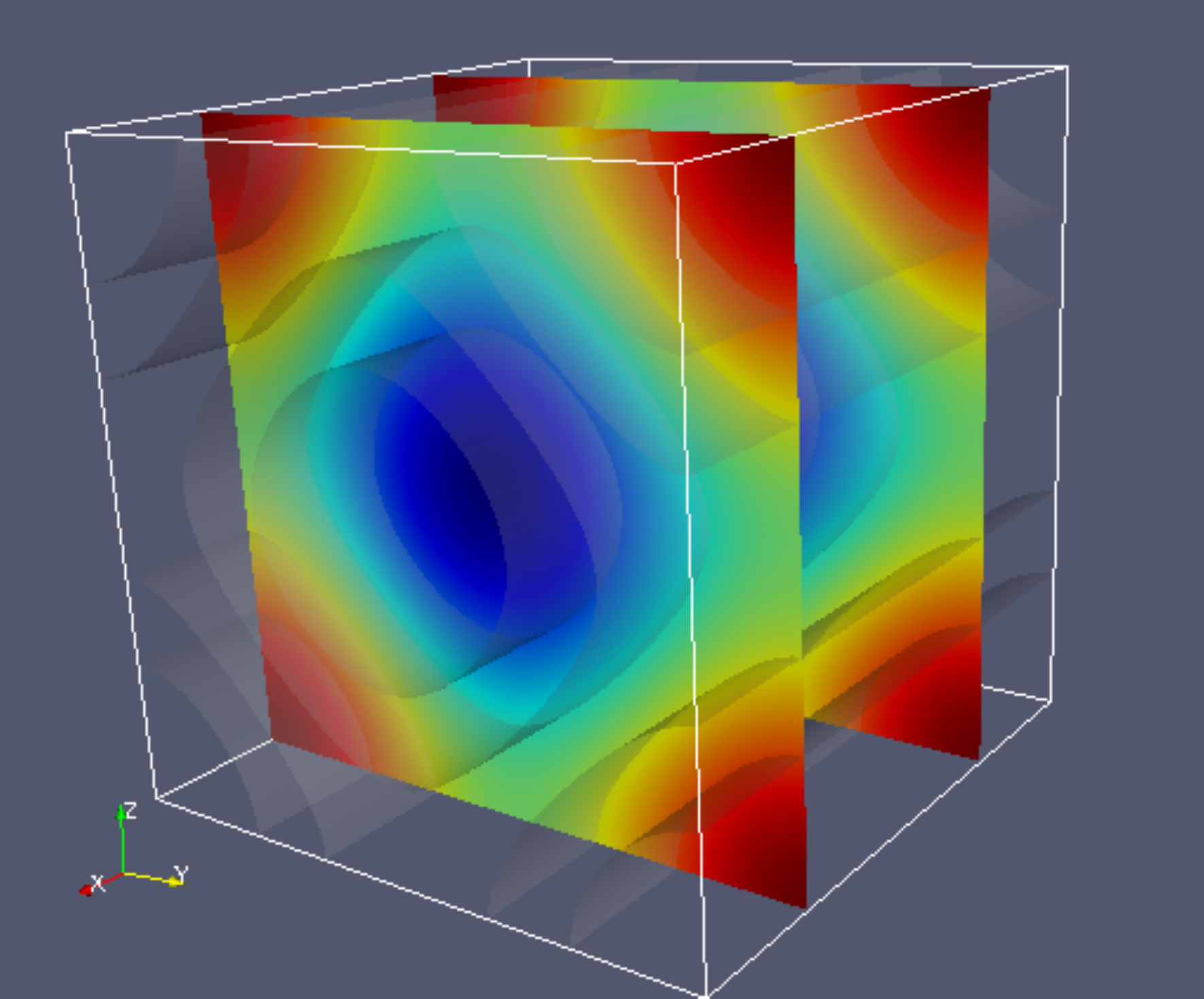}}
\subfigure[aligned ABC flow]{\includegraphics[width=0.32\textwidth]{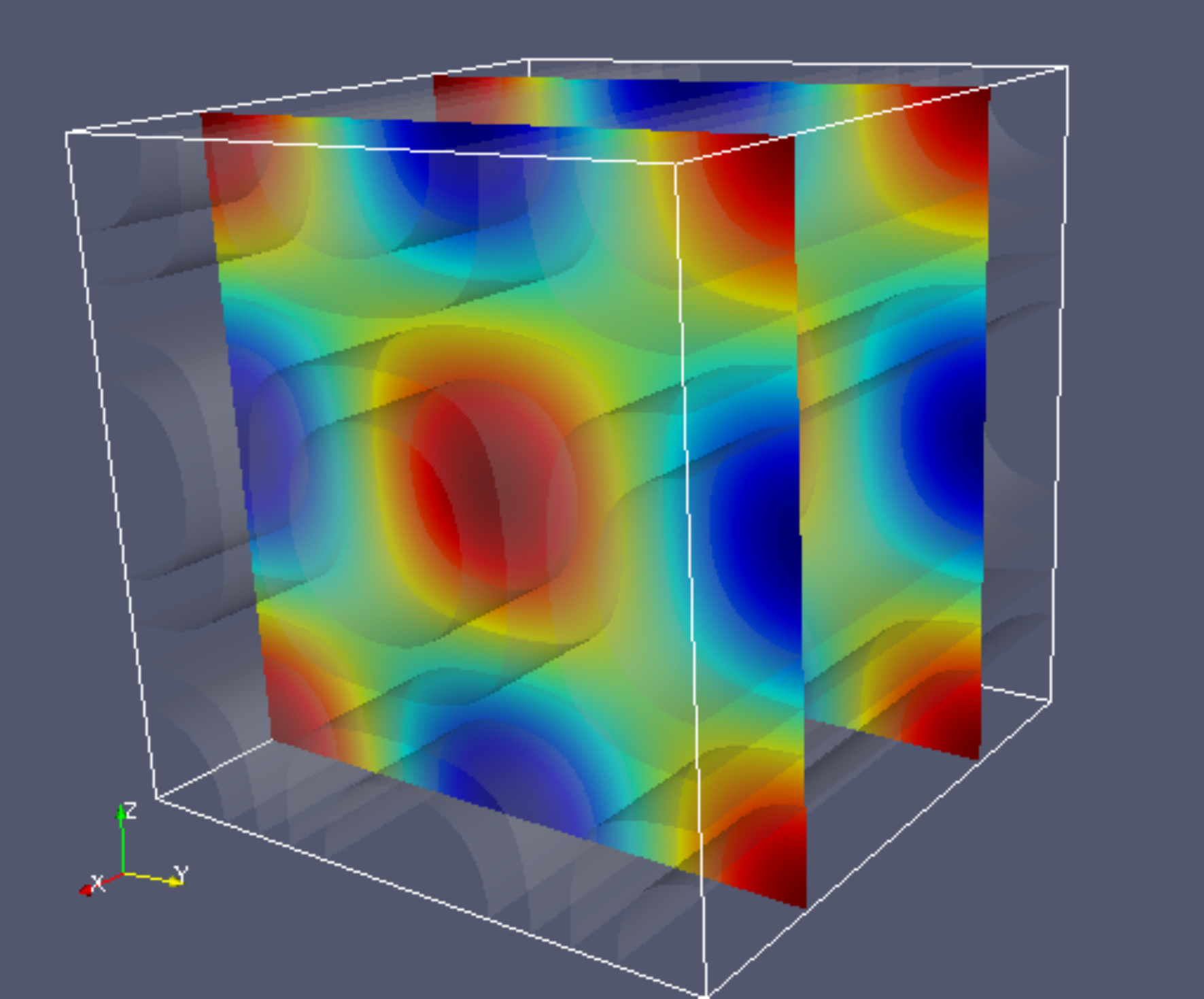}}
\subfigure[Taylor-Green flow]{\includegraphics[width=0.32\textwidth]{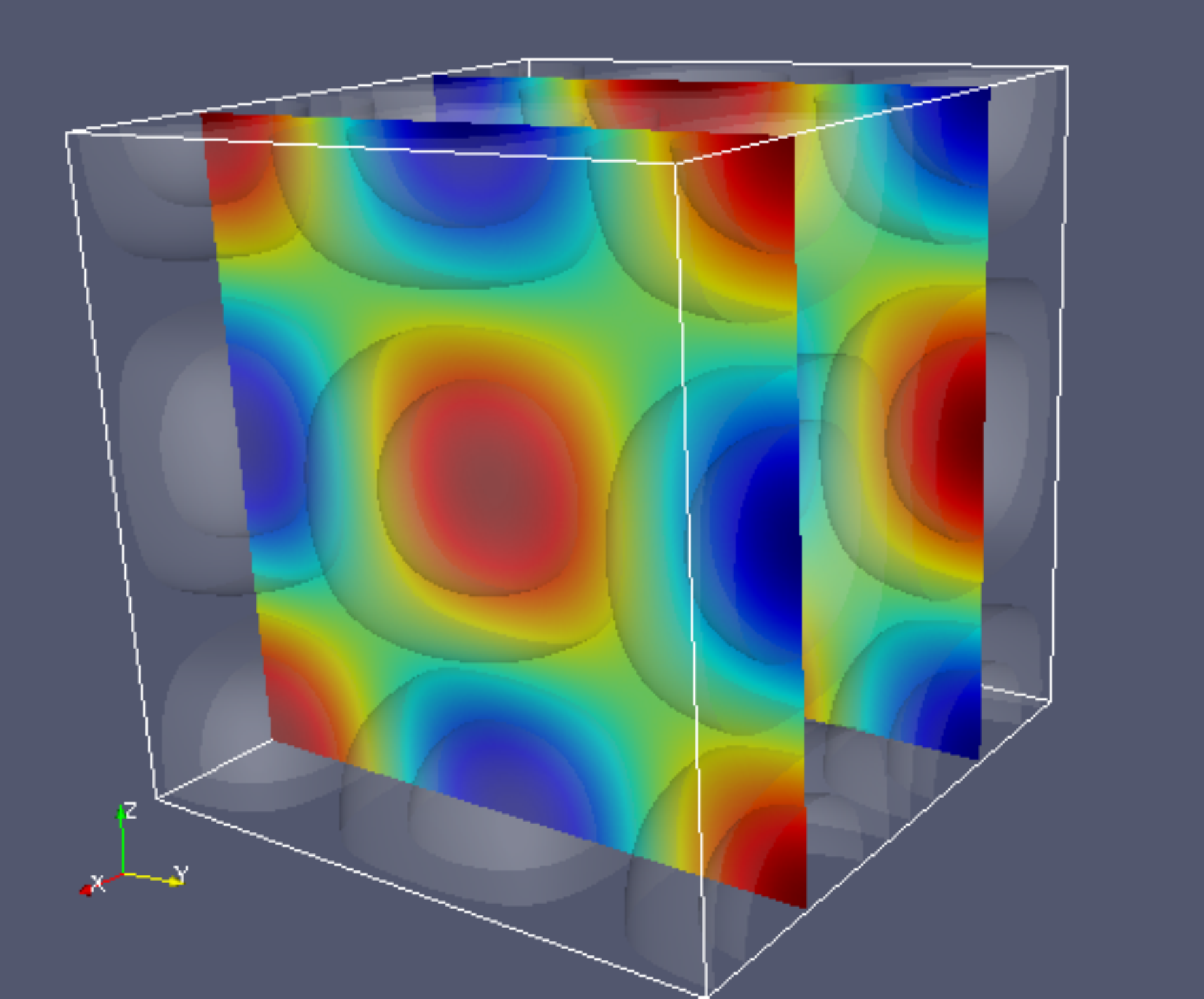}}
\caption[Optimal fields for $\E_0 \to 0$, (3D)]{Extreme vortex states
  obtained as solutions of maximization problem \ref{pb:maxdEdt_E} in
  the limit $\E_0 \to 0$ for different choices of set $\W_k$, as
  defined in equation \eqref{eq:defW}. Figures (a--c) represent the
  isosurfaces defined by the the relation $|\rot\uvec_1|(\xvec)
  = 0.95||\rot\uvec_1||_{L_\infty}$, whereas figures (d--f) depict the
  isosurfaces and cross-sectional distributions in the $y-z$
    plane of the $x_1$ component of the field $\uvec_1$. Case
  (\ref{c1}), cf.~\eqref{eq:uvec_3D_k1}, is presented in figures
  (a,d), case (\ref{c2}), cf.~\eqref{eq:uvec_3D_k2}, in figures (b,e),
  and case (iii), cf.~\eqref{eq:uvec_3D_k3}, in figures (c,f) (see
  also Table \ref{tab:E0}).}
\label{fig:maxdEdt_vortexCells}
\end{center}
\end{figure}

The three fields constructed in \eqref{eq:uvec_3D_k1},
\eqref{eq:uvec_3D_k2} and \eqref{eq:uvec_3D_k3} are visualized in
figure \ref{fig:maxdEdt_vortexCells}.  This analysis is performed
using the level sets $\Gamma_{s}(F)\subset\Omega$ defined as
\begin{equation}\label{eq:levelSets}
\Gamma_{s}(F) :=  \{\xvec\in\Omega : F(\xvec) = s \}, 
\end{equation}
for a suitable function $F:\Omega\to\mathbb{R}$. In figures
  \ref{fig:maxdEdt_vortexCells}(a--c) we choose $F(\xvec) =
  |\rot\uvec_1|(\xvec)$ with $s = 0.95||\rot\uvec_1||_{L_\infty}$.
To complement this information, in figures
\ref{fig:maxdEdt_vortexCells}(d--f) we also plot the isosurfaces and
cross-sectional distributions of the $x_1$ component of the field
$\uvec_1$.

The fields shown in figure \ref{fig:maxdEdt_vortexCells} reveal
  interesting patterns involving well-defined ``vortex cells''. More
  specifically, we see that in case (\ref{c1}), given by equation
\eqref{eq:uvec_3D_k1} and shown in figures
\ref{fig:maxdEdt_vortexCells}(a,d), the vortex cells are staggered
with respect to the orientation of the cubic domain $\Omega$ in
all three planes, whereas in case (\ref{c3}), given by equation
\eqref{eq:uvec_3D_k3} and shown in figures
\ref{fig:maxdEdt_vortexCells}(c,f), the vortex cells are aligned
with the domain $\Omega$ in all three planes. On the
  other hand, in case (\ref{c2}), given by equation
\eqref{eq:uvec_3D_k2} and shown in figures
\ref{fig:maxdEdt_vortexCells}(b,e), the vortex cells are staggered in
one plane and aligned in another with the arrangement in the third
plane resulting from the arrangement in the first two.  These
geometric properties are also reflected in the $x_1$-component of the
field $\uvec_1$ which is independent of $x_1$ in cases (\ref{c1}) and
(\ref{c2}), but exhibits, respectively, a staggered and aligned
arrangement of the cells in the $y-z$ plane in these two
cases. In case (\ref{c3}) the cells exhibit an aligned arrangement in
all three planes. The geometric properties of the extreme vortex
  states obtained in the limit $\E_0 \rightarrow 0$ are summarized in
  Table \ref{tab:E0}.  We remark that an analogous structure of the
optimal fields, featuring aligned and staggered arrangements of vortex
cells in the limiting case, was also discovered by \cite{ap13a} in
their study of the maximum palinstrophy growth in 2D. While due to a
smaller spatial dimension only two optimal solutions were found in
that study, the one characterized by the staggered arrangement also
lead to a larger (less negative) rate of palinstrophy production.

\begin{table}
  \begin{center}
    \begin{tabular}{l|c|c|c|c}      
        \Bmp{1.0cm} \small \begin{center} {\sc Case} \\ \smallskip \end{center} \Emp   
      & \Bmp{2.75cm} \small \begin{center} {\sc Formula for the velocity field}  \end{center} \Emp 
      & \Bmp{2.75cm} \small \begin{center} {\sc Arrangement of cells in $y-z$ plane}  \end{center} \Emp       
      & \Bmp{2.75cm} \small \begin{center} {\sc Dependence of $x_1$ component of $\uvec_1$ on $x_1$}  \end{center} \Emp       
      & \Bmp{3.0cm} \small \begin{center} {\sc Remarks}  \end{center} \Emp \\  
      \hline
        \Bmp{1.0cm} \small \begin{center} (\ref{c1}) \\ \smallskip \end{center} \Emp   
      & \Bmp{2.75cm} \small \begin{center} \eqref{eq:uvec_3D_k1}  \end{center} \Emp 
      & \Bmp{2.75cm} \small \begin{center} staggered \end{center} \Emp       
      & \Bmp{2.75cm} \small \begin{center} uniform  \end{center} \Emp       
      & \Bmp{3.0cm} \small \begin{center} staggered ABC flow  \end{center} \Emp \\
        \Bmp{1.0cm} \small \begin{center} (\ref{c2}) \\ \smallskip \end{center} \Emp   
      & \Bmp{2.75cm} \small \begin{center} \eqref{eq:uvec_3D_k2}  \end{center} \Emp 
      & \Bmp{2.75cm} \small \begin{center} aligned \end{center} \Emp       
      & \Bmp{2.75cm} \small \begin{center} uniform  \end{center} \Emp       
      & \Bmp{3.0cm} \small \begin{center} aligned ABC flow \end{center} \Emp \\
        \Bmp{1.0cm} \small \begin{center} (\ref{c3}) \\ \smallskip \end{center} \Emp   
      & \Bmp{2.75cm} \small \begin{center} \eqref{eq:uvec_3D_k3}  \end{center} \Emp 
      & \Bmp{2.75cm} \small \begin{center} aligned \end{center} \Emp       
      & \Bmp{2.75cm} \small \begin{center} cell-like  \end{center} \Emp       
      & \Bmp{3.0cm} \small \begin{center} Taylor-Green vortex  \end{center} \Emp \\
     \hline
    \end{tabular}
  \end{center}
  \caption{Summary of the properties of extreme vortex states $\uvec_1$ obtained as 
    solutions of optimization problem \ref{pb:maxdEdt_E} in the limit $\E_0 \to 0$.}
  \label{tab:E0}
\end{table}

It is also worth mentioning that the initial data for two
well-known flows, namely, the Arnold-Beltrami-Childress (ABC) flow
\citep{mb02} and the Taylor-Green flow \citep{tg37}, are in fact
particular instances of the optimal field $\uvec_1$ corresponding to,
respectively, cases (\ref{c1}) and (\ref{c3}).  Following the notation
of \citet{df86}, general ABC flows are characterized by the following
velocity field
\begin{equation}\label{eq:ABC_flow}
\begin{array}{r@{\,\,}c@{\,\,}l}
u_1(x_1,x_2,x_3) & = & A'\sin(2\pi x_3) + C'\cos(2\pi x_2), \\
u_2(x_1,x_2,x_3) & = & B'\sin(2\pi x_1) + A'\cos(2\pi x_3), \\
u_3(x_1,x_2,x_3) & = & C'\sin(2\pi x_2) + B'\cos(2\pi x_1),
\end{array}
\end{equation}
where $A'$, $B'$ and $C'$ real constants. The vector field in equation
\eqref{eq:ABC_flow} can be obtained from equation
\eqref{eq:uvec_3D_k1} by choosing $\mathbf{A} = (B'/2)[0,-i,1]$,
$\mathbf{B} = (C'/2)[1,0,-i]$ and $\mathbf{C} = (A'/2)[-i, 1,0]$.
By analogy, we will refer to the state described by
  \eqref{eq:uvec_3D_k2} as the ``aligned ABC flow''.  Likewise, the
well-known Taylor-Green vortex can be obtained as a particular
instance of the field $\uvec_1$ from equation \eqref{eq:uvec_3D_k3}
again using a suitable choice of the constants
$\mathbf{A},\mathbf{B},\mathbf{C},\mathbf{D}$.  Traditionally, the
velocity field $\uvec = [u_1,u_2,u_3]$ characterizing the Taylor-Green
vortex is defined as \citep{bmonmu83}
\begin{equation}\label{eq:TaylorGreen_vortex}
\begin{array}{r@{\,\,}c@{\,\,}l}
u_1(x_1,x_2,x_3) & = & \gamma_1\sin(2\pi x_1)\cos(2\pi x_2)\cos(2\pi x_3), \\
u_2(x_1,x_2,x_3) & = & \gamma_2\cos(2\pi x_1)\sin(2\pi x_2)\cos(2\pi x_3), \\
u_3(x_1,x_2,x_3) & = & \gamma_3\cos(2\pi x_1)\cos(2\pi x_2)\sin(2\pi x_3), \\
 0 & = & \gamma_1+\gamma_2+\gamma_3,  
\end{array}
\end{equation}
for $\gamma_1,\gamma_2,\gamma_3\in\mathbb{R}$. For given values of
$\gamma_1$,$ \gamma_2$ and $\gamma_3$ in
\eqref{eq:TaylorGreen_vortex}, the corresponding constants
$\mathbf{A},\mathbf{B},\mathbf{C},\mathbf{D}$ in \eqref{eq:uvec_3D_k3}
can be found by separating them into their real and imaginary
  parts denoted, respectively,
  $\Re{\mathbf{A}},\Re{\mathbf{B}},\Re{\mathbf{C}},\Re{\mathbf{D}}$
  and
  $\Im{\mathbf{A}},\Im{\mathbf{B}},\Im{\mathbf{C}},\Im{\mathbf{D}}$.
Then, after choosing
\begin{displaymath}
\Re{\mathbf{A}}=\Re{\mathbf{B}}=\Re{\mathbf{C}}=\Re{\mathbf{D}}=\mathbf{0} = [0,0,0], 
\end{displaymath}
the imaginary parts can be determined by solving the following
system of linear equations
\begin{displaymath}
2\left[
\begin{array}{cccc}
I_3 & -I_3 & -I_3 & -I_3 \\ 
-I_3 & I_3 & -I_3 & -I_3 \\ 
-I_3 & -I_3 & I_3 & -I_3 \\ 
-I_3 & -I_3 & -I_3 & I_3   
\end{array} 
\right]
\left[
\begin{array}{c}
\Im{\mathbf{A}} \\
\Im{\mathbf{B}} \\
\Im{\mathbf{C}} \\
\Im{\mathbf{D}}
\end{array}
\right] = 
\left[
\begin{array}{c}
\mathbf{0} \\
\gamma_1\mathbf{e}_1 \\
\gamma_2\mathbf{e}_2 \\
\gamma_3\mathbf{e}_3
\end{array}
\right],
\end{displaymath}
where $I_3$ is the $3 \times 3$ identity matrix. The values of
$\Im{\mathbf{A}},\ldots,\Im{\mathbf{D}}$ are thus given by
\begin{displaymath}
\Im{\mathbf{A}} = -\frac{1}{8}\left[ \begin{array}{c}\gamma_1 \\ \gamma_2 \\ \gamma_3 \end{array} \right],\, 
\Im{\mathbf{B}} = -\frac{1}{8}\left[ \begin{array}{c}-\gamma_1 \\ \gamma_2 \\ \gamma_3 \end{array} \right],\,
\Im{\mathbf{C}} = -\frac{1}{8}\left[ \begin{array}{c}\gamma_1 \\ -\gamma_2 \\ \gamma_3 \end{array} \right],\,
\Im{\mathbf{D}} = -\frac{1}{8}\left[ \begin{array}{c}\gamma_1 \\ \gamma_2 \\ -\gamma_3 \end{array} \right].
\end{displaymath}
A typical choice of the parameters used in the numerical studies
performed by \cite{bmonmu83} and \cite{b91} is $\gamma_1 = - \gamma_2
= 1$ and $\gamma_3 = 0$. 

We remark that the Taylor-Green vortex has been employed as the
initial data in a number of studies aimed at triggering singular
behaviour in both the Euler and Navier-Stokes systems
\citep{tg37,bmonmu83,b91,bb12}. It is therefore interesting to note
that it arises in the limit $\E_0 \rightarrow 0$ as one of the extreme
vortex states in the variational formulation considered in the present
study. It should be emphasized, however, that out of the three optimal
states identified above (see Table \ref{tab:E0}), the
Taylor-Green vortex is characterized by the smallest (i.e., the most
negative) instantaneous rate of enstrophy production $d\E/dt$.
On the other hand, we are not aware of any prior studies
  involving ABC flows in the context of extreme behaviour and potential
  singularity formation. The time evolution corresponding to these
states and some other initial data will be analyzed in detail
in \S \ref{sec:timeEvolution}.

\section{Extreme Vortex States with Finite $\E_0$}
\label{sec:3D_InstOpt_E}

In this section we analyze the optimal vortex states $\tuvecE$
obtained for finite values of the enstrophy in which we extend
the results obtained in the seminal study by \cite{ld08}.  As was also
the case in the analogous study in 2D \citep{ap13a}, there is a
distinct branch of extreme states $\tuvecE$ parameterized by the
enstrophy $\E_0$ and corresponding to each of the three limiting
states discussed in \S \ref{sec:3D_InstOpt_E0to0} (cf.~figure
\ref{fig:maxdEdt_vortexCells} and Table \ref{tab:E0}). Each of these
branches is computed using the continuation approach outlined in
  Algorithm \ref{alg:optimAlg}. As a key element of the
gradient-based maximization technique \eqref{eq:desc},
the gradient expressions \eqref{eq:gradRL2}--\eqref{eq:gradRH2}
are approximated pseudo-spectrally using standard dealiasing of
the nonlinear terms and with resolutions varying from $128^3$ in the
low-enstrophy cases to $512^3$ in the high-enstrophy cases, which
necessitated a massively parallel implementation using the Message
Passing Interface (MPI).  As regards the computation of the Sobolev
$H^2$ gradients, cf.~\eqref{eq:gradRH2}, we set $\ell_1 = 0$, whereas
the second parameter $\ell_2$ was adjusted during the optimization
iterations and was chosen so that $\ell_2 \in [\ell_{\min},
\ell_{\max}]$, where $\ell_{\min}$ is the length scale associated with
the spatial resolution $N$ used for computations and $\ell_{\max}$ is
the characteristic length scale of the domain $\Omega$, that is,
$\ell_{\min} \sim \O( 1/N) $ and $\ell_{\max} \sim \O(1)$. We remark
that, given the equivalence of the inner products \eqref{eq:ipH2}
corresponding to different values of $\ell_1$ and $\ell_2$ (as long as
$\ell_2 \neq 0$), these choices do not affect the maximizers found,
but only how rapidly they are approached by iterations
\eqref{eq:desc}. For further details concerning the computational
approach we refer the reader to the dissertation by \cite{a14}.  As
was the case in the analogous 2D problem studied by \citet{ap13a}, the
largest instantaneous growth of enstrophy is produced by the states
with vortex cells staggered in all planes, cf.~case (\ref{c1}) in
Table \ref{tab:E0}. Therefore, in our analysis we will focus
exclusively on this branch of extreme vortex states which has been
computed for $\E_0 \in [10^{-3},2\times10^2]$.

\begin{figure}
\linespread{1.1}
\setcounter{subfigure}{0}
\begin{center}
\subfigure[]{\includegraphics[width=0.49\textwidth]{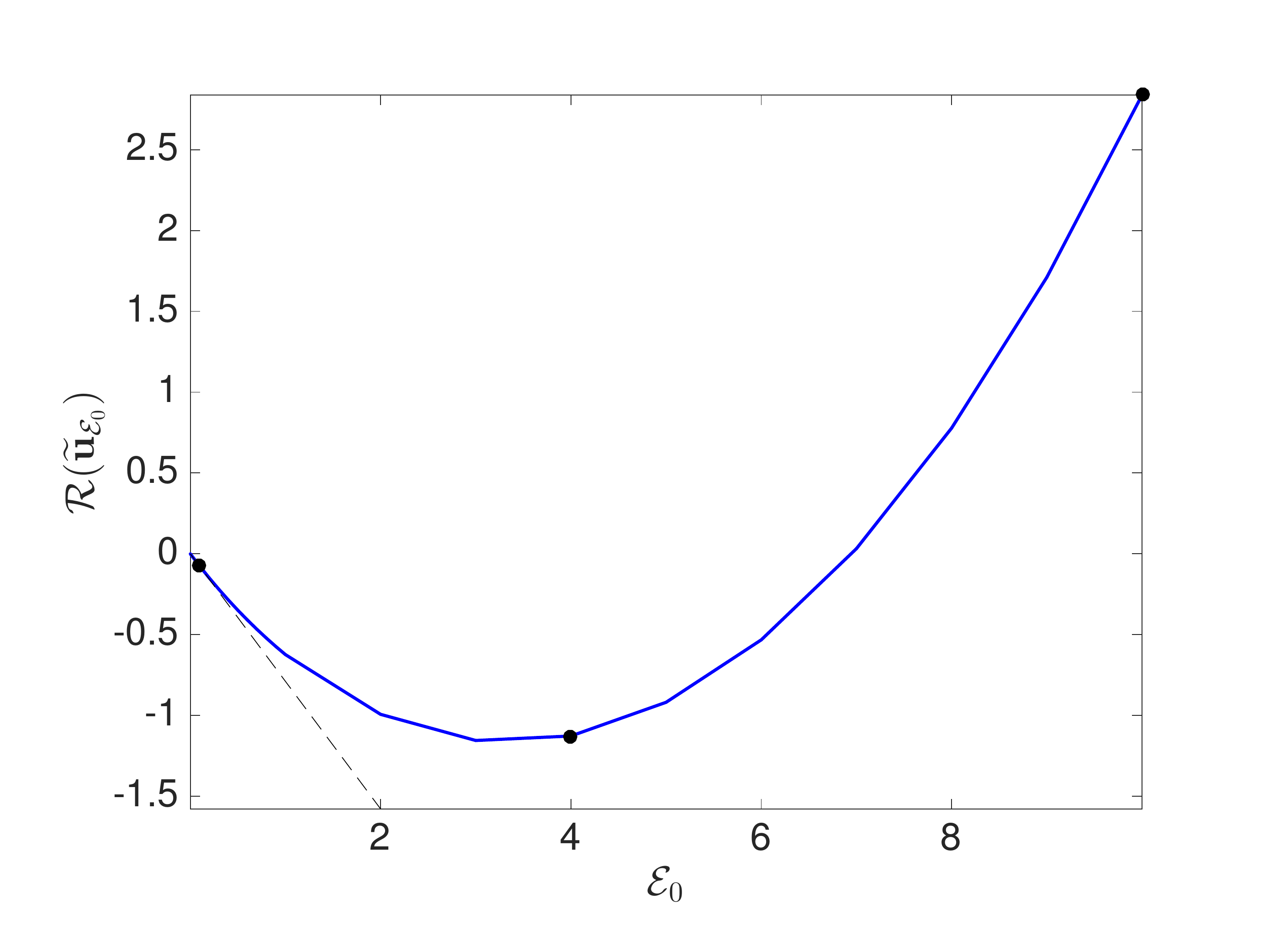}}
\subfigure[]{\includegraphics[width=0.49\textwidth]{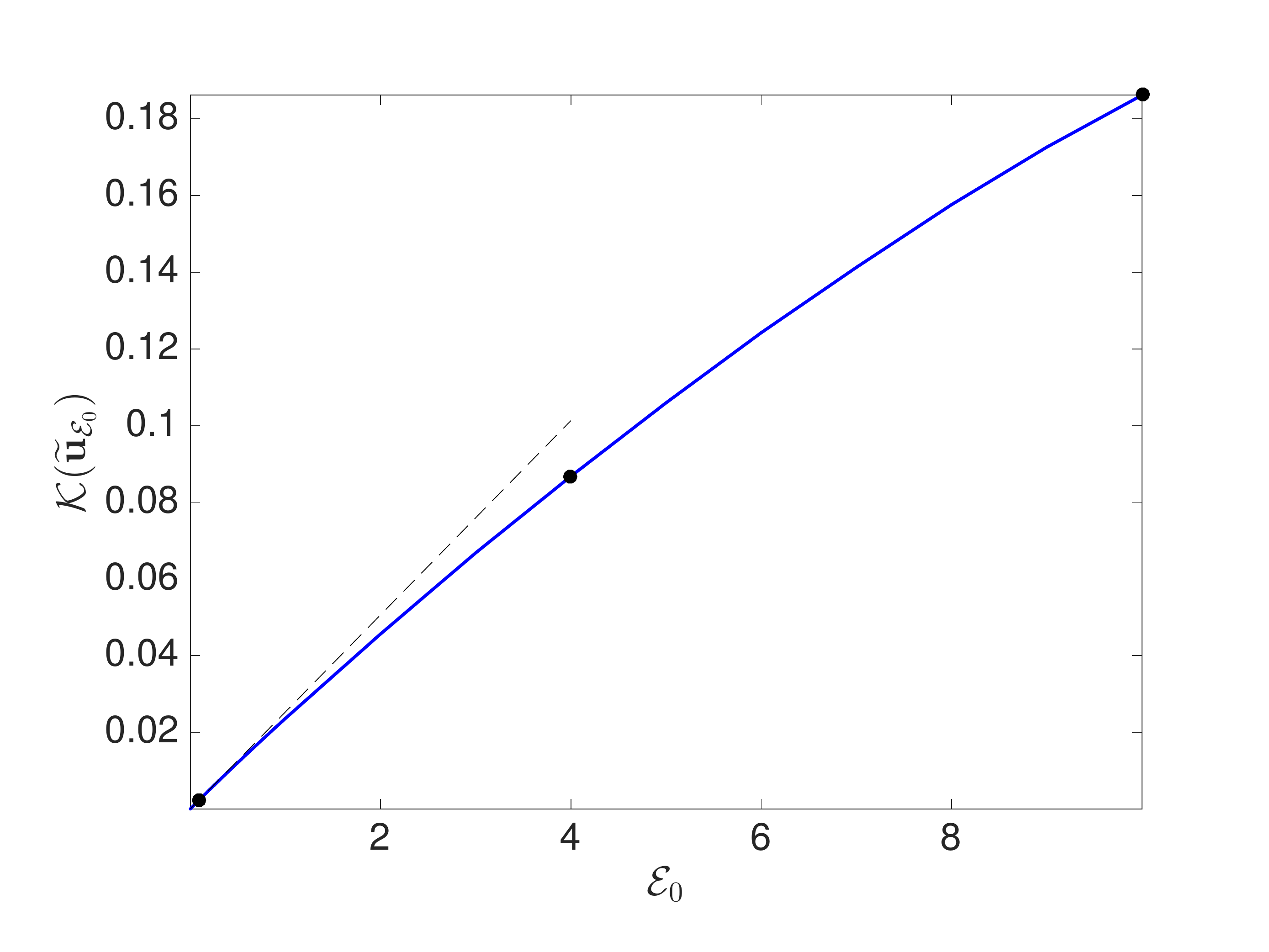}}\\
\subfigure[$\E_0 = 1\times10^{-2}$]{\includegraphics[width=0.3\textwidth]{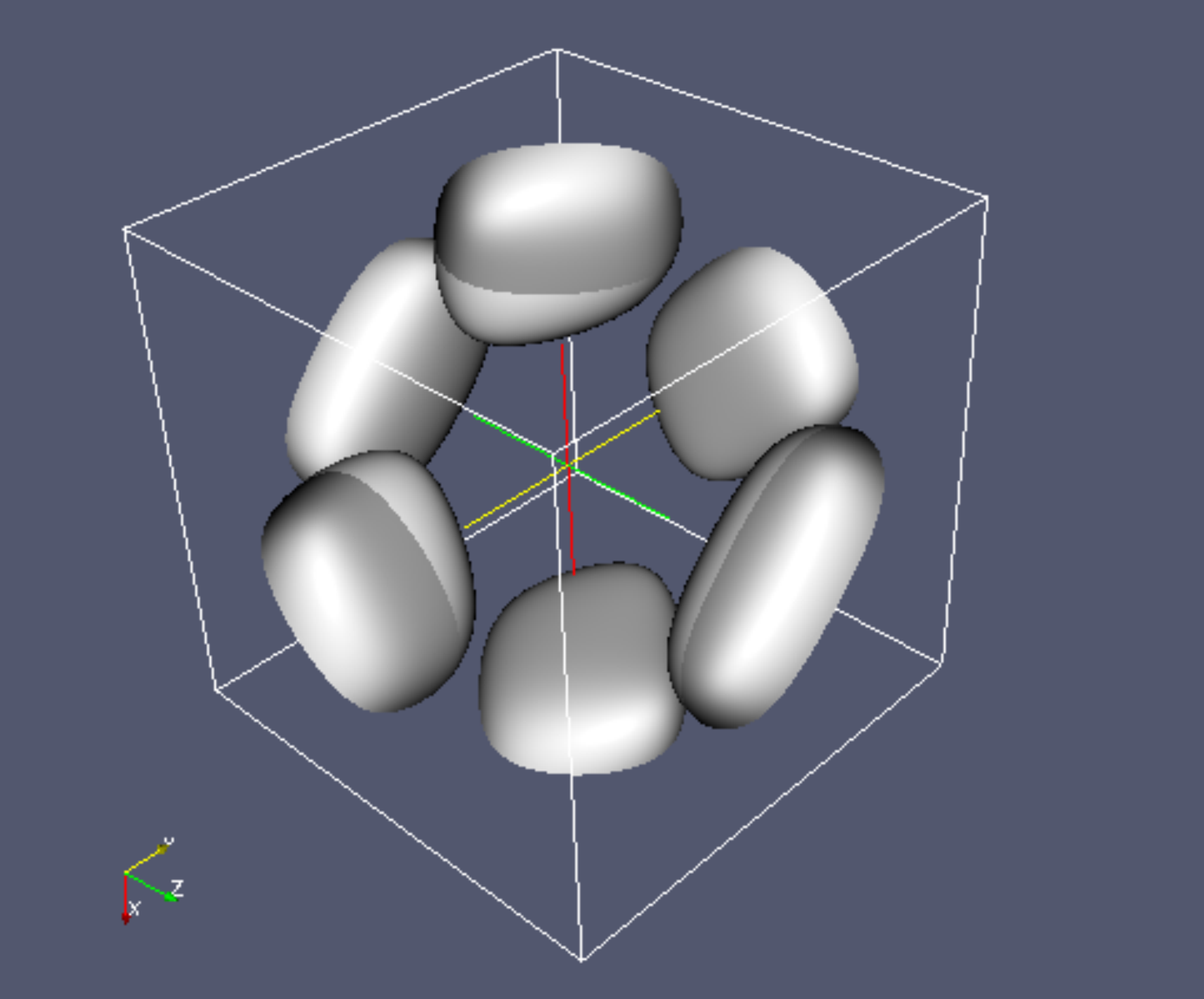}}
\subfigure[$\E_0 = 4$]{\includegraphics[width=0.3\textwidth]{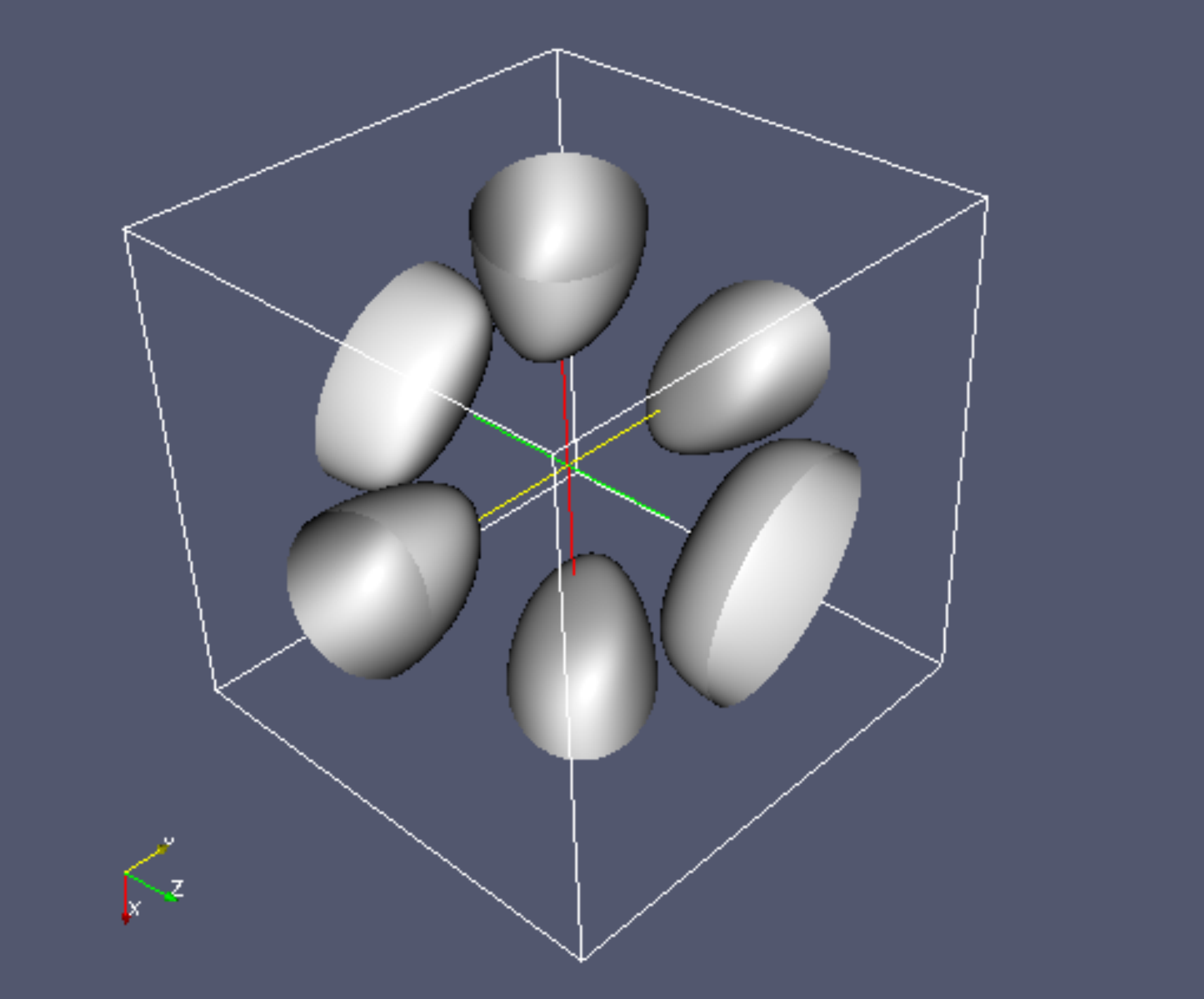}}
\subfigure[$\E_0 = 10$]{\includegraphics[width=0.3\textwidth]{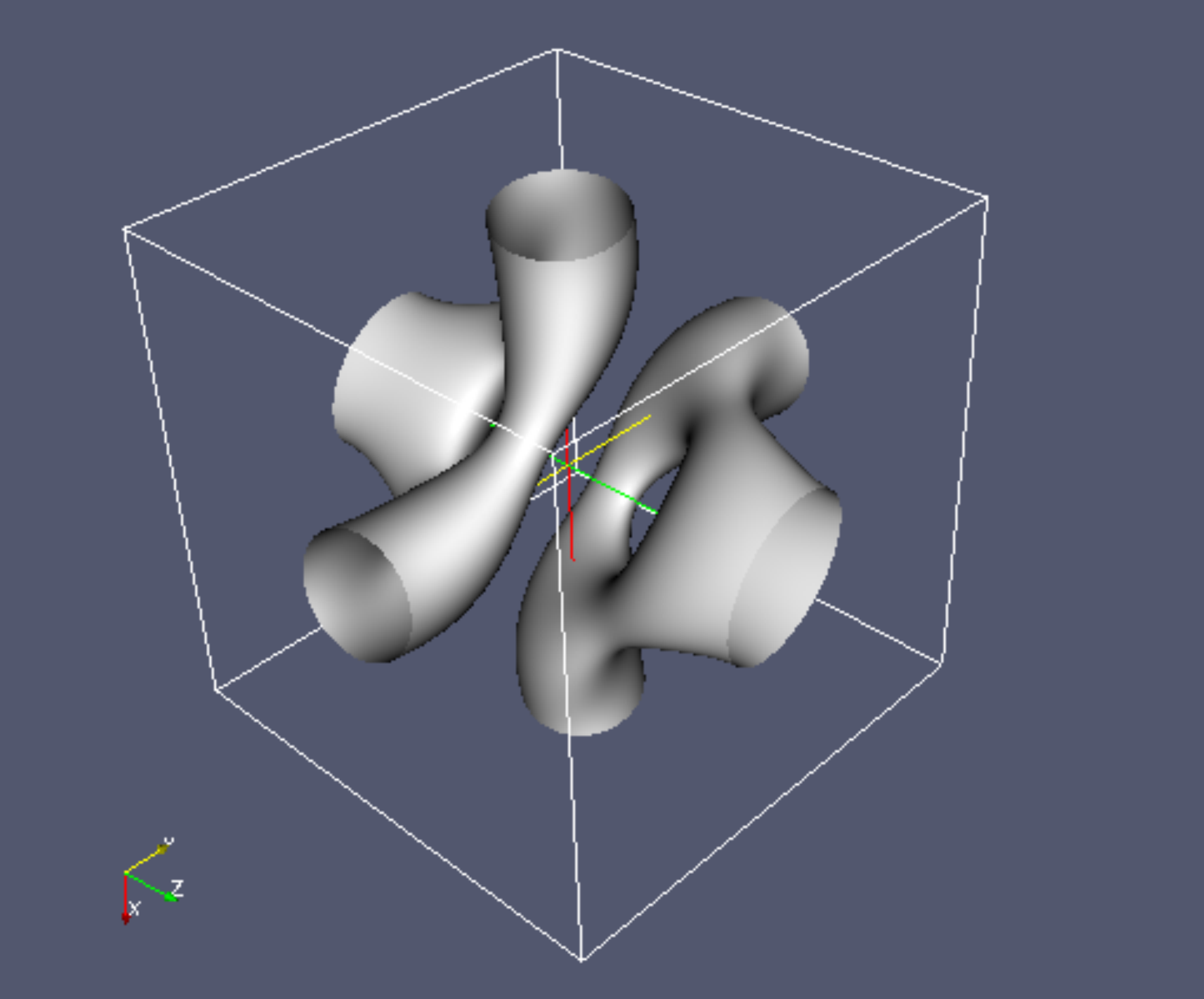}}\\
\subfigure[$\E_0 = 1\times10^{-2}$]{\includegraphics[width=0.3\textwidth]{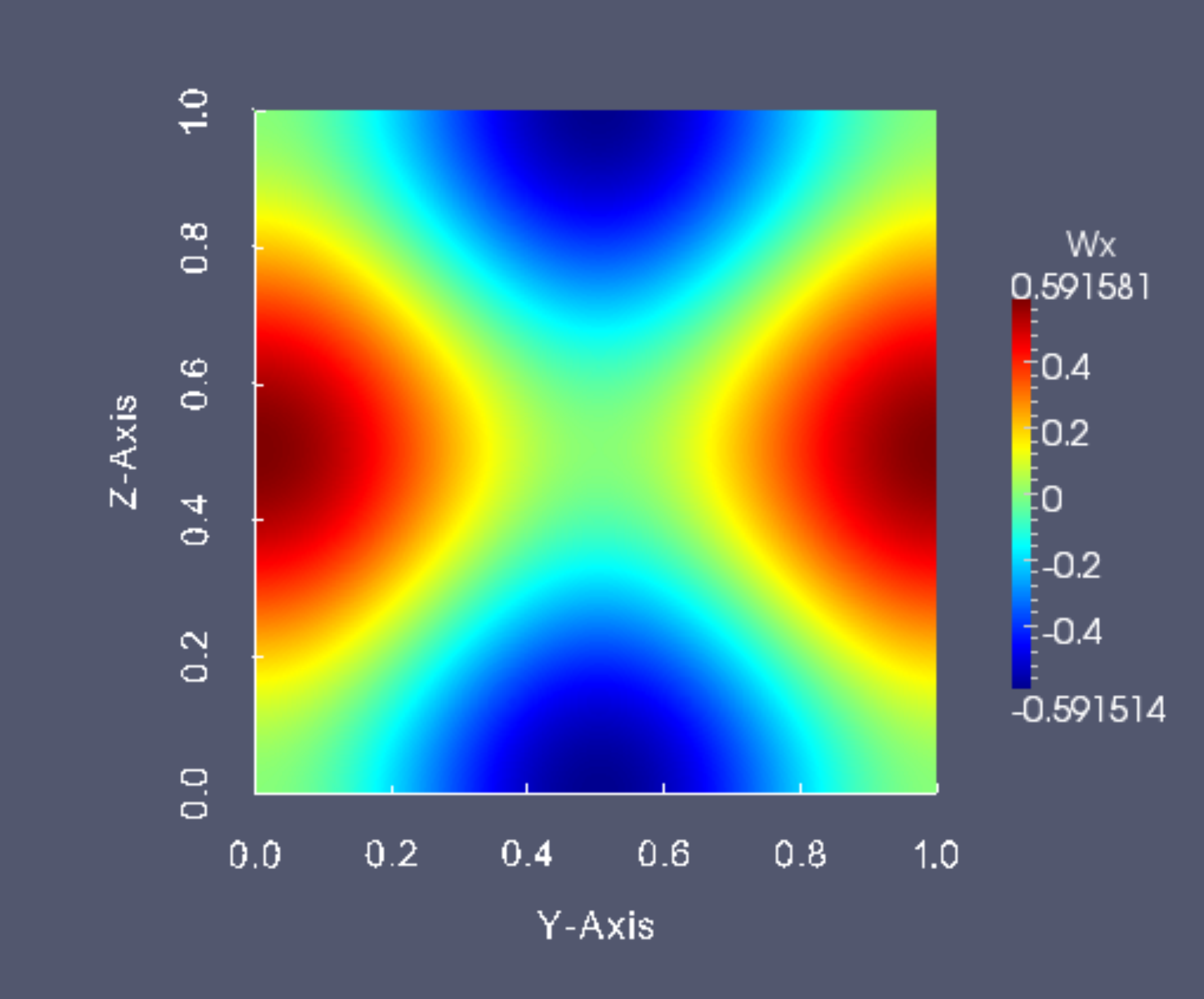}}
\subfigure[$\E_0 = 4$]{\includegraphics[width=0.3\textwidth]{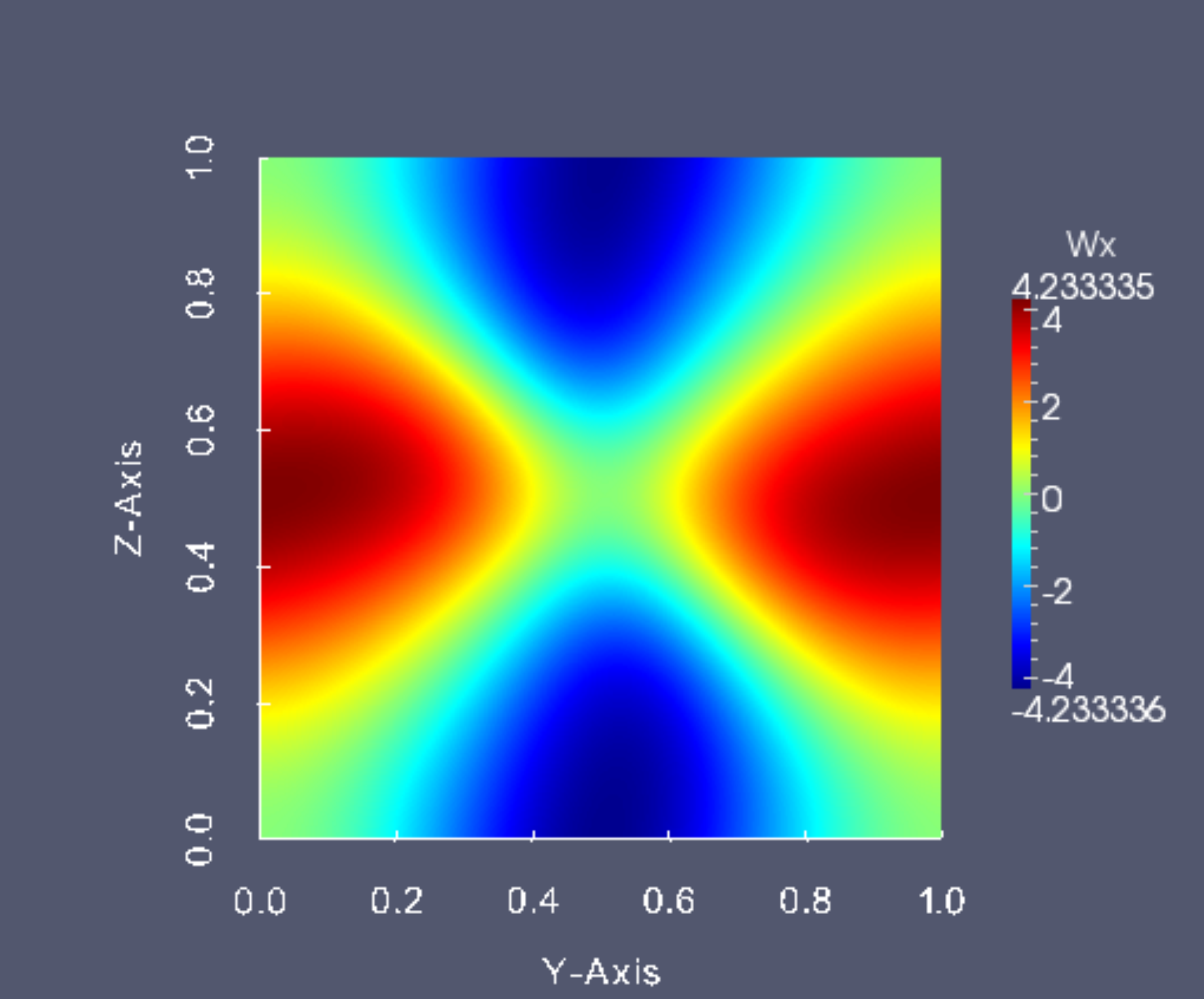}}
\subfigure[$\E_0 = 10$]{\includegraphics[width=0.3\textwidth]{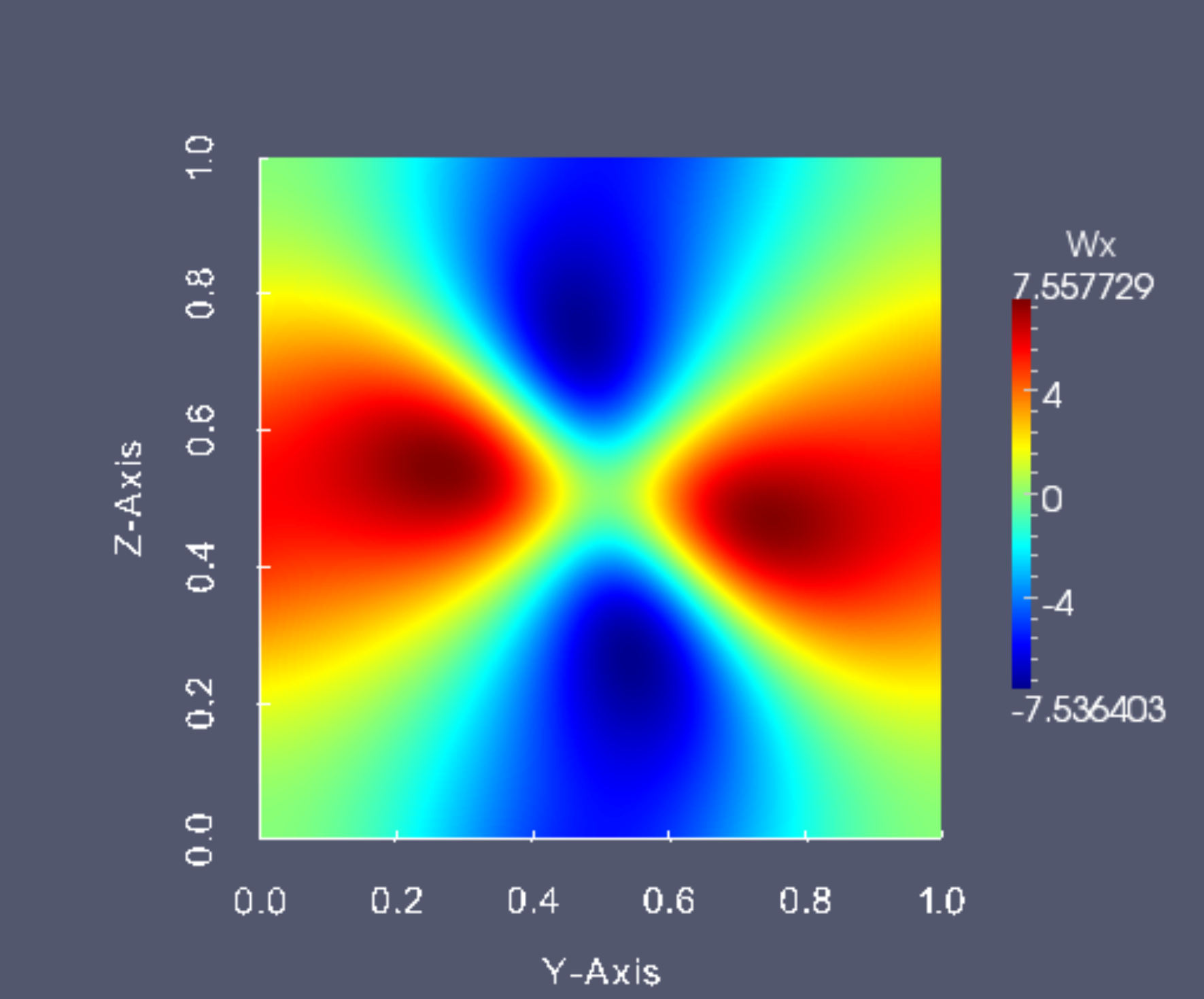}}
\caption[Maximum $d\E/dt$ for $\E_0 \to 0$, $(\E_0)$-Constraint (3D)]{
  (a) Maximum rate of growth of enstrophy $\R(\tuvecE)$ and (b) energy
  of optimal states $\K(\tuvecE)$ as functions of $\E_0$ for small
  values of enstrophy; the dashed lines represent the asymptotic
  relation \eqref{eq:R0_kvec_3D_k1} (a) and the Poincar\'e limit $\K_0
  = \E_0 / (2\pi)^2$ (b). (c)--(h) Extreme vortex states
    $\tuvecE$ obtained for the three values of enstrophy $\E_0$
    indicated with solid symbols in figures (a) and (b): panels
    (c)--(e) show the isosurfaces corresponding to $Q(\xvec) =
  \frac{1}{2}||Q||_{L_\infty}$ with $Q$ defined in equation
  \eqref{eq:Q_3D}, whereas panels (f)--(h) show the component
  of vorticity normal to the plane defined by $\mathbf{n}\cdot(\xvec -
  \xvec_0) = 0$, $\mathbf{n} = [1,0,0]$ and $\xvec_0 = [1/2,1/2,1/2]$.
}
\label{fig:RvsE0_FixE_small}
\end{center}
\end{figure}

The optimal instantaneous rate of growth of enstrophy $\R_{\E_0} =
\R(\tuvecE)$ and the energy of the optimal states $\K(\tuvecE)$ are
shown as functions of $\E_0$ for small $\E_0$ in figures
\ref{fig:RvsE0_FixE_small}(a) and \ref{fig:RvsE0_FixE_small}(b),
respectively. As indicated by the asymptotic form of $\R$ in
\eqref{eq:R0_kvec_3D_k1} and the Poincar\'e limit
{$\K_0=\E_0/(2\pi)^2$}, both of which are marked in these figures, the
behaviour of $\R_{\E_0}$ and $\K(\tuvecE)$ as $\E_0 \rightarrow 0$ is
correctly captured by the numerically computed optimal states.
In particular, we note that $\R_{\E_0}$ is negative for $0 \le
  \E_0 \lessapprox 7$ and exhibits the same trend as predicted in
  \eqref{eq:R03D} for $\E_0 \rightarrow 0$. For larger values of $\E_0$
  the rate of growth of enstrophy becomes positive. Likewise, the
asymptotic behaviour of the energy of the optimal fields does not come
as a surprise since, as discussed in \S\ref{sec:3D_InstOpt_E0to0}, in
the limit $\E_0 \to 0$ the maximizers of $\R$ are eigenfunctions of
the Laplacian, which also happen to saturate Poincar\'e's inequality.

The structure of the optimal vortex states $\tuvecE$ is analyzed
  next. They are visualized using \eqref{eq:levelSets} in which the
  vortex cores are identified as regions $\Sigma := \{\Gamma_{s}(Q): s
  \ge 0 \}$ for $Q$ defined as \citep{davidson:turbulence}
\begin{equation} \label{eq:Q_3D}
Q(\xvec) := \, \frac{1}{2}\left[ \tr(\bm{\Omega}\bm{\Omega}^T) - \tr(\mathbf{S}\mathbf{S}^T) \right], 
\end{equation}
where $\mathbf{S}$ and $\bm{\Omega}$ are the symmetric and
anti-symmetric parts of the velocity gradient tensor $\nabla\uvec$,
that is, $[\mathbf{S}]_{ij} = \frac{1}{2}(\partial_j u_i + \partial_i
u_j)$ and $[\bm{\Omega}]_{ij} = \frac{1}{2}(\partial_j u_i -
\partial_i u_j )$, $i,j=1,2,3$. The quantity $Q$ can be interpreted as
the local balance between the strain rate and the vorticity magnitude.
The isosurfaces $\Gamma_{0}(Q - 0.5||Q||_{L_\infty})$ representing the
optimal states $\tuvecE$ with selected values of $\E_0$ are shown in
figures \ref{fig:RvsE0_FixE_small}(c)-(e). For the smallest
values of $\E_0$, the optimal fields exhibit a cellular structure
already observed in figure \ref{fig:maxdEdt_vortexCells}(a).
For increasing values of $\E_0$ this cellular structure transforms
into a vortex ring, as seen in figure
\ref{fig:RvsE0_FixE_small}(e).  The component of vorticity normal to
the plane $P_x = \{\xvec\in\Omega : \mathbf{n}\cdot(\xvec-\xvec_0) = 0
\}$ for $\mathbf{n} = [1,0,0]$ and $\xvec_0 = [1/2,1/2,1/2]$ is shown
in figures \ref{fig:RvsE0_FixE_small}(f)-(h), where the transition
from cellular structures to a localized vortex structure as
  enstrophy increases is evident.

\begin{figure}
\linespread{1.1}
\setcounter{subfigure}{0}
\begin{center}
\subfigure[]{\includegraphics[width=0.49\textwidth]{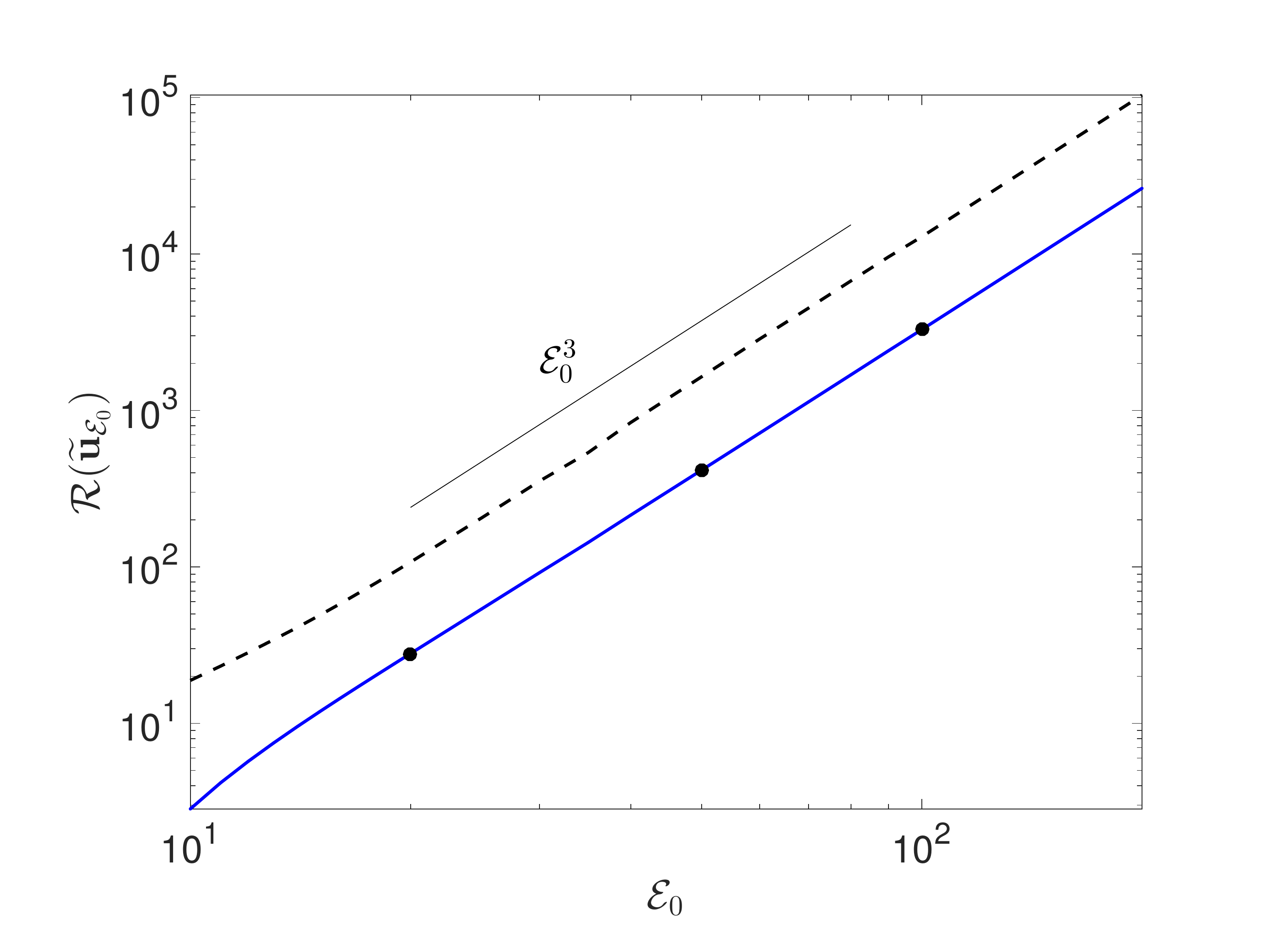}}
\subfigure[]{\includegraphics[width=0.49\textwidth]{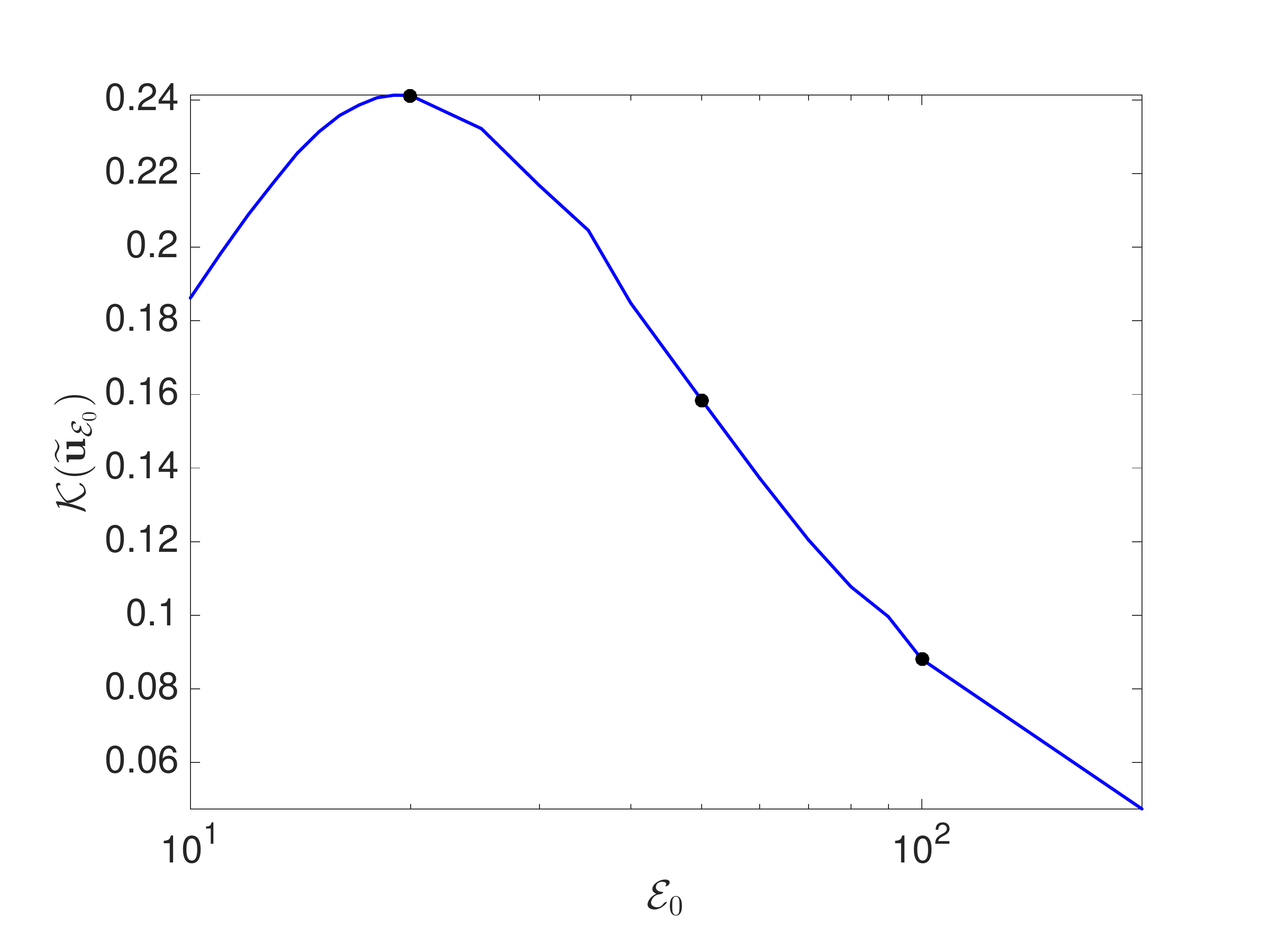}}\\
\subfigure[$\E_0 = 20$]{\includegraphics[width=0.3\textwidth]{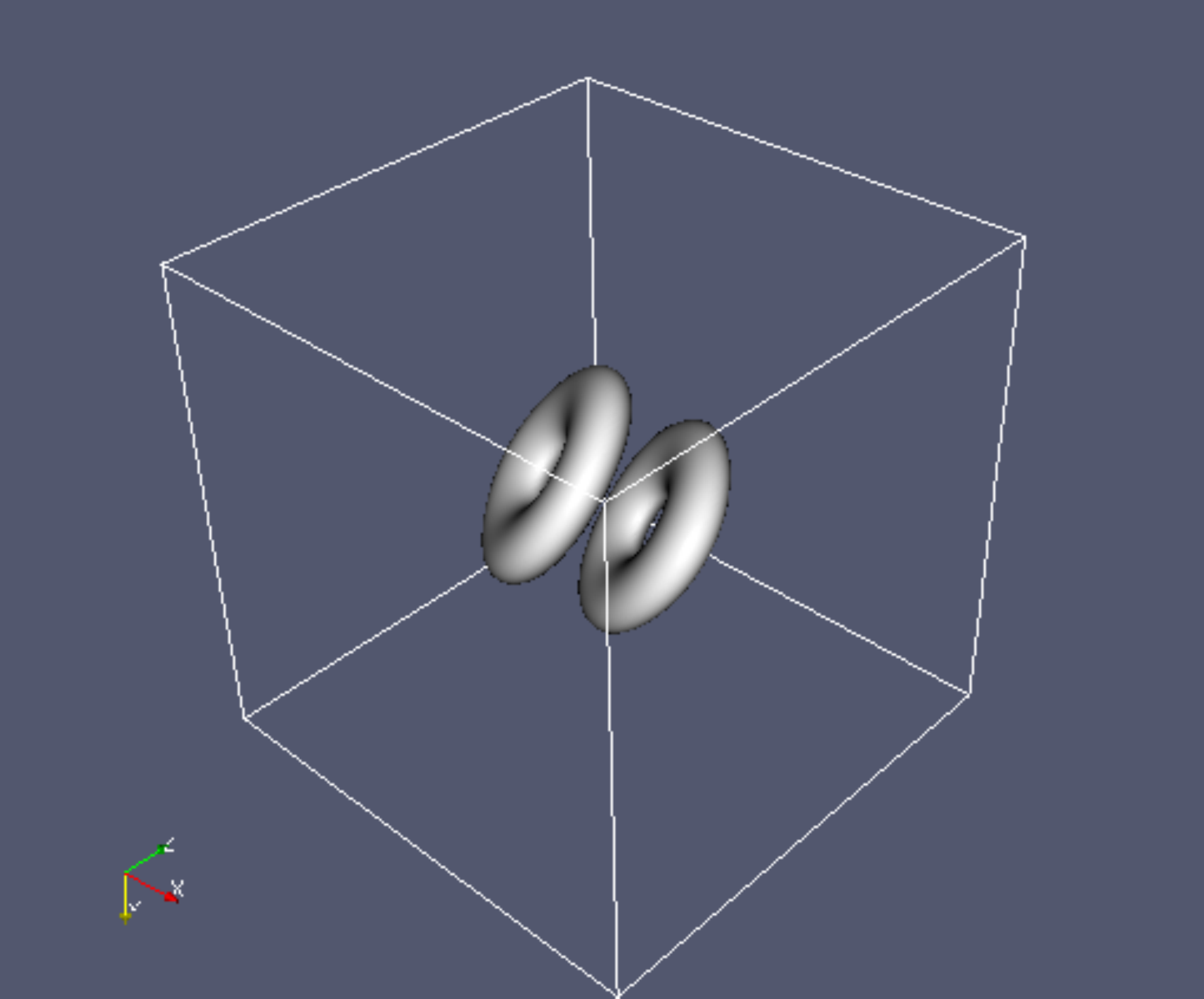}}
\subfigure[$\E_0 = 50$]{\includegraphics[width=0.3\textwidth]{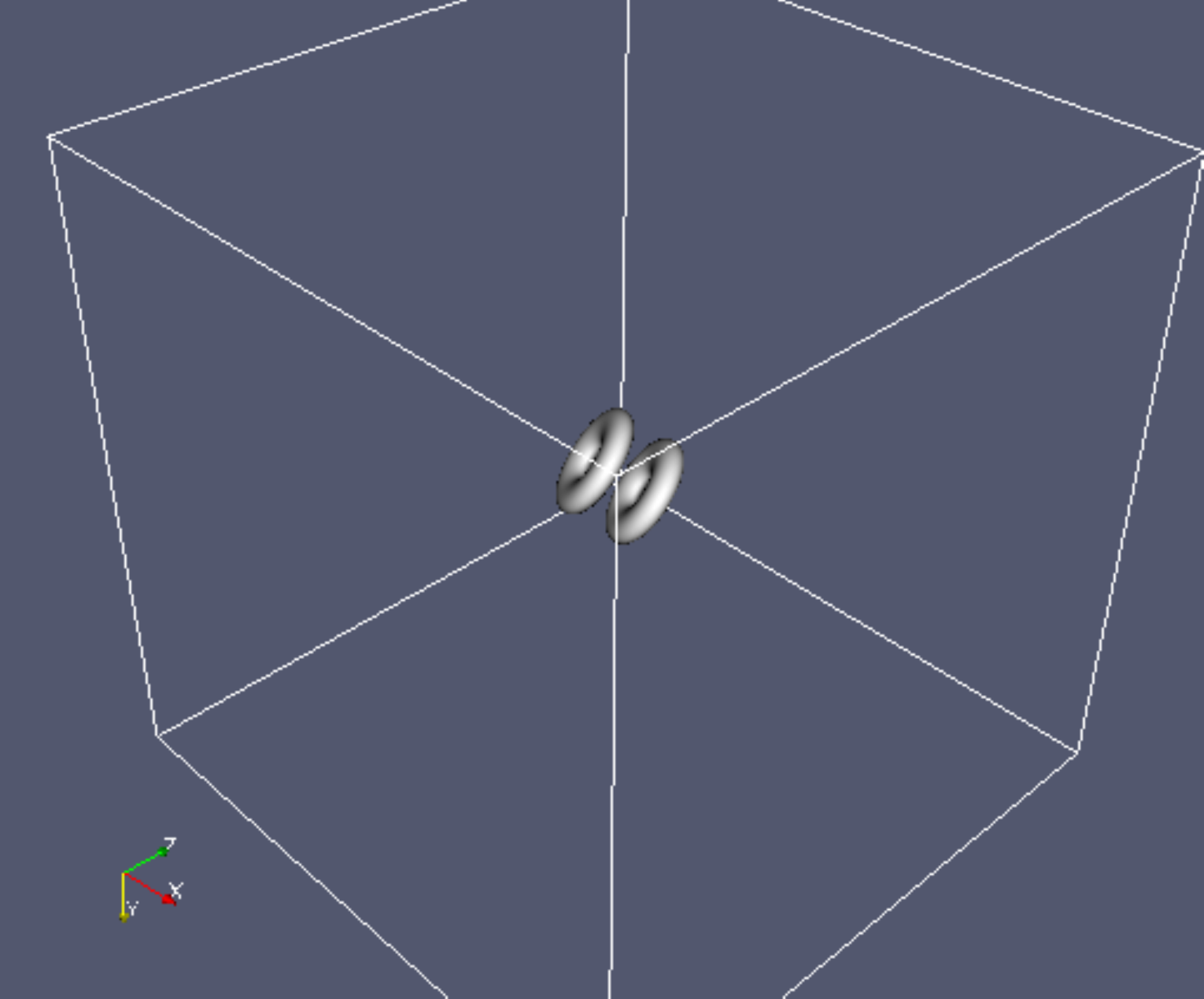}}
\subfigure[$\E_0 = 100$]{\includegraphics[width=0.3\textwidth]{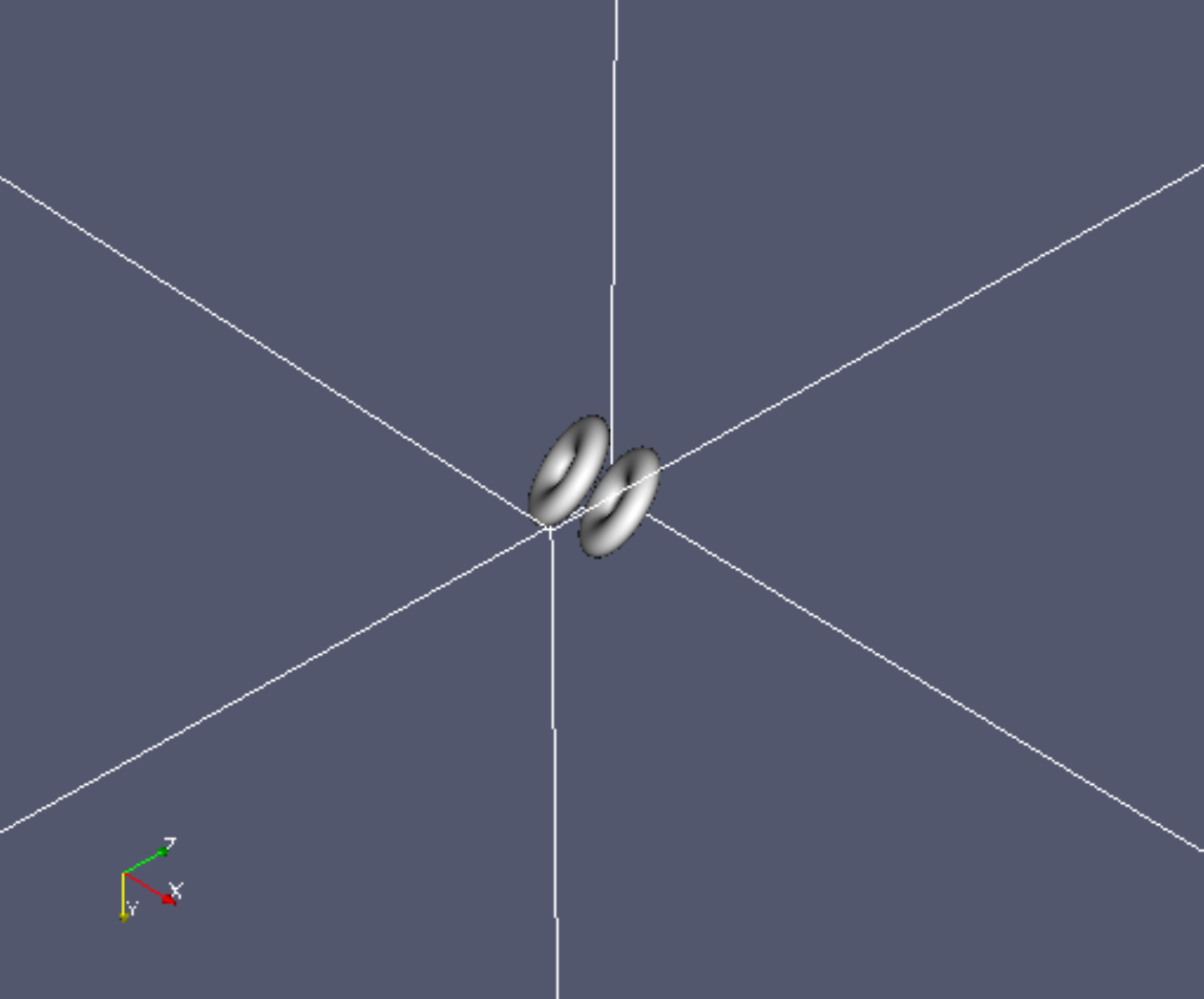}}\\
\subfigure[$\E_0 = 20$]{\includegraphics[width=0.3\textwidth]{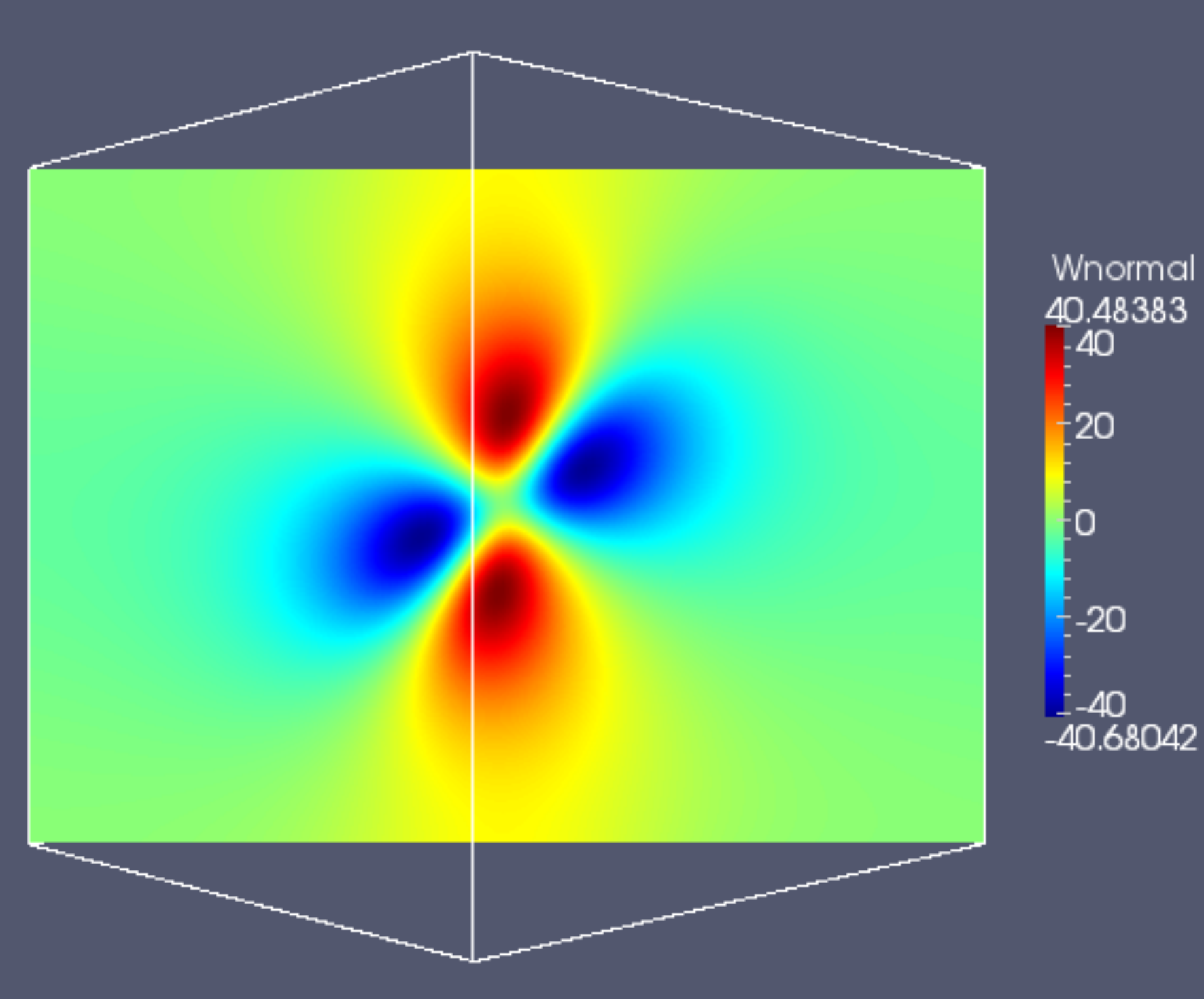}}
\subfigure[$\E_0 = 50$]{\includegraphics[width=0.3\textwidth]{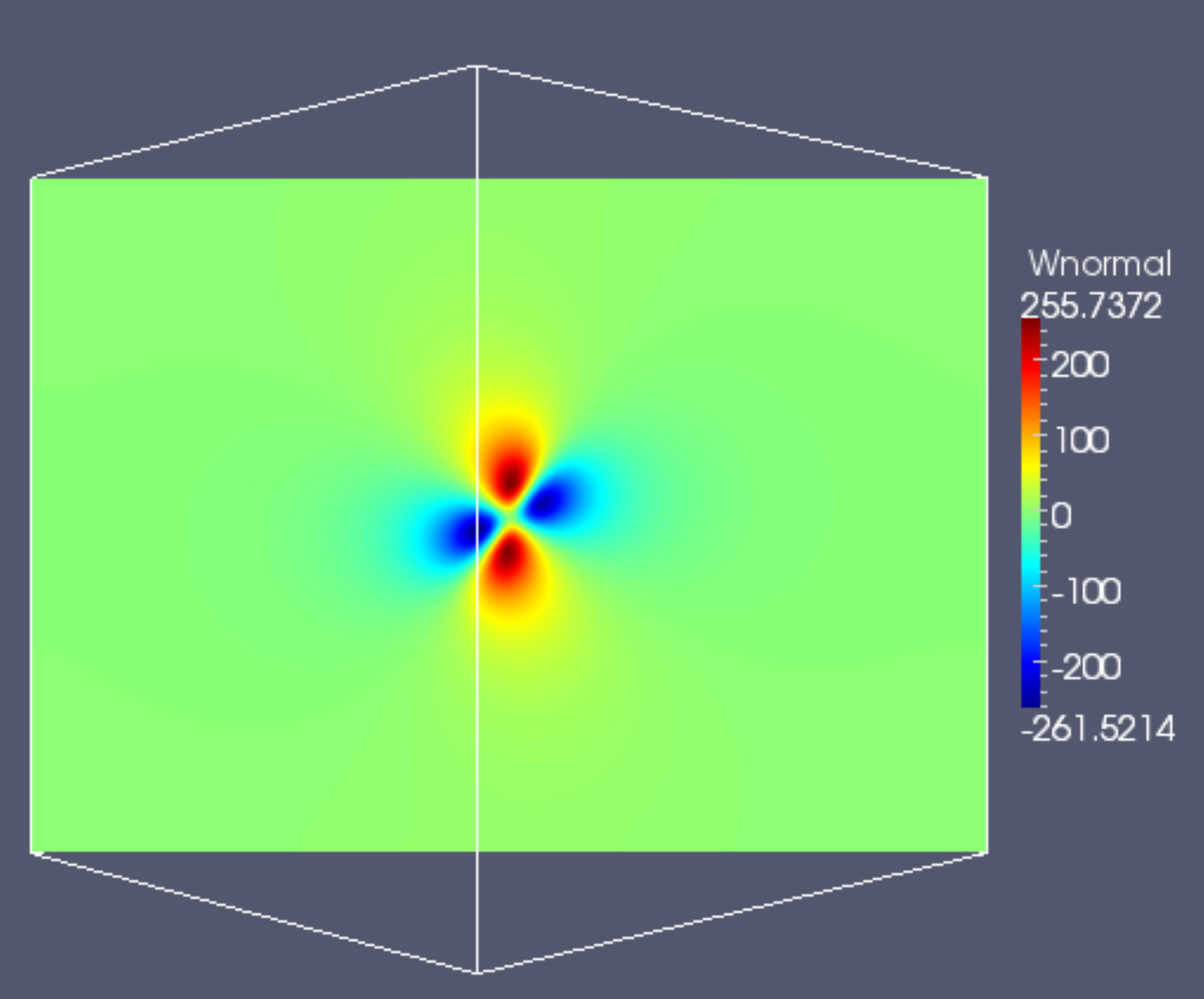}}
\subfigure[$\E_0 = 100$]{\includegraphics[width=0.3\textwidth]{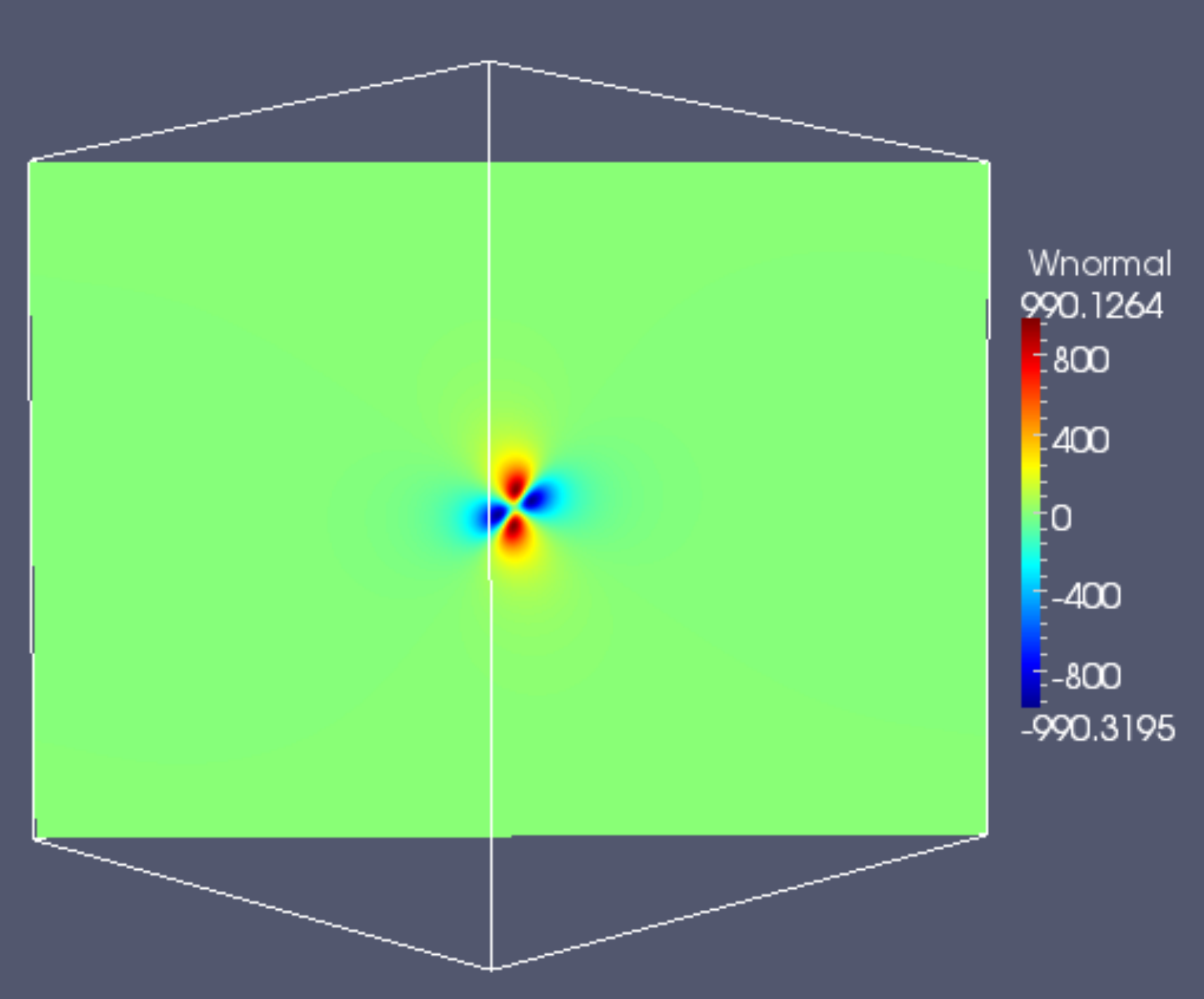}}
\caption[Maximum $d\E/dt$ for $\E_0\to\infty$, $(\E_0)$-Constraint
(3D)]{ (a) Maximum rate of growth of enstrophy $\R(\tuvecE)$ (solid
  line) and its cubic part $\R_{\textrm{cub}}(\tuvecE)$,
  cf.~\eqref{eq:Rcub}, (dotted line) and (b) energy of optimal states
  $\K(\tuvecE)$ as functions of $\E_0$ for large values of enstrophy.
  (c)--(h) Extreme vortex states $\tuvecE$ obtained for the
    three values of enstrophy $\E_0$ indicated with solid symbols in
    figures (a) and (b): panels (c)--(e) show the isosurfaces
  corresponding to $Q(\xvec) = \frac{1}{2}||Q||_{L_\infty}$ with $Q$
  defined in equation \eqref{eq:Q_3D}, whereas panels (f)--(h)
    show the component of vorticity normal to the plane defined by
  $\mathbf{n}\cdot(\xvec-\xvec_0) = 0$, $\mathbf{n} = [1,0,-1]$ and
  $\xvec_0 = [1/2,1/2,1/2]$. }  
\label{fig:RvsE0_FixE_large}
\end{center}
\end{figure}

  The results corresponding to large values of $\E_0$ are shown in
  figure \ref{fig:RvsE0_FixE_large} with the maximum rate of
  growth of enstrophy $\R_{\E_0}$ plotted as a function of
    $\E_0$ in figure \ref{fig:RvsE0_FixE_large}(a). We
    observe that, as $\E_0$ increases, this relation approaches a
    power law of the form $\R_{\E_0} = C_1' \,\E_0^{\alpha_1}$. In
    order to determine the prefactor $C_1'$ and the exponent
    $\alpha_1$ we perform a local least-squares fit of the power law
    to the actual relation $\R_{\E_0}$ versus $\E_0$ for increasing
    values of $\E_0$ starting with $\E_{0} = 20$ (this particular
    choice the starting value is justified below). Then, the exponent
    $\alpha_1$ is computed as the average of the exponents obtained
    from the local fits with their standard deviation providing the
    error bars, so that we obtain
\begin{equation}\label{eq:RvsE0_powerLaw}
\R_{\E_0} = C'_1\E_0^{\,\alpha_1}, \qquad C'_1 = 3.72 \times 10^{-3} , \ \alpha_1 = 2.97 \pm 0.02 
\end{equation}
(the same approach is also used to determine the exponents in
  other power-law relations detected in this study).  We note that
the exponent $\alpha_1$ obtained in \eqref{eq:RvsE0_powerLaw}
is in fact very close to 3 which is the exponent in estimate
\eqref{eq:dEdt_estimate_E}.  For the value of the viscosity
coefficient used in the computations ($\nu=0.01$), the constant factor
$C_1 = 27/(8\pi^4\nu^3)$ in estimate \eqref{eq:dEdt_estimate_E} has
the value $C_1 \approx 3.465 \times 10^4$ which is approximately seven
orders of magnitude larger than $C'_1$ given in
\eqref{eq:RvsE0_powerLaw}. To shed more light at the source of this
discrepancy, the objective functional $\R$ from equation
\eqref{eq:dEdt} can be separated into a negative-definite viscous part
$\R_{\nu}$ and a cubic part $\R_{\textrm{cub}}$ defined as
\begin{subequations}
\begin{align}
\R_{\nu}(\uvec) & :=  -\nu\int_\Omega |\laplacian\uvec|^2\,d\xvec, \\
\R_{\textrm{cub}}(\uvec) & :=  \int_{\Omega} \uvec\cdot\nabla\uvec\cdot\laplacian\uvec\, d\xvec, \label{eq:Rcub}
\end{align}
\end{subequations}
so that $\R(\uvec) = \R_{\nu}(\uvec) + \R_{\textrm{cub}}(\uvec)$.  The
values of $\R_{\textrm{cub}}(\tuvecE)$ are also plotted in
figure \ref{fig:RvsE0_FixE_large}(a) and it is observed that this
quantity exhibits a power-law behaviour of the form
\begin{equation}\label{eq:RcubvsE0_powerLaw}
\R_{\textrm{cub}}(\tuvecE) = C''_1\E_0^{\,\alpha_2}, \qquad C''_1 =  1.38\times 10^{-2},\ \alpha_2 = 2.99 \pm 0.05. 
\end{equation}
While the value of $C''_1$ is slightly larger than the value of $C'_1$
in \eqref{eq:RvsE0_powerLaw}, it is still some six orders of magnitude
smaller than the constant factor $C_1 = 27/(8\pi^4\nu^3)$ from
estimate \eqref{eq:dEdt_estimate_E}. These differences
notwithstanding, we may conclude that estimate
\eqref{eq:dEdt_estimate_E} is sharp in the sense of definition
\ref{def:NotionSharpness}. The power laws from equations
\eqref{eq:RvsE0_powerLaw} and \eqref{eq:RcubvsE0_powerLaw} are
consistent with the results first presented by \citet{l06,ld08}, where
the authors reported a power-law with exponent $\alpha_{LD} = 2.99$
and a constant of proportionality $C_{LD} = 8.97\times 10^{-4}$. The
energy of the optimal fields $\K(\tuvecE)$ for large values of $\E_0$
is shown in figure \ref{fig:RvsE0_FixE_large}(b) in which we observe
that the energy stops to increase at about $\E_0 \approx 20$.
This transition justifies using $\E_0 = 20$ as the lower bound
  on the range of $\E_0$ where the power laws are determined via
  least-square fits.

Figures \ref{fig:RvsE0_FixE_large}(c)-(e) show the isosurfaces
$\Gamma_{0}(Q - 0.5||Q||_{L_\infty})$ representing the optimal fields
$\tuvecE$ for selected large values of $\E_0$.  The formation of
these localized vortex structures featuring two rings as $\E_0$
increases is evident in these figures. The formation process of
localized vortex structures is also visible in figures
\ref{fig:RvsE0_FixE_large}(f)--(h), where the component of vorticity
normal to the plane $P_{xz} = \{\xvec\in\Omega : \mathbf{n}\cdot(\xvec
- \xvec_0) = 0 \}$ for $\mathbf{n} = [1,0,-1]$ and $\xvec_0 =
[1/2,1/2,1/2]$ is shown (we note that the planes used in figures
\ref{fig:RvsE0_FixE_small}(c)--(e) and
\ref{fig:RvsE0_FixE_large}(c)--(e) have different orientations).

\begin{figure}
\linespread{1.1}
\setcounter{subfigure}{0}
\begin{center}
\subfigure[]{\includegraphics[width=0.49\textwidth]{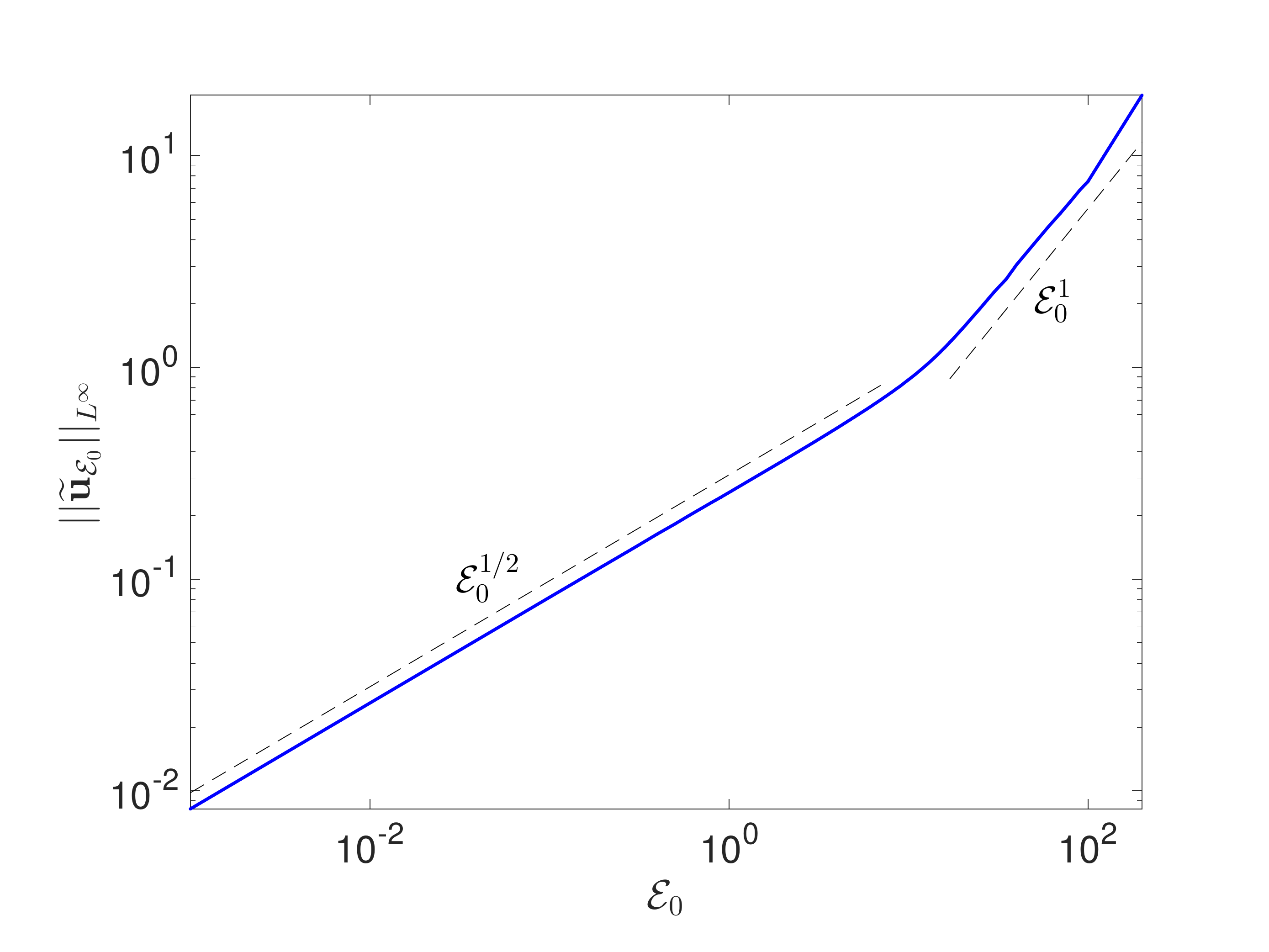}}
\subfigure[]{\includegraphics[width=0.49\textwidth]{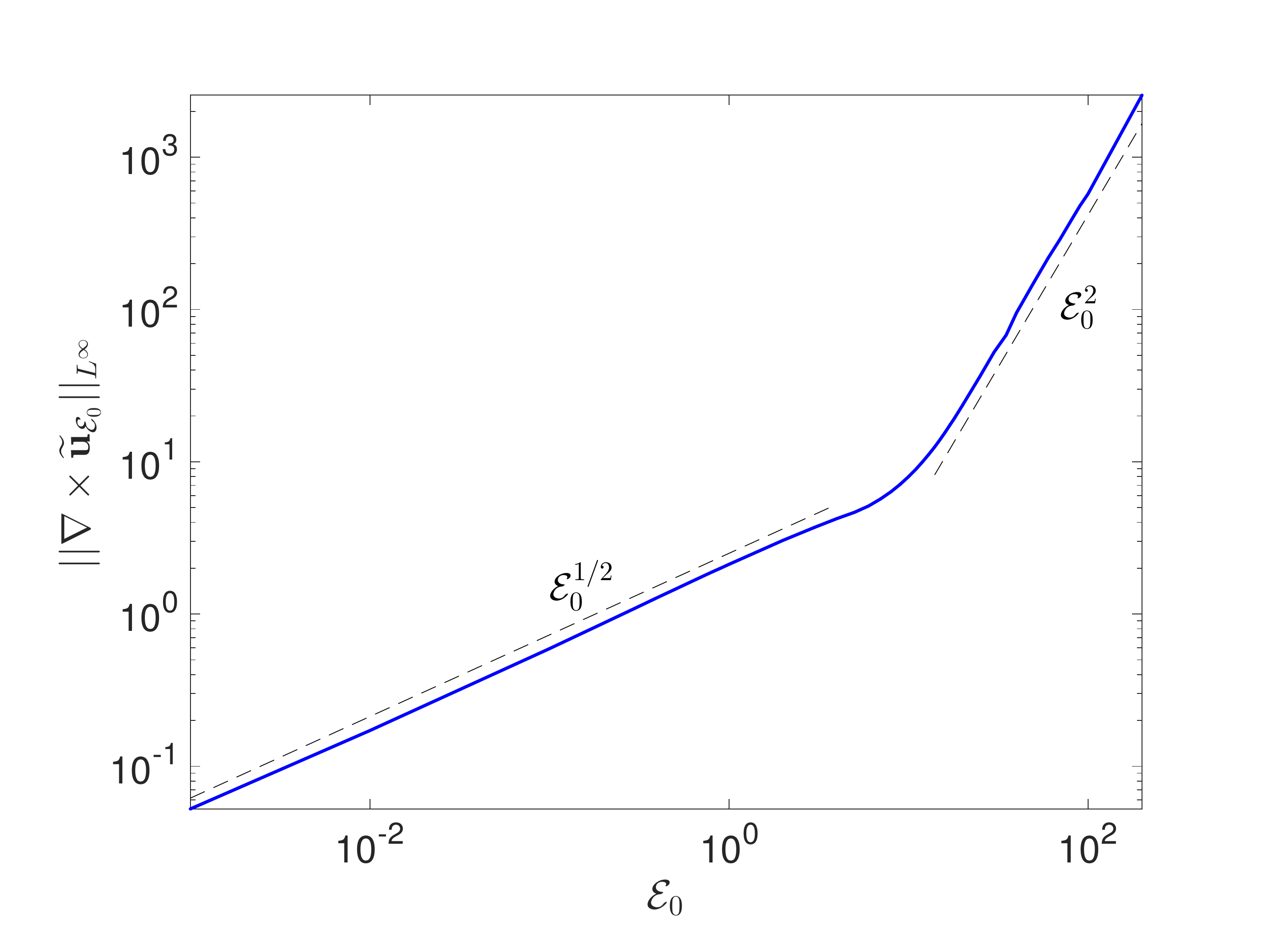}}\\
\subfigure[]{\includegraphics[width=0.49\textwidth]{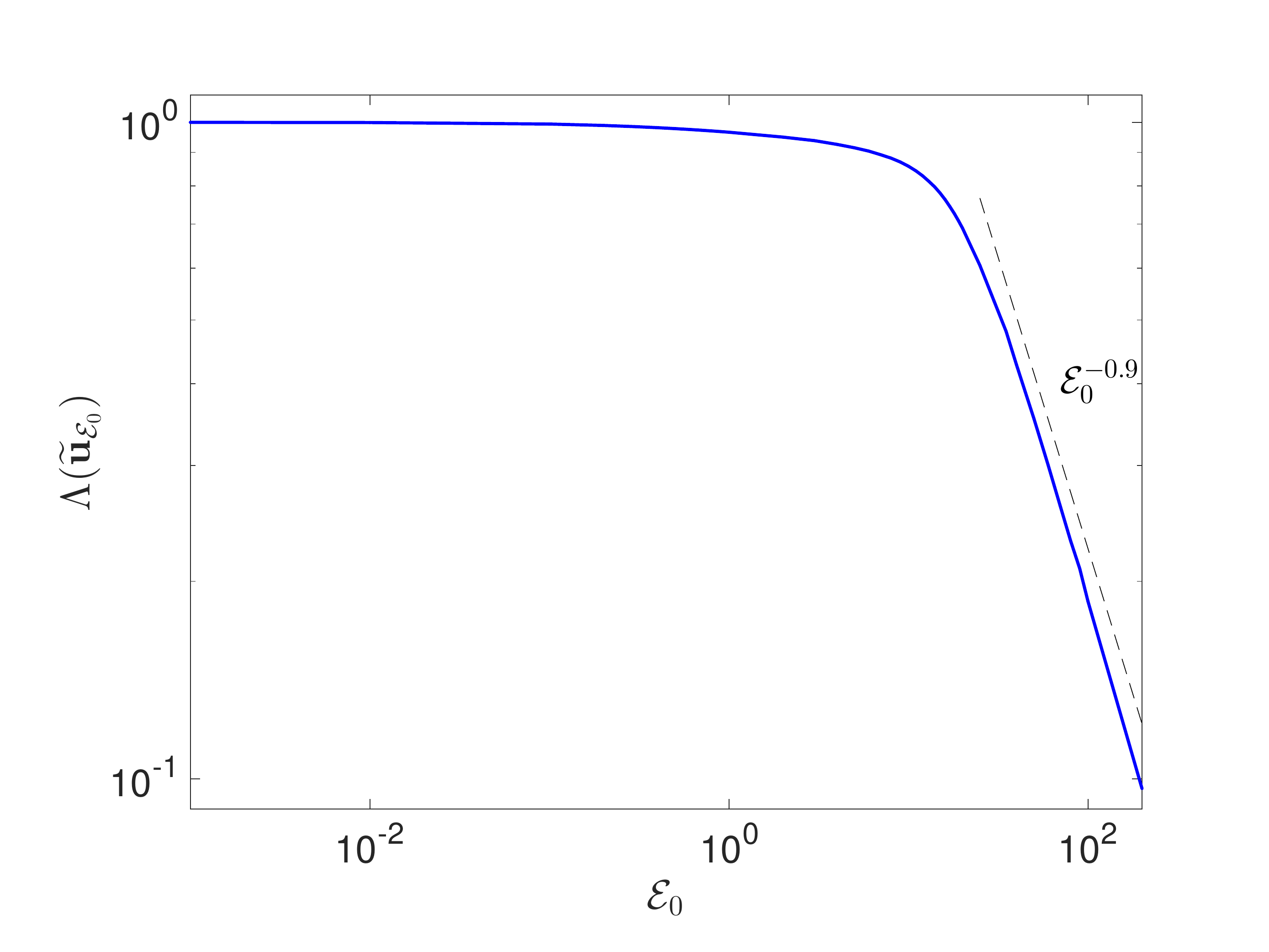}}
\subfigure[]{\includegraphics[width=0.49\textwidth]{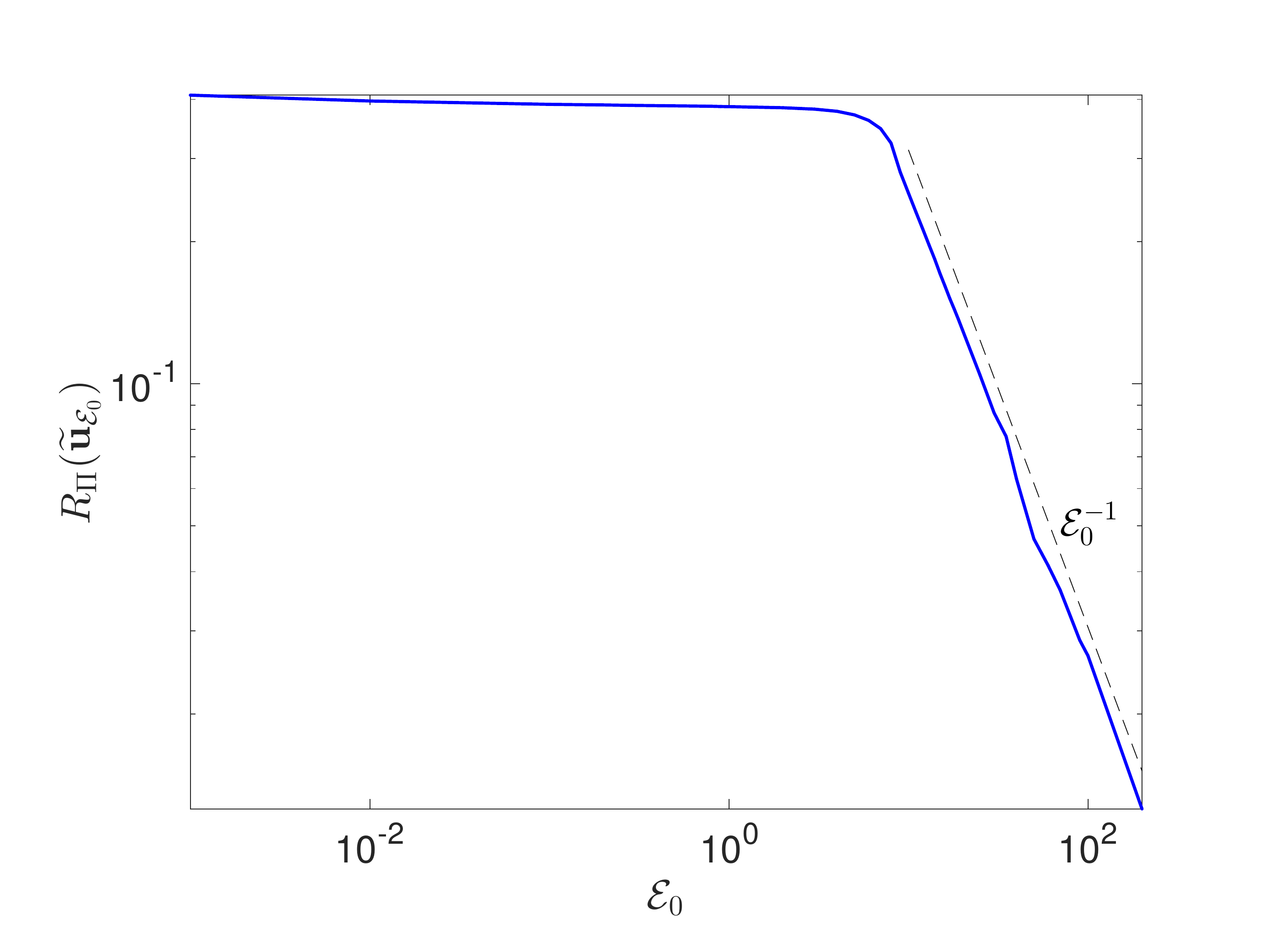}}
\caption[Scaling laws for optimal fields, $(\E_0)$-Constraint (3D)]{
  (a) Maximum velocity $||\tuvecE||_{L_\infty}$, (b) maximum vorticity
  $||\rot\tuvecE||_{L_\infty}$, (c) characteristic length scale
  $\Lambda$ and (d) characteristic radius $R_{\Pi}$ of the extreme
  vortex states as functions of $\E_0$ (all marked with blue
    solid lines). In all cases two distinct behaviours, corresponding
  to $\E_0 \to 0 $ and $\E_0\to\infty$, are evident with the
  corresponding approximate power laws indicated with
    black dashed lines.}
\label{fig:ScalingLaws_fixE}
\end{center}
\end{figure}

Next we examine the variation of different diagnostics
applied to the extreme states $\tuvecE$ as enstrophy $\E_0$ increases.
The maximum velocity $||\tuvecE||_{L_\infty}$ and maximum vorticity
$||\rot\tuvecE||_{L_\infty}$ of the optimal fields are shown,
respectively, in figures \ref{fig:ScalingLaws_fixE}(a) and
\ref{fig:ScalingLaws_fixE}(b) as functions of $\E_0$. For each
quantity, two distinct power laws are observed in the forms

\begin{subequations}
\begin{alignat}{4}
||\tuvecE||_{L_\infty} & \sim C_1\E^{\alpha_1}_0,& \quad 
C_1 &= 0.263,&  \alpha_1 & = 0.5 \pm 0.023,&  \quad & 
\mbox{as }\, \E_0\to 0, \\ 
||\tuvecE||_{L_\infty} & \sim C_2\E^{\alpha_2}_0,& \quad 
C_2 &= 6.3\times 10^{-2},& \quad \alpha_2 &= 1.04 \pm 0.13,&  \qquad &
\mbox{as }\, \E_0\to \infty, \label{eq:uLinf}
\end{alignat} 
\end{subequations}
and
\begin{subequations}
\begin{alignat}{4}
||\rot\tuvecE||_{L_\infty} & \sim C_1\E_0^{\alpha_1},& \quad 
C_1 &= 2.09,& \alpha_1 & = 0.54 \pm 0.03,&  \quad &
\mbox{as }\, \E_0\to 0, \\ 
||\rot\tuvecE||_{L_\infty} & \sim C_2\E_0^{\alpha_2},& \quad 
C_2 &= 6.03\times 10^{-2},& \quad \alpha_2 &= 1.99 \pm 0.17,&  \quad & 
\mbox{as }\, \E_0\to \infty.
\label{eq:omegaLinf}
\end{alignat} 
\end{subequations}

In order to quantify the variation of the
relative size of the vortex structures, we will introduce two
characteristic length scales.  The first one is based on the energy
and enstrophy, and was defined by \cite{dg95} as
\begin{equation}\label{eq:Lambda_def}
\Lambda := \frac{1}{2\pi}\left[ \frac{\K(\tuvecE)}{\E(\tuvecE)} \right]^{1/2}.
\end{equation}
It is therefore equivalent to the Taylor microscale $\lambda^2
= 15\int_\Omega |\uvec|^2 d\xvec / \int_\Omega |\wvec|^2 d\xvec$ used
in turbulence research \citep{davidson:turbulence}. Another length
scale, better suited to the ring-like vortex structures shown in
figures \ref{fig:RvsE0_FixE_large}(c)-(e), is the average radius
$R_{\Pi}$ of one of the vortex rings calculated as
\begin{equation}\label{eq:VortexRadius_def}
R_{\Pi} :=  \frac{ \int_\Omega r(\xvec)\chi_{\Pi}(\xvec) \,d\xvec }{ \int_\Omega \chi_{\Pi}(\xvec)d\xvec },
\ \text{where} \ \
r(\xvec) = |\xvec - \overline{\xvec}|,\quad \overline{\xvec} = \frac{\int_{\Omega} \xvec \chi_{\Pi}(\xvec) d\xvec}{\int_{\Omega} \chi_{\Pi}(\xvec)d\xvec}, 
\end{equation}
and $\chi_{\Pi}$ is the characteristic function of the set 
\begin{eqnarray*}
\Pi & = & \{ \Gamma_s( Q ) : s > 0.9|| Q ||_{L_\infty}\} \cap \\
 &  & \{ \xvec\in\Omega : \nvec\cdot(\xvec-\xvec_0) > 0, \ \nvec = [1,1,1], \ \xvec_0 = [1/2,1/2,1/2] \}.
\end{eqnarray*}
In the above definition of the set $\Pi$, the intersection of the two
regions is necessary to restrict the set $R_{\Pi}$ to
only one of the two ring structures visible in figures
\ref{fig:RvsE0_FixE_large}(c)--(e).  The quantity $\overline{\xvec}$
can be therefore interpreted as the geometric centre of one of the
vortex rings. The dependence of $\Lambda$ and $R_\Pi$ on $\E_0$ is
shown in figures \ref{fig:ScalingLaws_fixE}(c,d) in which the
following power laws can be observed

\begin{subequations}
\begin{alignat}{4}
 \Lambda &\sim \O(1)\quad\mbox{and}& \quad  R_\Pi &\sim \O(1)&&& \quad &
 \mbox{as } \,\E_0\to 0, \\
 \Lambda &\sim C_1\E_0^{\alpha_1},& \quad 
 C_1 &= 10.96,&   \alpha_1 &= -0.886 \pm 0.105,& \qquad &
 \mbox{as } \, \E_0\to \infty, \label{eq:Lambda_powerLaw_largeE0} \\ 
 R_\Pi &\sim C_2\E_0^{\alpha_2},& \quad 
 C_2 &= 2.692,& \quad \alpha_2 &= -1.01 \pm 0.16,& \quad &
 \mbox{as } \, \E_0\to \infty. 
\label{eq:Radius_powerLaw_largeE0}
\end{alignat} 
\end{subequations} 
By comparing the error bars in the key power laws
  \eqref{eq:RvsE0_powerLaw} and \eqref{eq:RcubvsE0_powerLaw} with the
  error bars in power-law relations \eqref{eq:uLinf},
  \eqref{eq:omegaLinf}, \eqref{eq:Lambda_powerLaw_largeE0} and
  \eqref{eq:Radius_powerLaw_largeE0}, we observe that there is less
  uncertainty in the first case, indicating that the quantities
  $||\tuvecE||_{L_\infty}$, $||\rot\tuvecE||_{L_\infty}$, $\Lambda$
  and $R_\Pi$ tend to be more sensitive to approximation errors than
  $\R_{\E_0}(\tuvecE)$. Non-negligible error bars may also indicate
  that, due to modest enstrophy values attained in our computations,
  the ultimate asymptotic regime corresponding to $\E_0 \rightarrow
  \infty$ has not been reached in some power laws.

A useful aspect of employing the average ring radius $R_{\Pi}$ as the
characteristic length scale is that its observed scaling with
respect to $\E_0$ can be used as an approximate indicator of the
resolution $1/N$ required to numerically solve problem
\ref{pb:maxdEdt_E} for large values of enstrophy. From the scaling in
relation \eqref{eq:Radius_powerLaw_largeE0}, it is evident that a
two-fold increase in the value of $\E_0$ will be accompanied by a
similar reduction in $R_\Pi$, thus requiring an eight-fold increase in
the resolution (a two-fold increase in each dimension). This is one of
the reasons why computation of extreme vortex states $\tuvecE$ for
large enstrophy values is a very challenging computational task.
In particular, this relation puts a limit on the largest value
  of $\E_0$ for which problem \ref{pb:maxdEdt_E} can be in principle
  solved computationally at the present moment: a value of $\E_0 =
  2000$, a mere order of magnitude above the largest value of $\E_0$
  reported here, would require a resolution of $8192^3$ used by some
  of the largest Navier-Stokes simulations to date.

To summarize, as the enstrophy increases from $\E_0 \approx 0$ to
$\E_0 = \O(10^2)$, the optimal vortex states change their structure
from cellular to ring-like. While with the exception of $\R(\tuvecE)$
and $\K(\tuvecE)$, all of the diagnostic quantities behave in a
monotonous manner, the corresponding power laws change at about $\E_0
\approx 20$, which approximately marks the transition from the
cellular to the ring-like structure (cf.~figure
\ref{fig:RvsE0_FixE_small}(e) vs.~\ref{fig:RvsE0_FixE_large}(c)). This
is also the value of the enstrophy beyond which the energy
$\K(\tuvecE)$ starts to decrease (figure
\ref{fig:RvsE0_FixE_large}(b)). This transition also coincides
  with a change of the symmetry properties of the extreme vortex
  states $\tuvecE$ --- while in the limit $\E_0 \rightarrow 0$ these
  fields feature reflection and discrete rotation symmetries (cf.~\S
  \ref{sec:3D_InstOpt_E0to0}), for $20 \lessapprox \E_0 \rightarrow
  \infty$ the optimal states are characterized by axial symmetry. The
  asymptotic (as $\E_0 \rightarrow \infty$) extreme vortex states on
  locally maximizing branches corresponding to the aligned ABC flow
  and the Taylor-Green vortex (cf.~Table \ref{tab:E0}) are similar to
  the fields shown in figures \ref{fig:RvsE0_FixE_large}(c)--(h),
  except for a different orientation of their symmetry axes with
  respect to the periodic domain $\Omega$ (these results are not shown
  here for brevity). The different power laws found here are
compared to the corresponding results obtained in 2D in \S
\ref{sec:discuss}.  It is also worth mentioning that, as shown
  by \cite{ad16}, all power laws discussed in this section, cf.
  \eqref{eq:RvsE0_powerLaw}, \eqref{eq:RcubvsE0_powerLaw},
  \eqref{eq:uLinf}, \eqref{eq:omegaLinf} and
  \eqref{eq:Radius_powerLaw_largeE0}, can be deduced rigorously using
  arguments based on dimensional analysis under the assumption of
  axisymmetry for the optimal fields $\tuvecE$.

\begin{figure}
\linespread{1.1}
\begin{center}
\includegraphics[width = 1.0\textwidth]{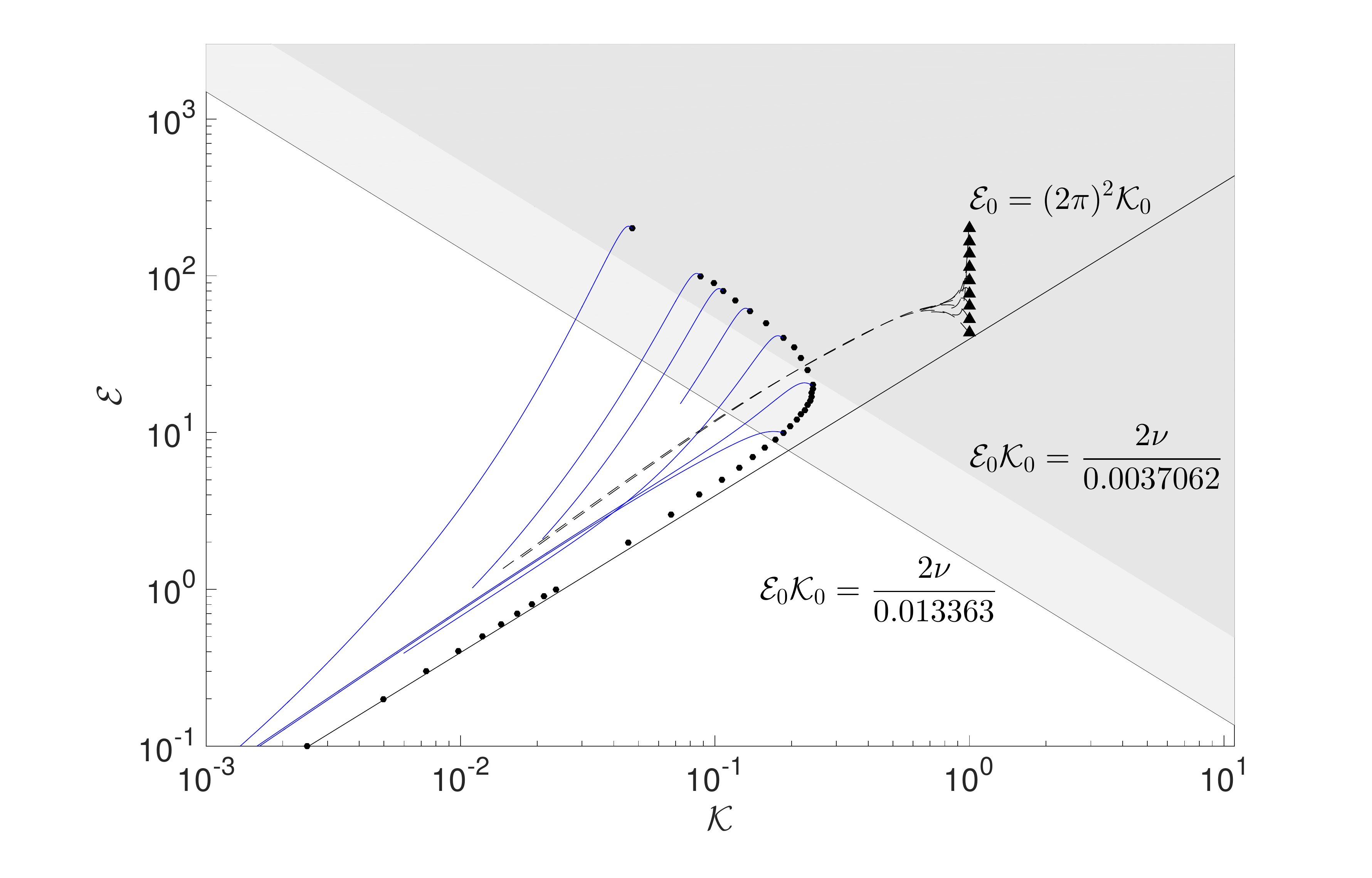}
\caption[Region of guaranteed smoothness, (3D)]{The phase space
  $\{\K,\E\}$. The solid circles and triangles represent,
  respectively, the instantaneously optimal fields $\tuvecE$
  and $\tuvecKE$, with the lines issuing from selected markers
  indicating the corresponding time-dependent trajectories (the
    optimal states $\tuvecKE$ are discussed in \S \ref{sec:final},
    cf.~problem \ref{pb:maxdEdt_KE}). The two lines with a negative
  slope represent condition \eqref{eq:K0E02} with the two different
  constants, whereas the line with a positive slope is the Poincar\'e
  limit $\K = (2\pi)^2 \E$.  The shaded areas represent regions of the
  phase space for which global regularity is not a priori guaranteed
  based on estimate \eqref{eq:dEdt_estimate_E} combined with fits
  \eqref{eq:RvsE0_powerLaw} and \eqref{eq:RcubvsE0_powerLaw}.  }
\label{fig:K0E0}
\end{center}
\end{figure}

Finally, the findings of this section allow us to shed some light on
the ``small data'' result \eqref{eq:K0E0} which provides the
conditions on the size of the initial data $\uvec_0$, given in
  terms of its energy $\K(0)$ and enstrophy $\E(0)$, in the
Navier-Stokes system \eqref{eq:NSE3D} guaranteeing that smooth
solutions exist globally in time.  The power-law fits
\eqref{eq:RvsE0_powerLaw} and \eqref{eq:RcubvsE0_powerLaw} allow us to
sharpen condition \eqref{eq:K0E0} be replacing the constant on the RHS
with either $2\nu / C'_1$ or $2\nu / C''_1$, so that we
obtain
\begin{equation}\label{eq:K0E02}
\K(0)\E(0) < \left\{ \frac{2\nu}{C'_1} \ \ \text{or} \ \ \frac{2\nu}{C''_1}  \right\}.
\end{equation}
The region of the ``phase space'' $\{\K,\E\}$ described by condition
\eqref{eq:K0E02} is shown in white in figure \ref{fig:K0E0}.  The gray
region represents the values of $\K(0)$ and $\E(0)$ for which
long-time existence of smooth solutions cannot be a priori guaranteed
(the two shades of gray correspond to the two constants which can be
used in \eqref{eq:K0E02}). Solid circles represent the
different extreme states found in this section, whereas the thin
curves mark the time-dependent trajectories which will be analyzed in
\S \ref{sec:timeEvolution}. We conclude from figure \ref{fig:K0E0}
that the change of the properties of the optimal states $\tuvecE$
discussed above occurs in fact for the values of enstrophy
($\E(0) \approx 20$) for which the states $\tuvecE$ are on the
boundary of the region of guaranteed long-time regularity.

\section{Time Evolution of Extreme Vortex States}
\label{sec:timeEvolution}

\begin{figure}
\linespread{1.1}
\setcounter{subfigure}{0}
\begin{center}
\subfigure[]{\includegraphics[width=0.32\textwidth]{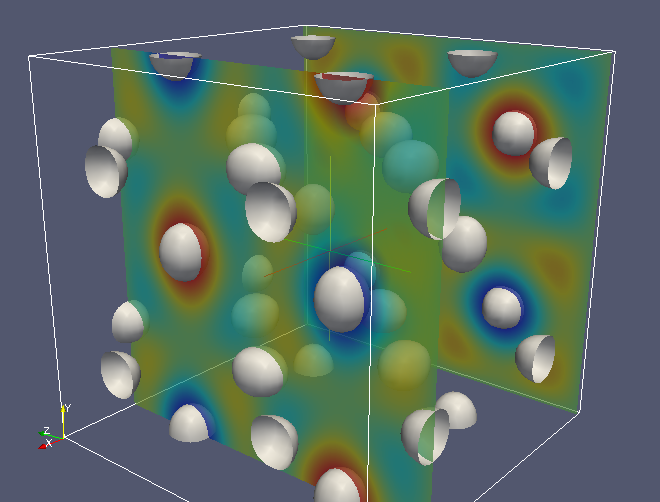}}
\subfigure[]{\includegraphics[width=0.32\textwidth]{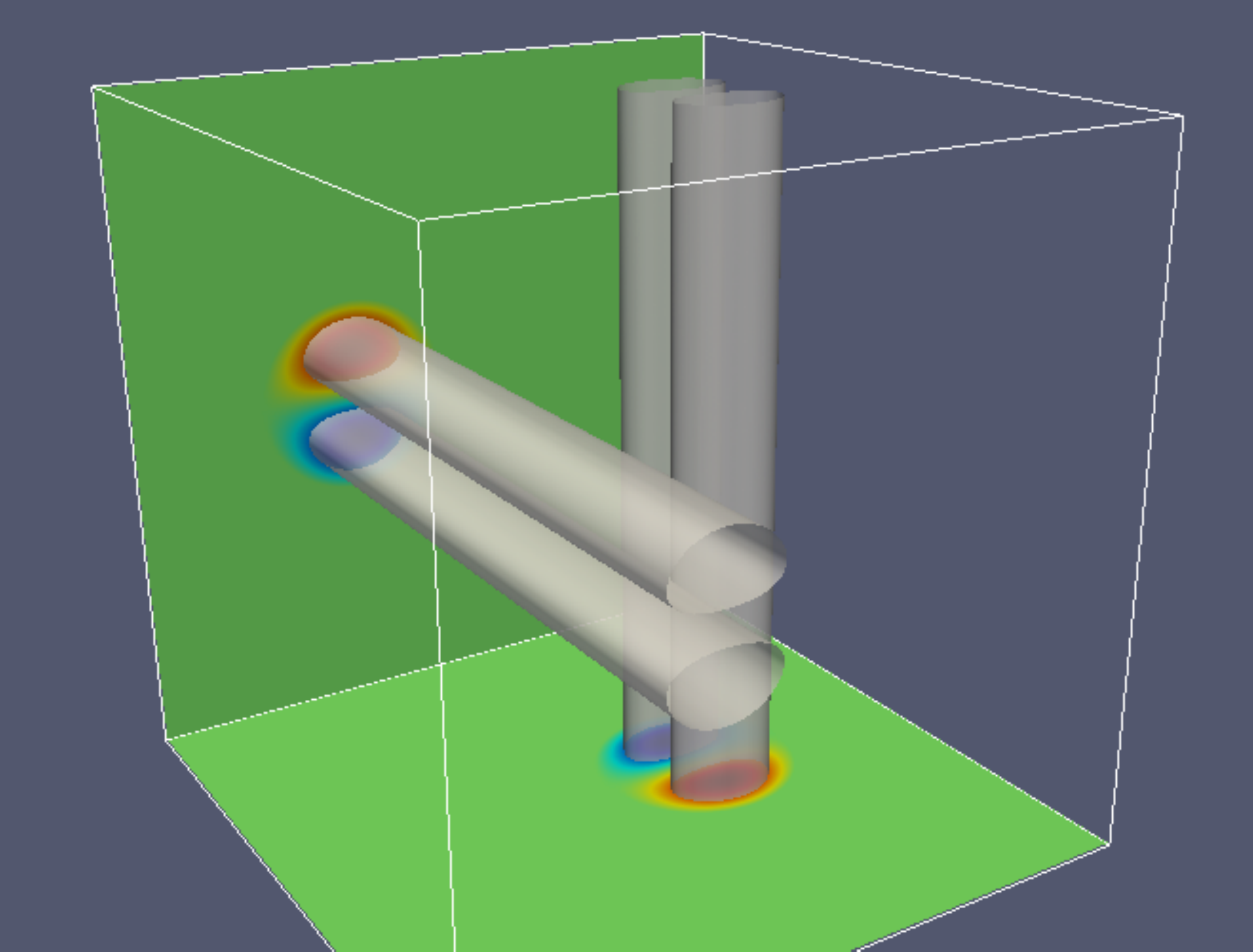}}
\subfigure[]{\includegraphics[width=0.32\textwidth]{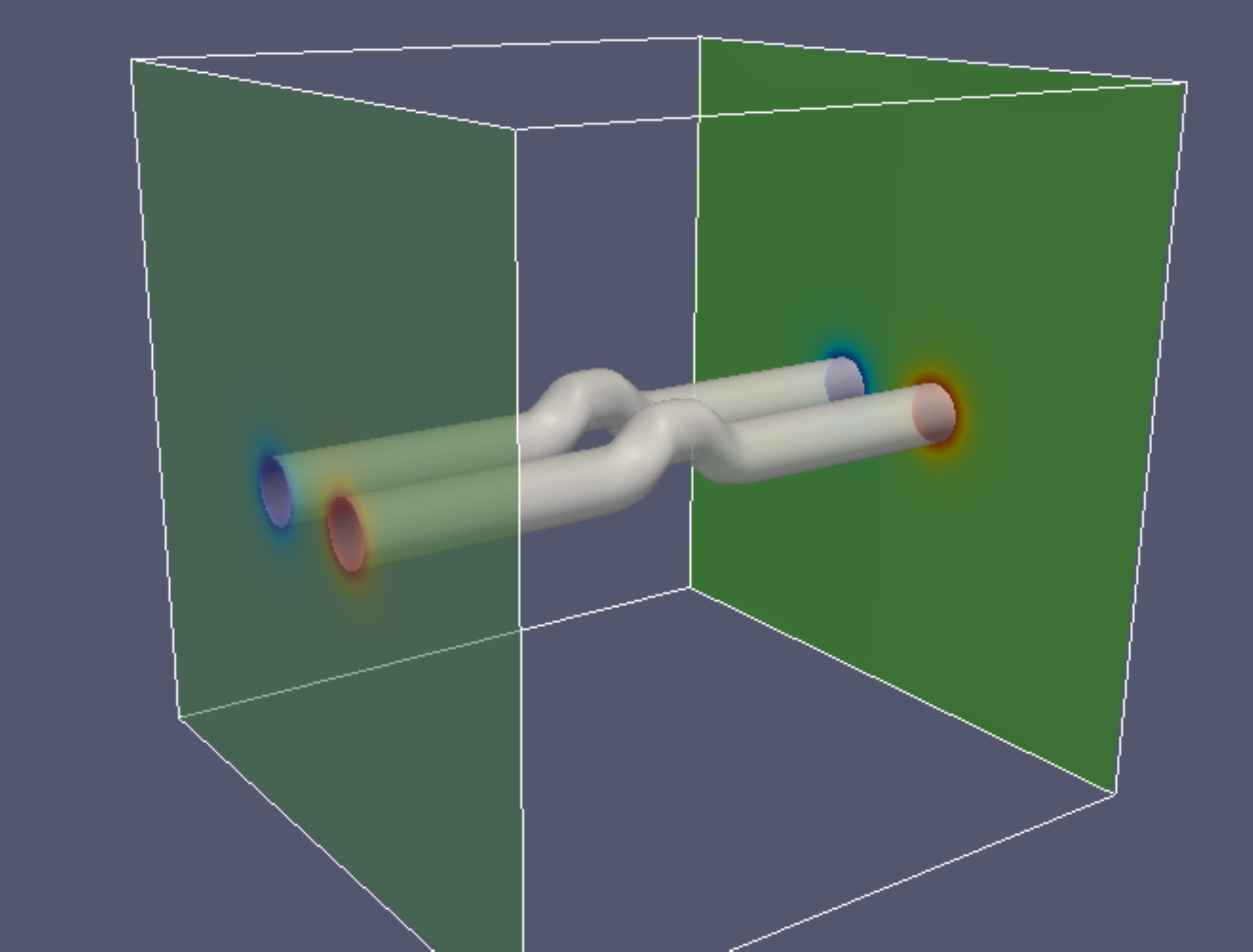}}
\caption[]{Isosurfaces corresponding to $Q(\xvec) =
  \tfrac{1}{2}||Q||_{L_\infty}$ for different initial conditions, all
  normalized to $\E_0 = 100$: (a) Kida-Pelz flow, (b) colliding
  Lamb-Chaplygin dipoles and (c) perturbed antiparallel vortex
  tubes.  Precise characterization of these different initial
  conditions is provided in Table \ref{tab:InitialConditions}.}
\label{fig:SummaryIC}
\end{center} 
\end{figure}

The goal of this section is to analyze the time evolution, governed by
the Navier-Stokes system \eqref{eq:NSE3D}, with extreme vortex states
identified in \S \ref{sec:3D_InstOpt_E} used as the initial data
$\uvec_0$. In particular, we are interested in the finite-time
growth of enstrophy $\E(t)$ and how it relates to estimates
\eqref{eq:dEdt_estimate_E}, \eqref{eq:Evs_t_fixE} and
\eqref{eq:Cond_for_globalReg}. We will compare these results with the
growth of enstrophy obtained using other types of initial data which
have also been studied in the context of the blow-up problem for both
the Euler and Navier-Stokes systems, namely, the Taylor-Green vortex
\citep{tg37,bmonmu83,b91,bb12}, the Kida-Pelz flow
\citep{bp94,p01,ghdg08}, colliding Lamb-Chaplygin dipoles
\citep{opc12} and perturbed antiparallel vortex tubes \citep{k93,k13}.
Precise characterization of these different initial conditions is
provided in Table \ref{tab:InitialConditions} and, for the sake
  of completeness, the last three states are also visualized in
figure \ref{fig:SummaryIC}. We comment that, with the exception of the
Taylor-Green vortex which was shown in \S \ref{sec:3D_InstOpt_E0to0}
to be a local maximizer of problem \ref{pb:maxdEdt_E} in the limit
$\E_0 \rightarrow 0$, all these initial conditions were postulated
based on rather ad-hoc physical arguments.  We also add that, in order
to ensure a fair comparison, the different initial conditions listed
in Table \ref{tab:InitialConditions} are rescaled to have the same
enstrophy $\E_0$, which is different from the enstrophy values
used in the original studies where these initial conditions
  were investigated \citep{opc12,k13,dggkpv13,opmc14}. As regards our
choices of the initial enstrophy $\E_0$, to illustrate different
possible behaviours, we will consider initial data located in the two
distinct regions of the phase space $\{\K,\E\}$ shown in figure
\ref{fig:K0E0}, corresponding to values of $\K_0$ and $\E_0$ for which
global regularity may or may not be a priori guaranteed according to
estimates \eqref{eq:Cond_for_globalReg}--\eqref{eq:K0E0}.

\begin{table}
\begin{center}
\begin{tabular}{c|c|c}
 & 
\Bmp{3.0in} \begin{displaymath} \uvec_0(\xvec) = [u, v, w] \end{displaymath} \Emp & 
\Bmp{1.3in} \centering {\bf Notes} \Emp \\
\hline
\Bmp{0.8in} Instantaneous \\ optimizer $\tuvecE$  \Emp & 
\Bmp{3.0in} \begin{displaymath}\uvec_0 = \mathop{\argmax}_{\uvec\in\mathcal{S}_{\E_0}} \R(\uvec) \end{displaymath}\Emp & 
\Bmp{1.3in} See problem \eqref{pb:maxdEdt_E} \Emp \\
\hline
\Bmp{0.8in} Taylor-Green \\ vortex \Emp & 
\Bmp{3.0in} \begin{eqnarray*} u(x,y,z) & = & A\sin(2\pi x)\cos(2\pi y)\cos(2\pi z) \\
                              v(x,y,z) & = & -A\cos(2\pi x)\sin(2\pi y)\cos(2\pi z) \\
                              w(x,y,z) & = & 0
            \end{eqnarray*}\Emp & 
\Bmp{1.3in} $\bm{\gamma}=(1,-1,0)$ in \\ equation \eqref{eq:TaylorGreen_vortex},\\
$A$ chosen so that $\E(\uvec_0) = \E_0$. \Emp \\
\hline
\Bmp{0.8in} Kida-Pelz \\ flow \Emp & 
\Bmp{3.0in} \begin{eqnarray*} u(x,y,z) & = & A\sin(2\pi x)[ \cos(6\pi y)\cos(2\pi z) - \\
                                               &   & \cos(2\pi y)\cos(6\pi z) ] \\
                              v(x,y,z) & = & A\sin(2\pi y)[ \cos(6\pi z)\cos(2\pi x) - \\
                                               &   & \cos(2\pi x_3)\cos(6\pi x_1) ] \\
                              w(x,y,z) & = & A\sin(2\pi z)[ \cos(6\pi x)\cos(2\pi y) - \\
                                               &   & \cos(2\pi x)\cos(6\pi y) ] \\
            \end{eqnarray*}\Emp & 
\Bmp{1.3in} Taken from \\ \citet{bp94}, \\ $A$ chosen so that $\E(\uvec_0) = \E_0$. \Emp \\
\hline
\Bmp{0.8in} Lamb-Chaplygin \\ dipoles \Emp & 
\Bmp{3.0in} \begin{displaymath} -\laplacian\uvec_0  =  \rot\wvec_0, \quad \wvec_0 = [\, 0, \, \omega(x,z), \,\omega(x,y)\,] \end{displaymath}
            \begin{displaymath} \omega(x(r,\theta),y(r,\theta)) = \left\{ 
            \begin{array}{c@{\,\,}r} 
            -2U\kappa \frac{J_1(\kappa r)}{J_0(\kappa a)}\sin(\theta) & (r \leq a) \\ 
                             0 & (r > a)
            \end{array} 
            \right.         
            \end{displaymath} \Emp & 
\Bmp{1.3in} Taken from \\ \citet{opc12}. \\ 
$a = 0.15$, $\kappa a = z_1$, \\the first zero of $J_1$ \\ $U = \sqrt{\frac{\E_0}{2\pi z^2_1}}$ \Emp \\
\hline
\Bmp{0.8in} Perturbed \\ anti-parallel \\ vortex tubes \Emp & 
\Bmp{3.0in} \begin{displaymath} -\laplacian\uvec_0  =  \rot\wvec_0, \quad \wvec_0 = \omega(x,y)\frac{\bm{\sigma'}}{|\bm{\sigma}'|}(s) \end{displaymath} 
            \begin{displaymath} \omega(x(r,\theta),y(r,\theta)) =  \frac{A}{(r/a)^{4} + 1} \end{displaymath}
            \begin{displaymath} \bm{\sigma}(s) =  [2a, 2b/\cosh(s^2/c^2)-b, s] \end{displaymath}\Emp &
\Bmp{1.3in} Taken from \\ \citet{k13}. \\ $a= 0.05$, $b=a/2$, \\ $c=a$, $s$ is the arc-length parameter. \\ $A$ chosen so that  $\E(\uvec_0) = \E_0$. \Emp \\
\hline
\end{tabular}
\caption{Characterization of the different initial data used in time
  evolution studies in \S\ref{sec:timeEvolution}.}
\label{tab:InitialConditions}
\end{center} 
\end{table}

System \eqref{eq:NSE3D} is solved numerically with an approach
combining a pseudo-spectral approximation of spatial derivatives with
a third-order semi-implicit Runge-Kutta method \citep{NumRenaissance}
used to discretize the problem in time. In the evaluation of the
nonlinear term dealiasing was used based on the $2/3$ rule
together with the Gaussian filtering proposed by \cite{hl07}.
Massively parallel implementation based on MPI and using the {\tt
  fftw} routines \citep{fftw} to perform Fourier transforms allowed us
to use resolutions varying from $256^3$ to $1024^3$ in the
low-enstrophy and high-enstrophy cases, respectively. A number of
different diagnostics were checked to ensure that all flows discussed
below are well resolved. We refer the reader to the dissertation by
\cite{a14} for additional details and a validation of this approach.

\begin{figure}
\begin{center}
\subfigure[$\E_0 = 10$]{\includegraphics[width=0.45\textwidth]{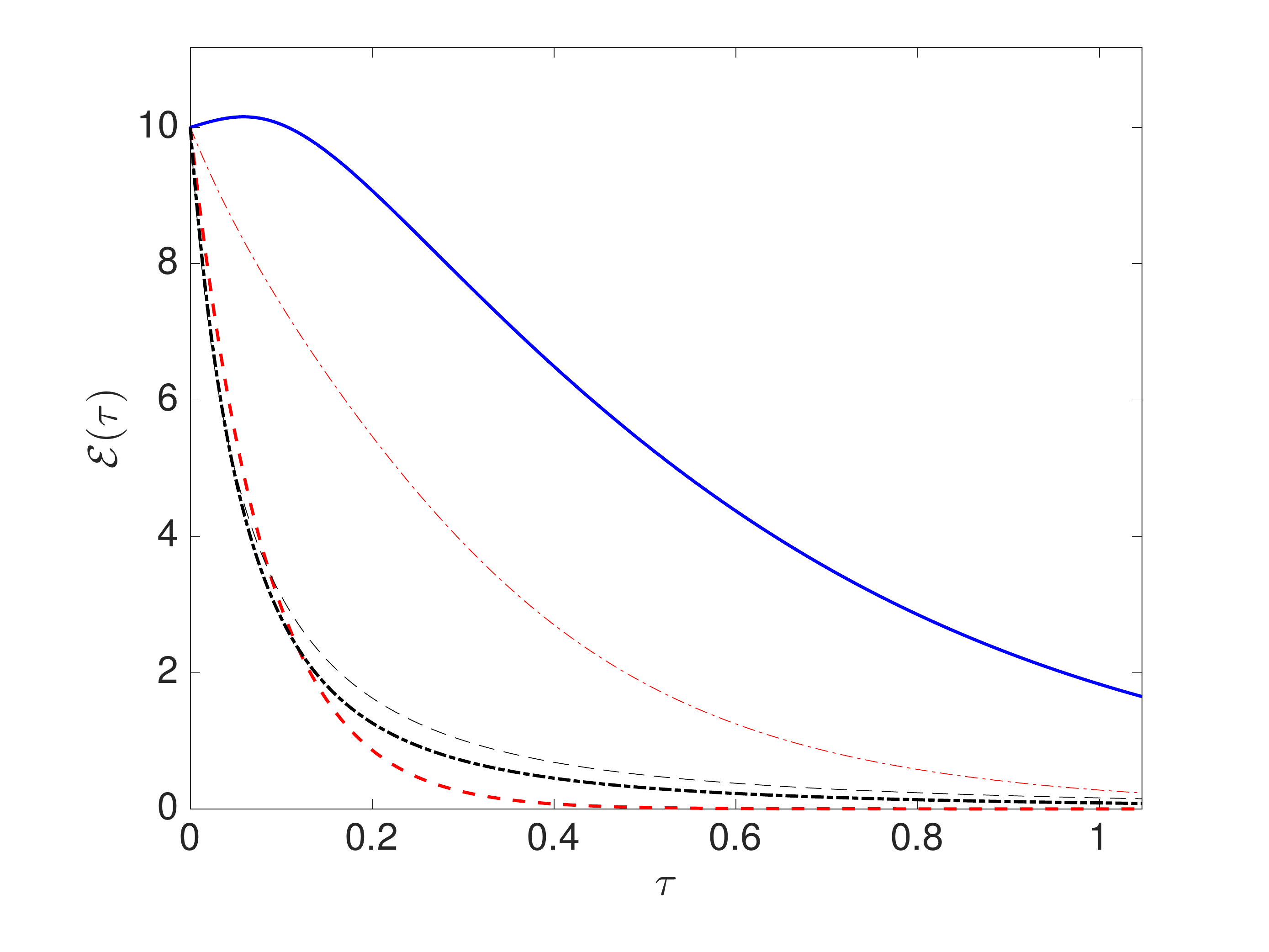}}
\subfigure[$\E_0 = 100$]{\includegraphics[width=0.45\textwidth]{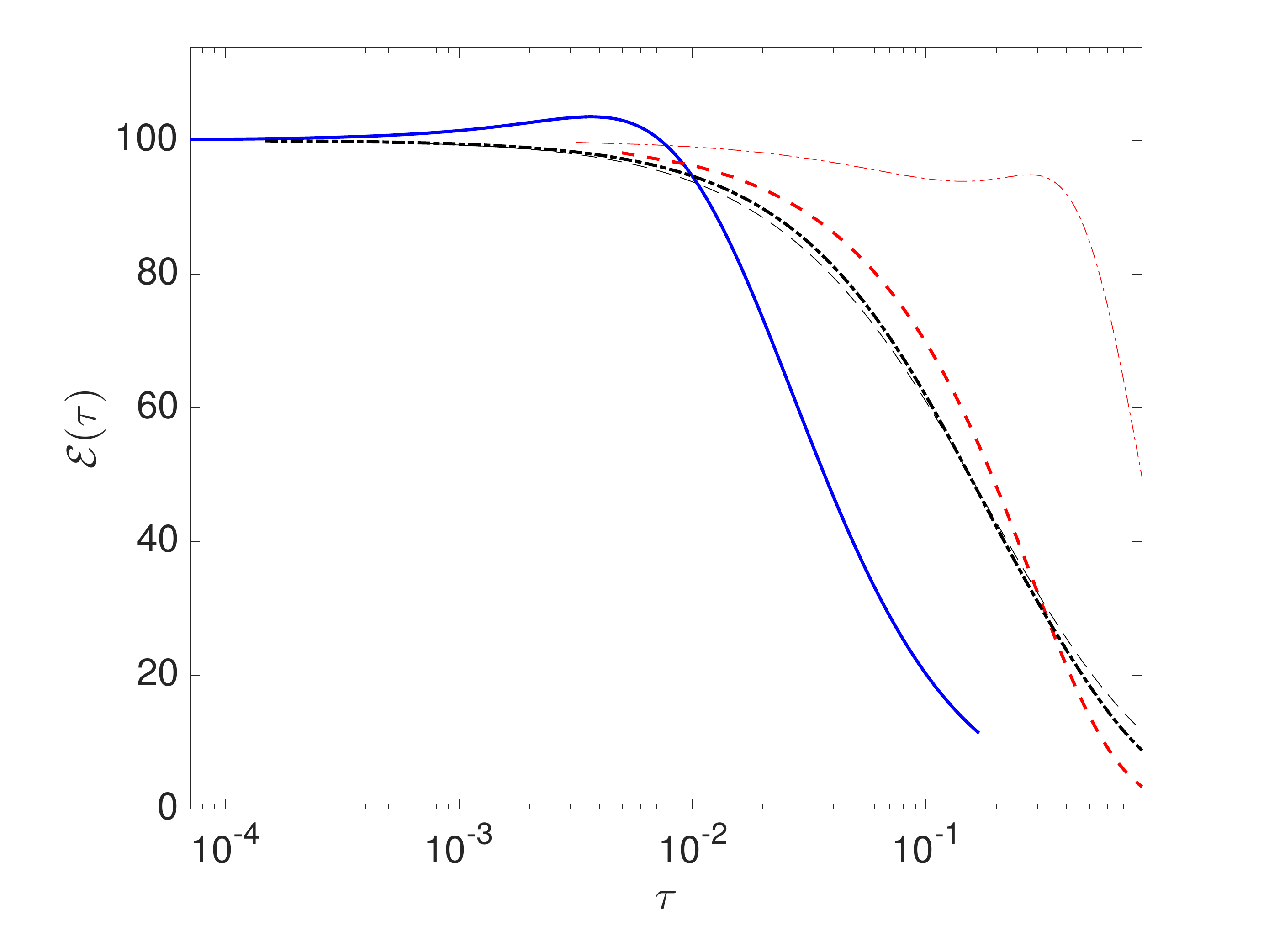}}
\caption{ Time evolution of enstrophy $\E(\tau)$ for the initial
  enstrophy (a) $\E_0 = 10$ and (b) $\E_0 = 100$ with the initial
  condition $\uvec_0$ corresponding to the instantaneous optimizer
  $\tuvecE$ (blue solid line), the Taylor-Green vortex (red
  dashed-dotted line), the Kida-Pelz vortex (red dashed line), the
  colliding Lamb-Chaplygin dipoles (black dashed-dotted lines)
  and the perturbed antiparallel vortex tubes (black dashed lines).
  Precise characterization of these different initial conditions is
  provided in Table \ref{tab:InitialConditions}.}
\label{fig:Fvs_t_fixE}
\end{center}
\end{figure}

The time-dependent results will be shown with respect to a normalized
time defined as $\tau := U_c t/\ell_c$ with $U_c := \|\tuvecE\|_{L_2}$
and $\ell_c = \Lambda$ (cf.~equation \eqref{eq:Lambda_def}) playing
the roles of the characteristic velocity and length scale. We begin by
showing the time evolution of the enstrophy $\E(\tau)$ corresponding
to the five different initial conditions listed in Table
\ref{tab:InitialConditions} with $\E_0 = 10$ and $\E_0 = 100$ in
figures \ref{fig:Fvs_t_fixE}(a) and \ref{fig:Fvs_t_fixE}(b),
respectively (because of the faster time-scale, the time axis in the
latter figure is scaled logarithmically). We see that the maximizers
$\tuvecE$ of problem \ref{pb:maxdEdt_E} are the only initial data
which triggers growth of enstrophy for these values of the initial
enstrophy and, as expected, this growth is larger for $\E_0 = 100$
than for $\E_0 = 10$. The other initial condition which exhibits some
tendency for growth when $\E_0 = 100$ is the Taylor-Green vortex. In
all cases the enstrophy eventually decays to zero for large times.

\begin{figure}
\begin{center}
  \subfigure[$\E_0 = 10$]{\includegraphics[width=0.45\textwidth]
    {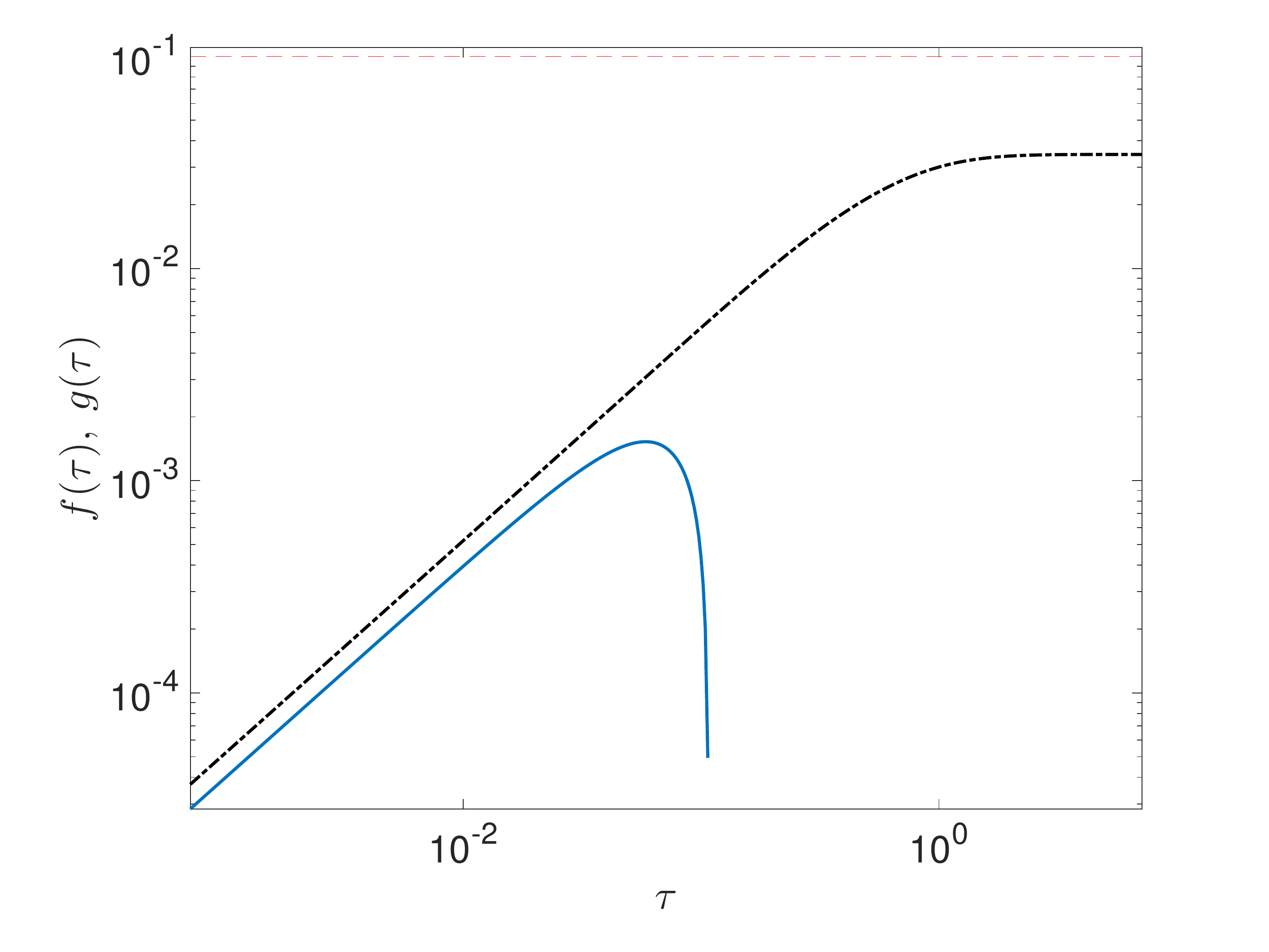}} \subfigure[$\E_0 =
  100$]{\includegraphics[width=0.45\textwidth]
    {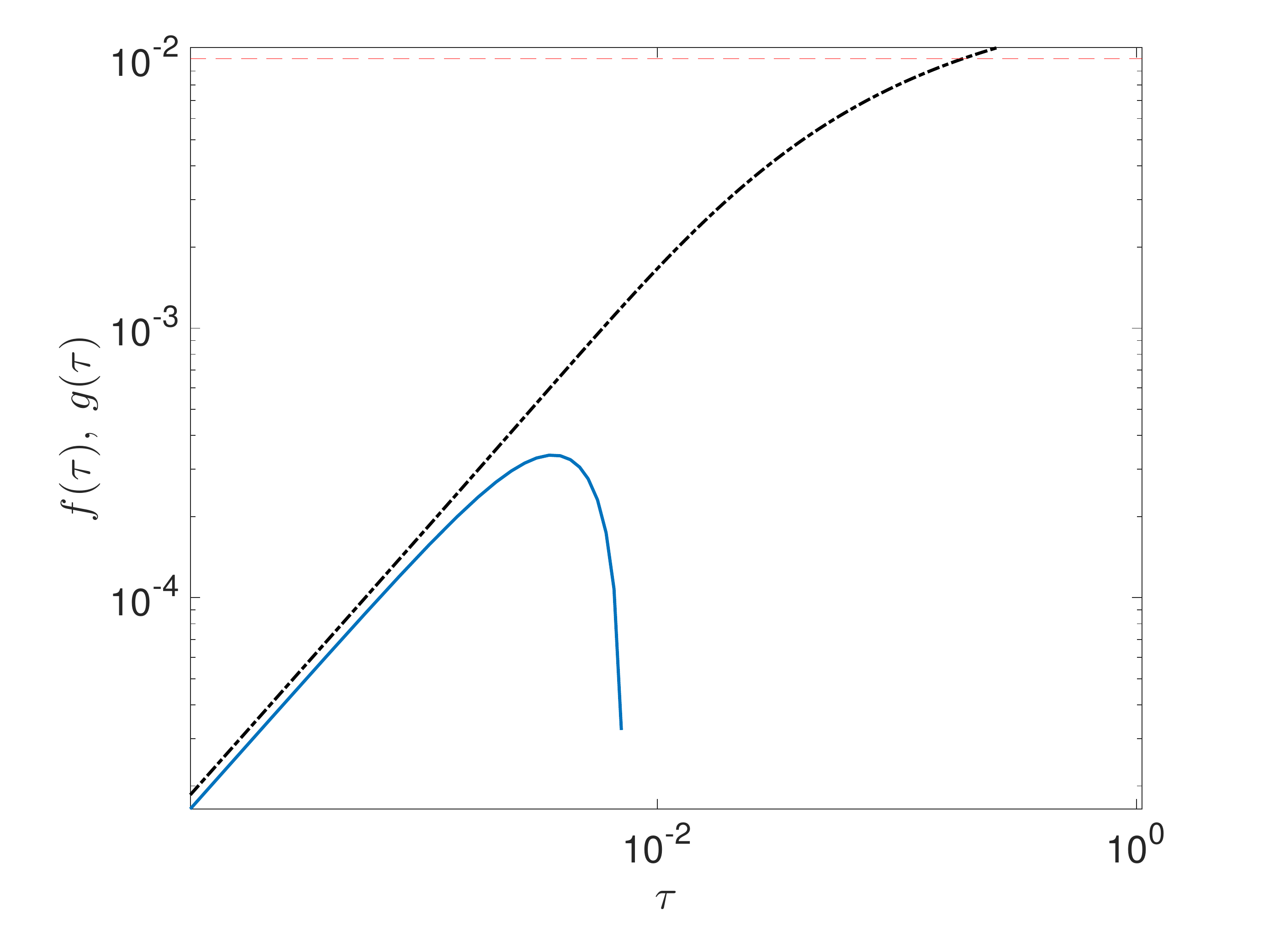}}
  \caption{Evolution of functions $f(\tau)$ (blue solid lines)
      and $g(\tau)$ (black dashed-dotted lines), cf.~equations
      \eqref{eq:fga}--\eqref{eq:fgb}, for flows with the optimal
      initial condition $\tuvecE$ with (a) $\E_0 = 10$ and (b) $\E_0 =
      100$. The value $1/\E_0$ which must be attained in a
      hypothetical blow-up event is marked by the horizontal dashed
      line.}
\label{fig:fg}
\end{center}
\end{figure}

Next we examine whether the flow evolutions starting from the
instantaneous maximizers $\tuvecE$ as the initial data saturate the
finite-time estimate \eqref{eq:Evs_t_fixE}. We do this by defining
functions
\begin{subequations}
\label{eq:fg}
\begin{align}
f(\tau) & := \frac{1}{\E(0)} - \frac{1}{\E(\tau)}\qquad\mbox{and} \label{eq:fga} \\
g(\tau) & := \frac{C}{2\nu}\left[ \K(0) - \K(\tau) \right] \label{eq:fgb}
\end{align} 
\end{subequations}
representing, respectively, the left- and right-hand side of the
estimate and then plotting them with respect to the normalized time
$\tau$, which is done in figures \ref{fig:fg}(a) and \ref{fig:fg}(b)
for $\E_0 = 10$ and $\E_0 = 100$, respectively. The constant $C>0$ in
the definition of $g(\tau)$ is numerically computed from the power-law
fit in \eqref{eq:RvsE0_powerLaw}. It follows from estimate
\eqref{eq:Evs_t_fixE} that $f(\tau) \leq g(\tau)$ pointwise in time.
The hypothetical extreme event of a finite-time blow-up can be
represented graphically by an intersection of the graph of $f(\tau)$
with the horizontal line $y = 1/\E_0$, which is also shown in figures
\ref{fig:fg}(a)--(b). The behaviour of $g(\tau)$, representing the
upper bound in estimate \eqref{eq:Evs_t_fixE}, is quite distinct in
figures \ref{fig:fg}(a) and \ref{fig:fg}(b) reflecting the fact that
the initial data $\tuvecE$ in the two cases comes from different
regions of the phase diagram in figure \ref{fig:K0E0}. In figure
\ref{fig:fg}(a), corresponding to $\E_0 = 10$, the upper bound
$g(\tau)$ never reaches $1/\E_0$, in agreement with the fact that the
finite-time blow-up is a priori ruled out in this case. On the other
hand, in figure \ref{fig:fg}(b), corresponding to $\E_0 = 100$, the
upper bound $g(\tau)$ does intersect $1/\E_0$ implying that, in
principle, finite-time blow-up might be possible in this case.
The sharpness of estimate \eqref{eq:Evs_t_fixE} can be assessed
  by analyzing how closely the behaviour of $f(\tau)$ matches
  that of $g(\tau)$. In both figures \ref{fig:fg}(a) and
\ref{fig:fg}(b) we observe that for a short period of time $f(\tau)$
exhibits a very similar growth to the upper bound $g(\tau)$, but then
this growth slows down and $f(\tau)$ eventually starts to decrease
short of ever approaching the limit $1/\E_0$.

\begin{figure}
\linespread{1.1}
\setcounter{subfigure}{0}
\begin{center}
\subfigure[]{\includegraphics[width=0.48\textwidth]{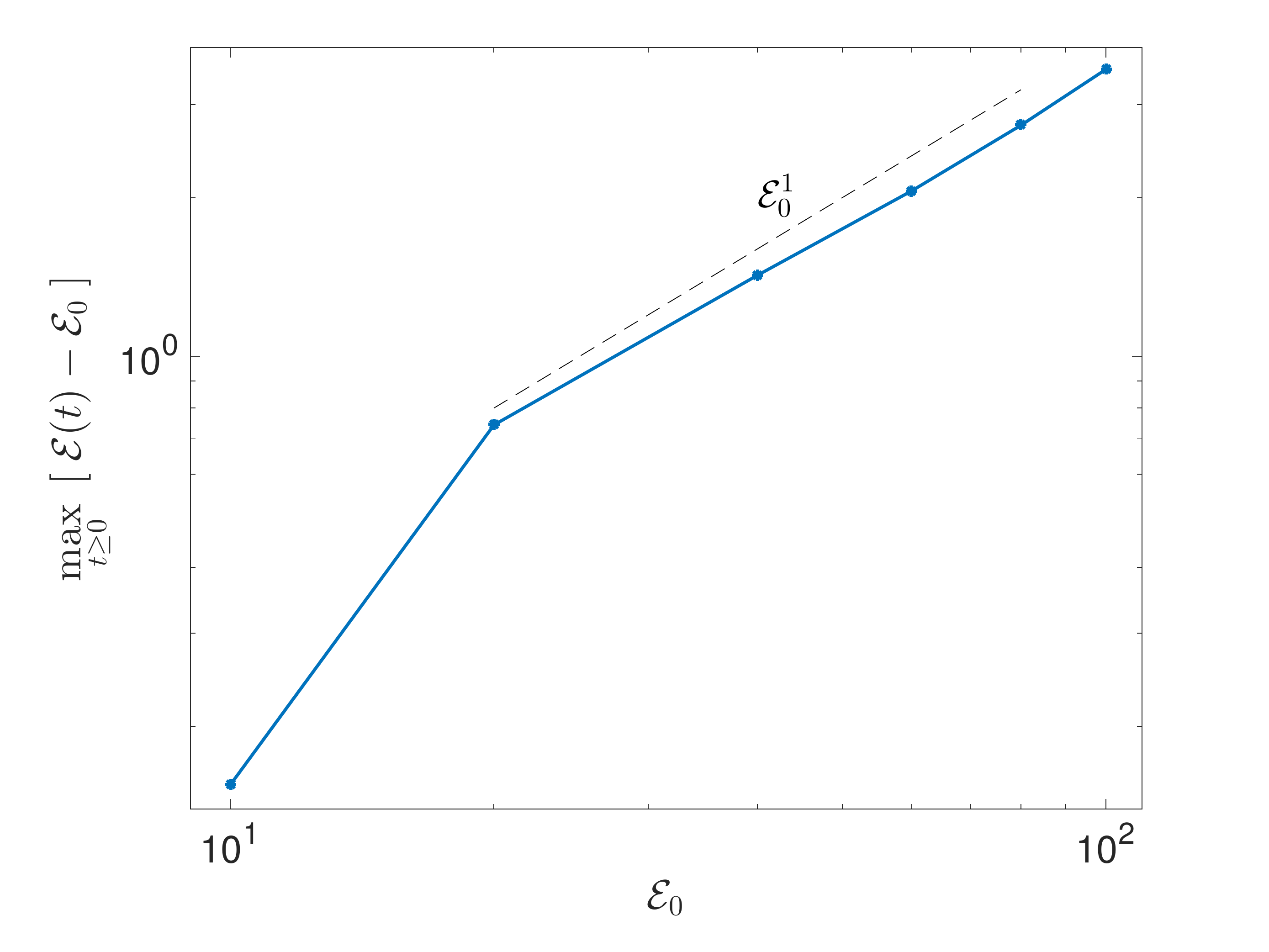}}
\subfigure[]{\includegraphics[width=0.48\textwidth]{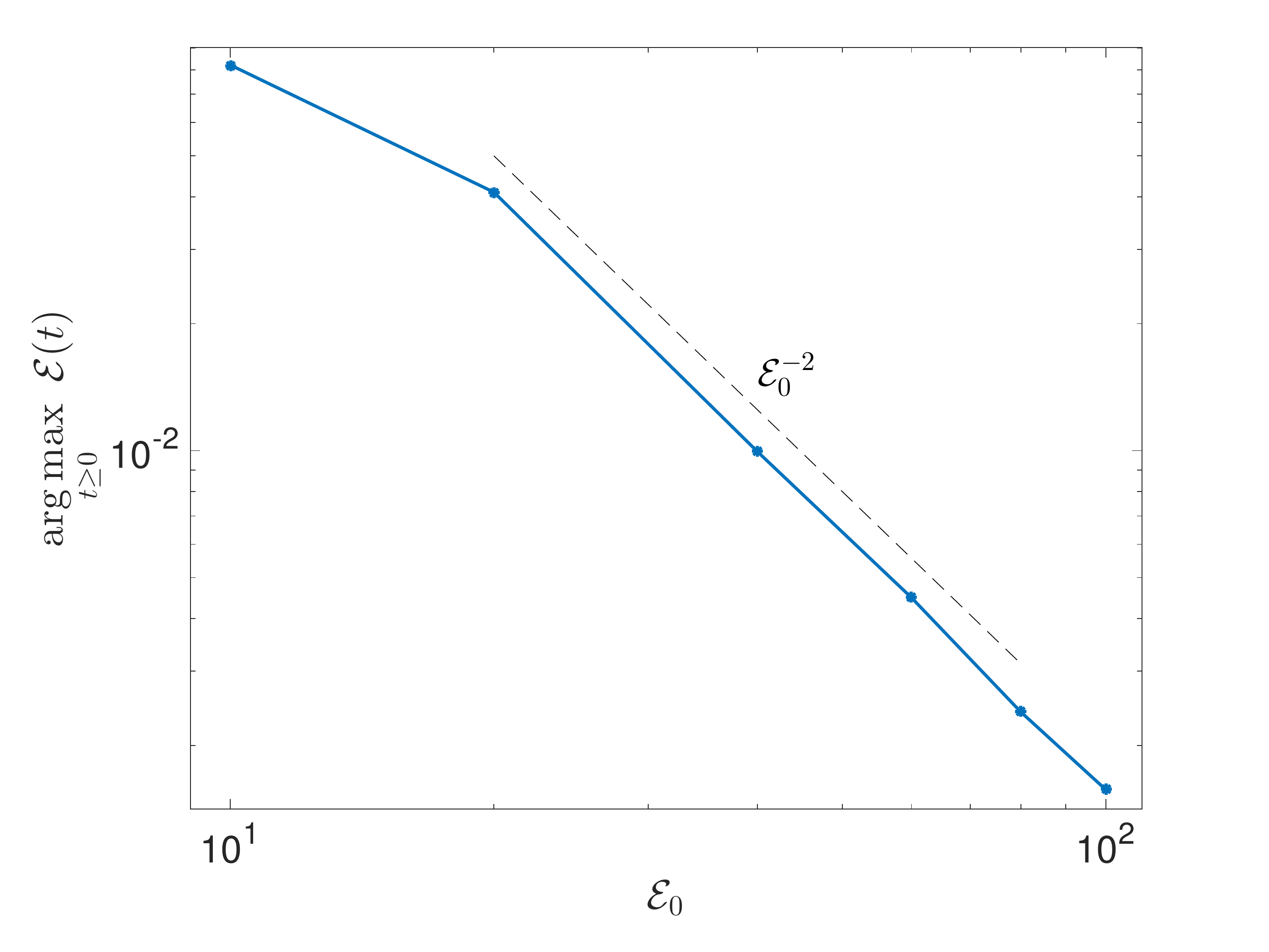}}\\
\subfigure[]{\includegraphics[width=0.48\textwidth]{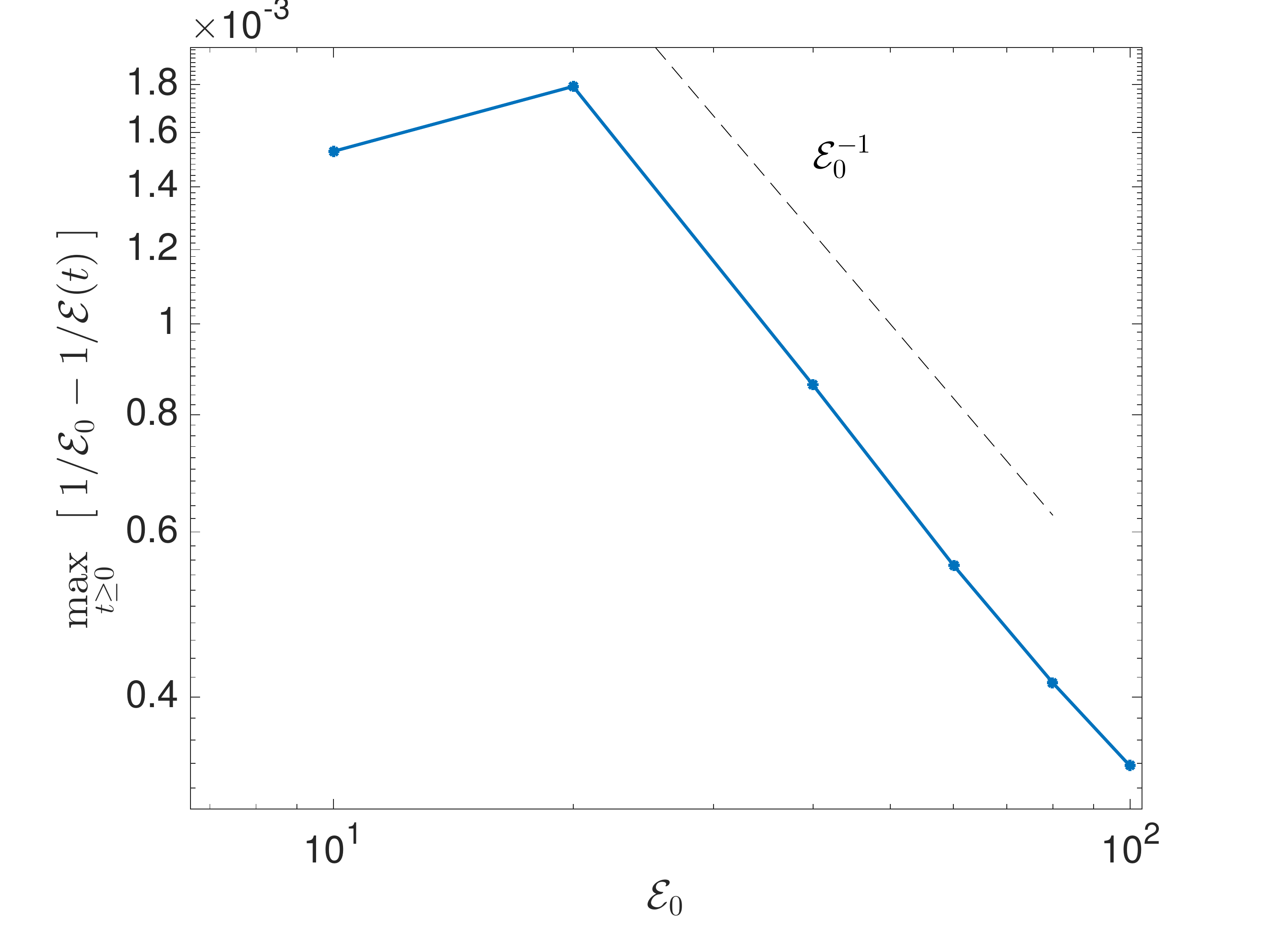}}
\subfigure[]{\includegraphics[width=0.48\textwidth]{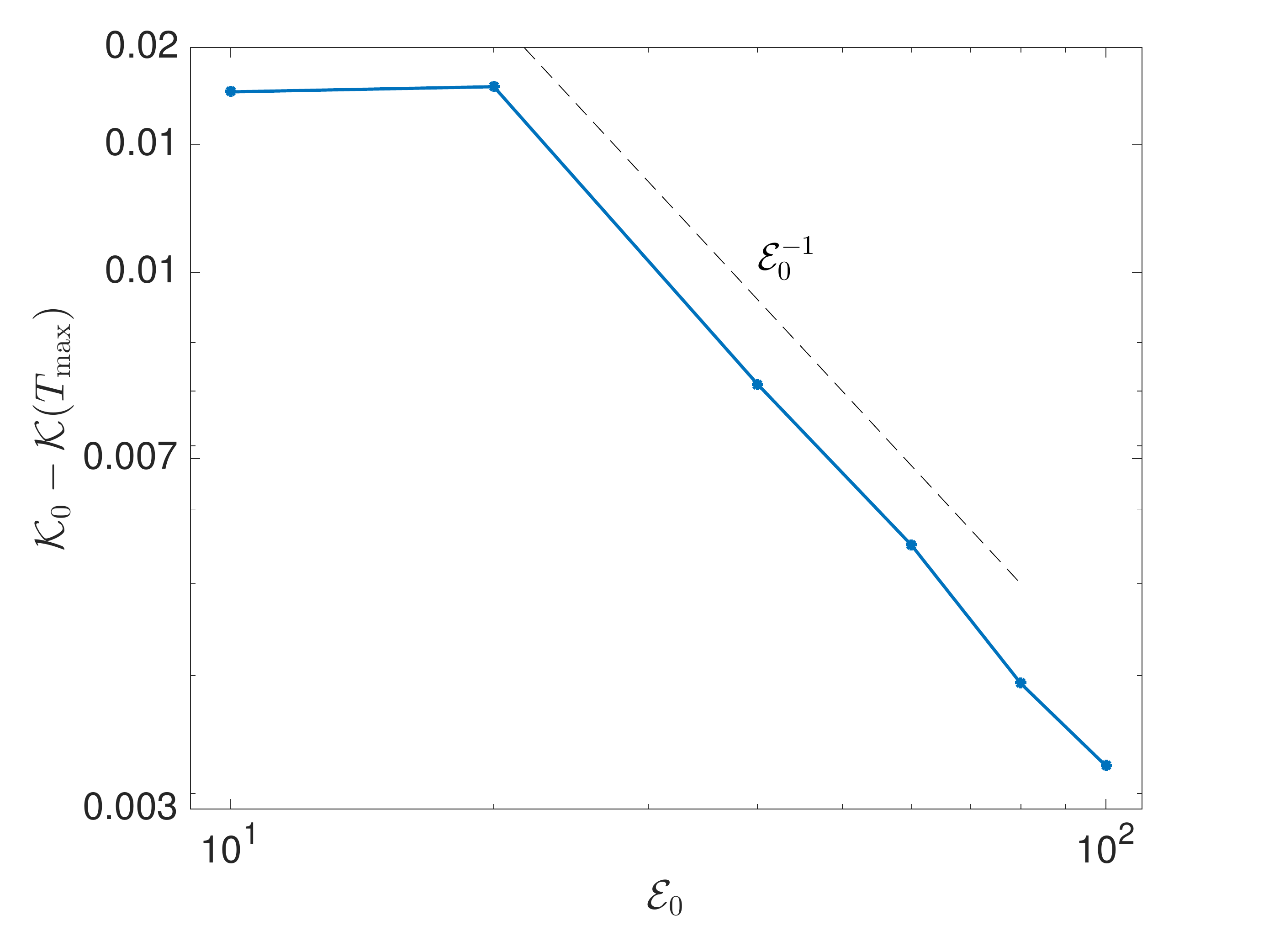}}
\caption[Maximum $\E(t)$ in finite time, $(\E_0)$-Constraint (3D)]{
  Dependence on $\E_0$ of (a) the maximum enstrophy increase over
  finite time $\delta\E$, (b) the time $T_{\max}$ when the enstrophy
  maximum is attained, (c) the maximum achieved by the LHS of estimate
  \eqref{eq:Evs_t_fixE}, cf.~\eqref{eq:fga}, and (d) the energy
  dissipation during $[0,T_{\max}]$; all data corresponds to the
    time evolution starting from the extreme vortex states
  $\tuvecE$.}
\label{fig:Emax_vsE0_fixE}
\end{center}
\end{figure}

We further characterize the time evolution by showing the maximum
enstrophy increase $\delta\E := \mathop{\max}_{t \geq 0} \, \{ \E(t) -
\E(0) \}$ and the time when the maximum is achieved $T_{\max} :=
\mathop{\arg\max}_{t \geq 0}\, \E(t)$ as functions of $\E_0$ in
figures \ref{fig:Emax_vsE0_fixE}(a) and \ref{fig:Emax_vsE0_fixE}(b),
respectively. In both cases approximate power laws in the
form
\begin{displaymath}
\delta\E  \sim  \E^{\alpha_1}_0, \quad \alpha_1 = 0.95 \pm 0.06 \qquad\mbox{and}\qquad 
T_{\max}  \sim  \E^{\alpha_2}_0, \quad \alpha_2  = -2.03 \pm 0.02   
\end{displaymath}
are detected in the limit $\E_0 \rightarrow \infty$ (as
regards the second result, we remark that $T_{\max}$ is not equivalent
to the time until which the enstrophy grows at the sustained rate
proportional to $\E_0^3$, cf.~figure \ref{fig:fg}).  To complete
presentation of the results, the dependence of the quantities
\begin{displaymath}
\mathop{\max}_{t \geq 0} \, \left\{\frac{1}{\E_0} - \frac{1}{\E(t)}\right\} \qquad\mbox{and}\qquad [\K(0) - \K(T_{\max})]
\end{displaymath}
on the initial enstrophy $\E_0$ is shown in figures
\ref{fig:Emax_vsE0_fixE}(c) and \ref{fig:Emax_vsE0_fixE}(d),
respectively. It is observed that both quantities
approximately exhibit a power-law behaviour of the form
$\E^{-1}_0$.  Discussion of these results in the context of the
estimates recalled in \S \ref{sec:intro} is presented in the next
section.

\section{Discussion}
\label{sec:discuss}

In this section we provide some comments about the results reported in
\S\S \ref{sec:3D_InstOpt_E0to0}, \ref{sec:3D_InstOpt_E} and
\ref{sec:timeEvolution}. First, we need to mention that our
gradient-based approach to the solution of optimization problem
\ref{pb:maxdEdt_E} can only yield local maximizers and, due to
nonconvexity of the problem, it is not possible to guarantee a priori
that the maximizers found are global.  To test for the possible
  presence of branches other than the ones found using the
  continuation approach described in \S \ref{sec:3D_InstOpt}, cf.
  Algorithm \ref{alg:optimAlg}, we tried to find new maximizers by
  initializing the gradient iterations \eqref{eq:desc} with different
  initial guesses $\uvec^0$. They were constructed as solenoidal
  vector fields with prescribed regularity and random structure, which
  was achieved by defining the Fourier coefficients of $\uvec^0$ as
  $\widehat{\uvec}^0(\kvec) = F(|\kvec|)e^{i\phi(\kvec)}$ with the
  amplitude $F(|\kvec|) \sim 1/|\kvec|^2$ and the phases $\phi(\kvec)$
  chosen as random numbers uniformly distributed in $[0,2\pi]$.
  However, in all such tests conducted for $\E_0 = O(1)$ the gradient
  optimization algorithm \eqref{eq:desc} would always converge to
  maximizers $\tuvecE$ belonging to one of the branches discussed in
  \S \ref{sec:3D_InstOpt_E} (modulo possible translations in the
  physical domain). While far from settling this issue definitely,
  these observations lend some credence to the conjecture that the
  branch identified in \S \ref{sec:3D_InstOpt_E} corresponds in fact
  to the global maximizers. These states appear identical to the
  maximizers found by \cite{ld08} and our search has also yielded two
  additional branches of locally maximizing fields, although we did
  not capture the lower branch reported by \cite{ld08}.  However,
  since that branch does not appear connected to any state in the
  limit $\E_0 \rightarrow 0$, we speculate that it might be an
  artifact of the ``discretize-then-optimize'' formulation used by
  \cite{ld08}, in contrast to the ``optimize-then-discretize''
  approach employed in our study which provides a more direct control
  over the analytic properties of the maximizers.  We add that the
  structure of the maximizing branches found here is in fact quite
  similar to what was discovered by \cite{ap13a} in an analogous
  problem in 2D. Since the 2D problem is more tractable from the
  computational point of view, in that case we were able to undertake
  a much more thorough search for other maximizers which did not
  however yield any solutions not associated with the main branches.

\begin{figure}
\begin{center}
\includegraphics[width=0.9\textwidth]{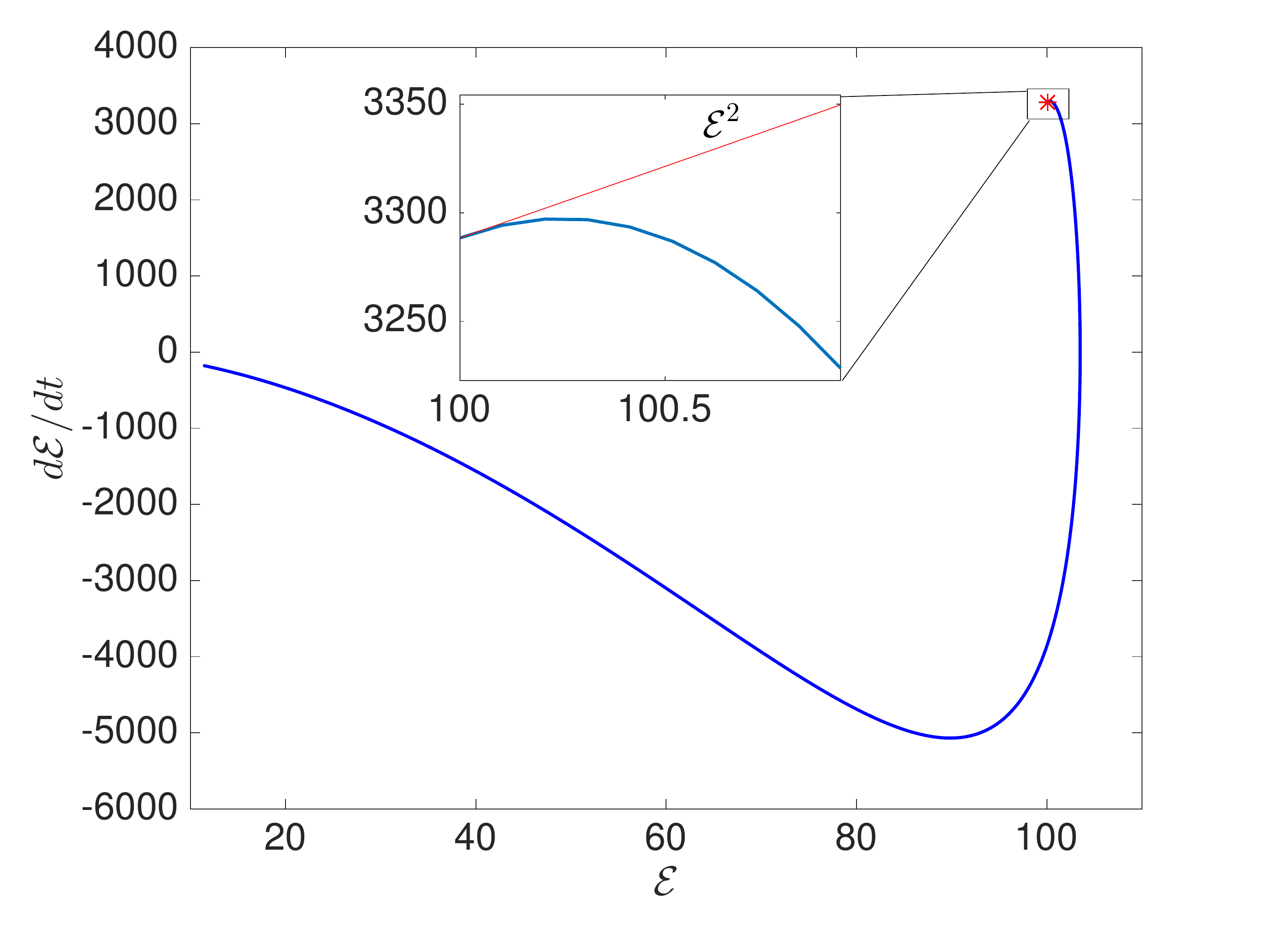}
\caption{Trajectory of the flow corresponding to the initial condition
  $\tuvecE$ with $\E_0 = 100$ in the coordinates $\{\E,d\E/dt\}$.
  For comparison, in the inset the thin line represents the
    borderline growth at the rate $d\E/dt \sim \E^2$}.
\label{fig:dEdtE}
\end{center}
\end{figure}

The results reported in \S \ref{sec:3D_InstOpt_E} and
\S\ref{sec:timeEvolution} clearly exhibit two distinct behaviours,
depending on whether or not global-in-time regularity can be
guaranteed a priori based on estimates
\eqref{eq:Evs_t_fixE}--\eqref{eq:K0E0}.  These differences are
manifested, for example, in the power laws evident in figures
\ref{fig:ScalingLaws_fixE} and \ref{fig:Emax_vsE0_fixE}, as well as in
the different behaviours of the RHS of estimate \eqref{eq:Evs_t_fixE}
with respect to time in figures \ref{fig:fg}(a) and \ref{fig:fg}(b).
However, for the initial data for which global-in-time regularity
cannot be ensured a priori there is no evidence of sustained growth of
enstrophy strong enough to signal formation of singularity in finite
time. Indeed, in figure \ref{fig:Emax_vsE0_fixE}(c) one sees that the
quantity $\mathop{\max}_{t \geq 0} \, \left\{1/\E_0 - 1/\E(t)\right\}$
behaves as $C_1 / \E_0$, where $C_1 < 1$, when $\E_0$ increases,
revealing no tendency to approach $1/\E_0$ which is a necessary
precursor of a singular behaviour (cf.~discussion in \S
\ref{sec:timeEvolution}). To further illustrate how the rate of growth
of enstrophy achieved initially by the maximizers $\tuvecE$ is
depleted in time, in figure \ref{fig:dEdtE} we show the flow evolution
corresponding to $\tuvecE$ with $\E_0 = 100$ as a trajectory in the
coordinates $\{\E, d\E/dt\}$.  From the discussion in Introduction we
know that in order for the singularity to occur in finite time, the
enstrophy must grow at least at a sustained rate $d\E / dt \sim
\E^\alpha$ for some $\alpha > 2$.  In other words, a ``blow-up
trajectory'' will be realized only if the trajectory of the flow,
expressed in $\{\E,d\E/dt \}$ coordinates, is contained in the region
$\mathcal{M} = \{ (\E,d\E/dt) \; : \; C_1\E^2 < d\E/dt \leq C_2\E^3
\}$, for some positive constants $C_1$ and $C_2$. For the flow
corresponding to the instantaneous optimizers $\tuvecE$, the initial
direction of a trajectory in $\{\E,d\E/dt \}$ coordinates is
determined by the vector $\vvec = \left[1,
  \left.\tfrac{d\R}{d\E}\right|_{\E_0} \right]$ and, for initial
conditions $\uvec_0$ satisfying $\R(\uvec_0) = C\E^3(\uvec_0)$, it
follows that
\begin{equation*}
\left.\frac{d\R}{d\E}\right|_{\E_0}  = 3C\E_0^2.
\end{equation*}
Since the optimal rate of growth is sustained only over a short
interval of time, the trajectory of the flow in the $\{\E,d\E/dt\}$
coordinates approaches the region $\mathcal{M}$ only tangentially
following the direction of the lower bound $C_1\E^2$, and remains
outside $\mathcal{M}$ for all subsequent times.  This behaviour is
clearly seen in the inset of figure \ref{fig:dEdtE}.

\begin{figure}
\begin{center}
\includegraphics[width=0.6\textwidth]{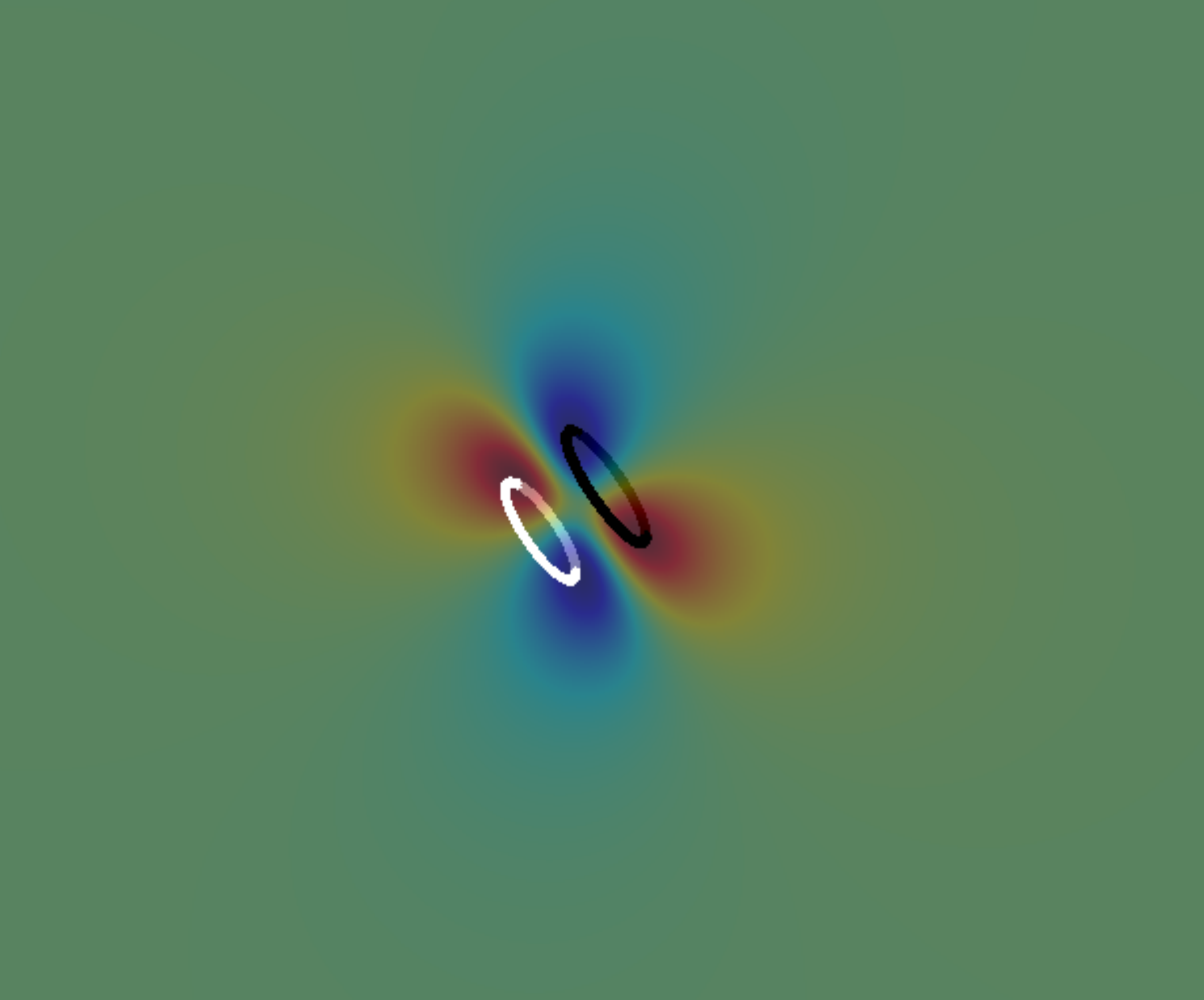}
\caption{Vortex lines inside the region with the strongest
  vorticity in the extreme vortex state $\tuvecE$ with $\E_0 = 100$.
  The colour coding of the vortex lines is for identification purposes
  only.}
\label{fig:ring}
\end{center}
\end{figure}

An interpretation of this behaviour can be proposed based on equation
\eqref{eq:dKdt_system} from which it is clear that the evolution of
the flow energy is closely related to the growth of enstrophy. In
particular, if the initial energy $\K(0)$ is not sufficiently
large, then its depletion due to the initial growth of enstrophy may
render the flow incapable of sustaining this growth over a longer
period of time. This is in fact what seems to be happening in the
present problem as evidenced by the data shown in figure
\ref{fig:K0E0}. We remark that, for a prescribed enstrophy $\E(0)$,
the flow energy cannot be increased arbitrarily as it is upper-bounded
by Poincar\'e's inequality {$\K(0) \le (2\pi)^2 \E(0)$}. This behaviour
can also be understood in terms of the geometry of the extreme vortex
states $\tuvecE$.  Figure \ref{fig:ring} shows a magnification of the
pair of vortex rings corresponding to the optimal field $\tuvecE$ with
$\E_0=100$. It is observed that the vorticity field $\rot\tuvecE$
inside the vortex core has an azimuthal component only which exhibits
no variation in the azimuthal direction. Thus, in the limit $\E_0
\rightarrow \infty$ the vortex ring shrinks with respect to the domain
$\Omega$ (cf.~figure \ref{fig:ScalingLaws_fixE}(d)) and the field
$\tuvecE$ ultimately becomes axisymmetric (i.e., in this limit
boundary effects vanish). At the same time, it is known that the 3D
Navier-Stokes problem on an unbounded domain and with axisymmetric
initial data is globally well posed \citep{k03}, a results which is a
consequence of the celebrated theorem due to \citet*{ckn83}.

We close this section by comparing the different power laws
characterizing the maximizers $\tuvecE$ and the corresponding flow
evolutions with the results obtained in analogous studies of extreme
behaviour in 1D and 2D (see also Table \ref{tab:estimates}).
First, we note that the finite-time growth of enstrophy $\delta\E$ in
3D, cf.~figure \ref{fig:Emax_vsE0_fixE}(a), exhibits the same
dependence on the enstrophy $\E_0$ of the instantaneously optimal
initial data as in 1D, i.e., is directly proportional to $\E_0$ in
both cases \citep{ap11a}.  This is also analogous to the maximum
growth of palinstrophy $\P$ in 2D which was found by \cite{ap13a} to
scale with the palinstrophy $\P_0$ of the initial data, when the
instantaneously optimal initial condition was computed subject to {\em
  one} constraint only (on $\P_0$). When the instantaneously optimal
initial data was determined subject to {\em two} constraints, on
$\K_0$ and $\P_0$, then the maximum finite-time growth of palinstrophy
was found to scale with $\P_0^{3/2}$ \citep{ap13b}. On the other hand,
the time $T_{\max}$ when the maximum enstrophy is attained, cf.~figure
\ref{fig:Emax_vsE0_fixE}(b), scales as $\E_0^{-2}$, which should be
contrasted with the scalings $\E_0^{-1/2}$ and $\P_0^{-1/2}$ found in
the 1D and 2D cases, respectively.  This implies that the time
interval during which the extremal growth of enstrophy is
sustained in 3D is shorter than the corresponding intervals in 1D and
2D.

\section{Conclusions and Outlook}
\label{sec:final}

By constructing the initial data to exhibit the most extreme behaviour
allowed for by the mathematically rigorous estimates, this study
offers a fundamentally different perspective on the problem of
searching for potentially singular solutions from most earlier
investigations.  Indeed, while the corresponding flow evolutions did
not reveal any evidence for finite-time singularity formation, the
initial data obtained by maximizing $d\E/dt$ produced a significantly
larger growth of enstrophy in finite time than any other candidate
initial conditions (cf.~Table \ref{tab:InitialConditions} and figure
\ref{fig:Fvs_t_fixE}). Admittedly, this observation is limited to the
initial data with $\E_0 \le 100$, which corresponds to Reynolds
  numbers $Re = \sqrt{\E_0\, \Lambda} / \nu \lessapprox 450$ lower
  than the Reynolds numbers achieved in other studies concerned with
  the extreme behaviour in the Navier-Stokes flows
  \citep{opc12,k13,dggkpv13,opmc14}. Given that the definitions of the
  Reynolds numbers applicable to the various flow configurations
  considered in these studies were not equivalent, it is rather
  difficult to make a precise comparison in terms of specific
  numerical values, but it is clear that the largest Reynolds numbers
  attained in these investigations were at least an order of magnitude
  higher than used in the present study; for Euler flows such a
  comparison is obviously not possible at all.  However, from the
mathematical point of view, based on estimates
\eqref{eq:dEdt_estimate_E}--\eqref{eq:K0E0}, there is no clear
indication that a very large initial enstrophy $\E_0$ (or,
equivalently, a high Reynolds number) should be a necessary condition
for singularity formation in finite time. In fact, blow-up cannot be a
priori ruled out as soon as condition \eqref{eq:K0E0} is violated,
which happens for all initial data lying on the gray region of the
phase space in figure \ref{fig:K0E0}.  We remark that additional
results were obtained (not reported in this paper) by studying the
time evolution corresponding to the optimal initial data $\tuvecE$,
but using smaller values of the viscosity coefficient $\nu=10^{-3},
10^{-4}$, thereby artificially increasing the Reynolds number at the
price of making the initial data suboptimal. Although these
  attempts did increase the amplification of enstrophy as
    compared to what was observed in figures \ref{fig:fg} and
    \ref{fig:Emax_vsE0_fixE}, no signature of finite-time singularity
  formation could be detected either.

Our study confirmed the earlier findings of \cite{ld08} about the
sharpness of the instantaneous estimate \eqref{eq:dEdt_estimate_E}. We
also demonstrated that the finite-time estimate \eqref{eq:Evs_t_fixE}
is saturated by the flow evolution corresponding to the optimal
initial data $\tuvecE$, but only for short times, cf.~figure
\ref{fig:fg}, which are not long enough to trigger a
singular behaviour.

In \S \ref{sec:discuss} we speculated that a relatively small initial
energy $\K(0)$, cf.~figure \ref{fig:RvsE0_FixE_large}(b), might be the
property of the extreme vortex states $\tuvecE$ preventing the
resulting flow evolutions from sustaining a significant growth of
enstrophy over long times. On the other hand, in Introduction we
showed that estimate \eqref{eq:dEdt_estimate_E} need not be saturated
for blow-up to occur in finite time and, in fact, sustained growth at
the rate $d\E/dt = C \, \E^\alpha$ with any $\alpha > 2$ will also
produce singularity in finite time. Thus, another strategy to
construct initial data which could lead to a more sustained growth of
enstrophy in finite time might be to increase its kinetic energy by
allowing for a smaller instantaneous rate of growth (i.e., with an
exponent $2 < \alpha \le 3$ instead of $\alpha = 3$).  This can be
achieved by prescribing an additional constraint in the formulation of
the variational optimization problem, resulting in
\begin{problem}\label{pb:maxdEdt_KE}
\begin{eqnarray*}
\tuvecKE & = & \mathop{\arg\max}_{\uvec\in\M{\K_0,\E_0}} \, \R(\uvec) \\
\M{\K_0,\E_0} & = & \left\{\uvec\in H_0^2(\Omega)\,\colon\,\nabla\cdot\uvec = 0, \; \K(\uvec) = \K_0, \; \E(\uvec) = \E_0 \right\}.
\end{eqnarray*} 
\end{problem}
\noindent
It differs from problem \ref{pb:maxdEdt_E} in that the maximizers are
sought at the intersection of the original constraint manifold
$\M{\E_0}$ and the manifold defined by the condition $\K(\uvec)
  = \K_0$, where $\K_0 \le (2\pi)^2 \E_0$ is the prescribed energy.
While computation of such maximizers is more complicated, robust
techniques for the solution of optimization problems of this
  type have been developed and were successfully used in the 2D
setting by \cite{ap13a}. Preliminary results obtained in the
  present setting by solving problem \ref{pb:maxdEdt_KE} for $\K_0 =
1$ are indicated in figure \ref{fig:K0E0}, where we see that the flow
evolutions starting from $\tuvecKE$ do not in fact produce a
significant growth of enstrophy either. An alternative, and arguably
more flexible approach, is to formulate this problem in terms of
multiobjective optimization \citep{k99} in which the objective
function $\R(\uvec)$ in problem \ref{pb:maxdEdt_E} would be replaced
with
\begin{equation}
\R_{\eta}(\uvec) := \eta\, \R(\uvec) + (1-\eta)\, \K(\uvec),
\label{eq:Rm}
\end{equation}
where $\eta \in [0,1]$. Solution of such a multiobjective optimization
problem has the form of a ``Pareto front'' parameterized by $\eta$.
Clearly, the limits $\eta \rightarrow 1$ and $\eta \rightarrow 0$
correspond, respectively, to the extreme vortex states already
found in \S \ref{sec:3D_InstOpt_E0to0} and \S \ref{sec:3D_InstOpt_E},
and to the Poincar\'e limit.  Another interesting possibility
is to replace the energy $\K(\uvec)$ with the helicity {$\H(\uvec) :=
  \int_{\Omega} \uvec\cdot(\bnabla\times\uvec)\,d\Omega$} in the
multiobjective formulation \eqref{eq:Rm}, as this might allow one to
obtain extreme vortex states with a more complicated topology (i.e., a
certain degree of ``knottedness''). We note that all the
extreme vortex states found in the present study were ``unknotted'',
i.e., were characterized by $\H(\tuvecE) = 0$, as the vortex rings
were in all cases disjoint (cf.~figure \ref{fig:ring}).

Finally, another promising possibility to find initial data producing
a larger growth of enstrophy is to solve a {\em finite-time}
optimization problem of the type already studied by \cite{ap11a} in
the context of the 1D Burgers equation, namely
\begin{problem}\label{pb:maxdE}
\begin{equation*}
\tilde{\uvec}_{0;\E_0,T}  = \mathop{\arg\max}_{\uvec_0\in\M{\E_0}} \, \E(T), 
\end{equation*} 
\end{problem}
\noindent
where $T>0$ is the length of the time interval of interest and
$\uvec_0$ the initial data for the Navier-Stokes system
\eqref{eq:NSE3D}. In contrast to problems \ref{pb:maxdEdt_E} and
\ref{pb:maxdEdt_KE}, solution of problem \ref{pb:maxdE} is more
complicated as it involves flow evolution. It represents therefore a
formidable computational task for the 3D Navier-Stokes system.
However, it does appear within reach given the currently available
computational resources and will be studied in the near future.

\section*{Acknowledgements}

The authors are indebted to Charles Doering for many enlightening
discussions concerning the research problems studied in this work. The
authors are grateful to Nicholas Kevlahan for making his parallel
Navier-Stokes solver available, which was used to obtain the results
reported in \S\ref{sec:timeEvolution}. Anonymous referees
  provided many insightful comments which helped us improve this
  work.  This research was funded through an Early Researcher Award
(ERA) and an NSERC Discovery Grant, whereas the computational time was
made available by SHARCNET under its Dedicated Resource Allocation
Program.  Diego Ayala was funded in part by NSF Award DMS-1515161 at
the University of Michigan and by the Institute for Pure and Applied
Mathematics at UCLA.



\end{document}